\documentclass[a4paper,fleqn,usenatbib]{mnras}
\pdfoutput=1
\usepackage{newtxtext,newtxmath}
\usepackage{array}

\usepackage[T1]{fontenc}
\usepackage{ae,aecompl}
\usepackage{slantsc}

\usepackage{subfig}
\usepackage{graphicx}	
\usepackage{amsmath}	
\usepackage{amssymb}	
\let\oldAA\AA
\renewcommand{\AA}{\text{\normalfont\oldAA}}

\title[Feedback in a high metallicity galaxy at $z = 2.9$]{The MUSE 3D view of feedback in a high-metallicity radio galaxy at $z = 2.9$}

\author[Silva, M. et al.]{
	M. Silva$^{1,2}$\thanks{E-mail: marckelson.silva@astro.up.pt},
	A. Humphrey$^{1}$,
	P. Lagos$^{1}$,
	M. Villar-Mart\'in$^{3}$,
	S. G. Morais$^{1,2}$,
	\newauthor
	S. di Serego Alighieri$^{4}$,
	A. Cimatti$^{5}$,
	R. Fosbury$^{6}$,
	R. A. Overzier$^{7}$,
	\newauthor
	J. Vernet$^{8}$,
	and L. Binette$^{9}$
	\\
	$^{1}$Institute of Astrophysics and Space Sciences, Universidade do Porto, CAUP, Rua das Estrelas, 4150-762 Porto, Portugal.\\
	$^{2}$Departamento de F\'isica e Astronomia, Faculdade de Ci\^encias, Universidade do Porto, R. Campo Alegre 687, 4169-007 Porto, Portugal\\
	$^{3}$Departamento de Astrof\'isica, Centro de Astrobiolog\'ia, Ctra de Torrej\'on a Ajalvir km 4, 28850, Madrid, Spain\\
	$^{4}$INAF - Osservatorio Astrofisico di Arcetri, Largo E. Fermi 5, I-50125 Firenze, Italy\\
	$^{5}$Dipartimento di Astronomia, Universit\`a di Bologna, Via Ranzani 1, I-40127 Bologna, Italy\\
	$^{6}$European Southern Observatory, Karl Schwarzschild Str. 2, D-85748 Garching, Germany\\
	$^{7}$Observat\'orio Nacional, Rua Jos\'e Cristino 77, S\~{a}o Cristov\~{a}o, Rio de Janeiro-RJ, 20921-400, Brazil\\
	$^{8}$European Southern Observatory, Karl-Schwarzschild-Str. 2, D-85748 Garching \\
	$^{9}$Instituto de Astronom\'ia, Universidad Nacional Aut\'onomo de M\'exico, Ap. 70-264, 04510 M\'exico D.F., M\'exico
}

\date{Accepted XXX. Received YYY; in original form ZZZ}

\pubyear{2017}

\begin{document}
\label{firstpage}
\pagerange{\pageref{firstpage}--\pageref{lastpage}}
\maketitle

\begin{abstract}
We present a detailed study of the kinematic, chemical and excitation properties of the giant Ly$\alpha$ emitting nebula and the giant \ion{H}{I} absorber associated with the $z = 2.92$ radio galaxy MRC 0943--242, using spectroscopic observations from VLT/MUSE, VLT/X-SHOOTER and other instruments. Together, these data provide a wide range of rest-frame wavelength (765 \AA$\,$ -- 6378 \AA$\,$ at $z = 2.92$) and 2D spatial information. 
We find clear evidence for jet gas interactions affecting the kinematic properties of the nebula, with evidence for both outflows and inflows being induced by radio-mode feedback. We suggest that the regions of relatively lower ionization level, spatially correlated with the radio hotspots, may be due to localised compression of photoionized gas by the expanding radio source, thereby lowering the ionization parameter, or due to a contribution from shock-heating.
We find that photoionization of super-solar metallicity gas ($Z/Z_{\odot}$ = 2.1) by an AGN-like continuum ($\alpha$=--1.0) at a moderate ionization parameter ($U$ = 0.018) gives the best overall fit to the complete X-SHOOTER emission line spectrum. 
We identify a strong degeneracy between column density and Doppler parameter such that it is possible to obtain a reasonable fit to the \ion{H}{I} absorption feature across the range log N(\ion{H}{I}/cm$^{-2}$) = 15.20 and 19.63, with the two best-fitting occurring near the extreme ends of this range. The extended \ion{H}{I} absorber is blueshifted relative to the emission line gas, but shows a systematic decrease in blueshift towards larger radii, consistent with a large scale expanding shell.  
\end{abstract}

\begin{keywords}
galaxies: evolution -- galaxies: high-redshift -- galaxies: individual: MRC 0943--242 -- galaxies: active -- galaxies: ISM -- galaxies: quasars: absorption lines -- galaxies: quasars: emission lines
\end{keywords}



\section{Introduction}

The study of high-redshift galaxies ($z \ga 2$) allows us to look back to the young Universe, giving us the opportunity to witness and understand the processes by which galaxies form and evolve. In particular, powerful active galaxies such as high-z radio galaxies (HzRGs) or quasars offer the chance to examine a crucial phase in the evolution of massive galaxies, when the host galaxy is still relatively gas-rich and when radio and quasar mode feedback is also taking place. As such, HzRGs represent laboratories for studying the interplay between the "ingredients" of galaxy formation, such as active galactic nuclei (AGN), gas, stars, feedback and other processes.

Many powerful HzRGs (or quasars) lie within giant haloes\footnote{These haloes are sometimes also known as the 'extended emission line region' (EELR), the 'extended narrow line region' (ENLR), or 'Ly$\alpha$ blobs' (LAB)} of gas which are prodigious sources of Ly$\alpha$ (and other emission lines), and which have sizes often exceeding 100 kpc \citep{fosbury1982,serego1988,Mc90,Vo,Pe98,Fr2001,Re,Hu1,Cantalupo2014,swinbank,borisova2016,Cai2017}.

These haloes have typical masses in gas of $\sim 10^{9-10}$ M$_{\sun}$, Ly$\alpha$ luminosities up to $\sim 10^{45}$ erg s$^{-1}$ and estimated electron densities of a few to several hundred cm$^{-3}$ \citep{Mc93,Re,VM1,Sa,Hu1,Cantalupo2014,Cai2017}. Typically the haloes can be divided into two kinematic and structural components: a high surface brightness region showing a clumpy, irregular morphology \citep{Re} often aligned with the radio jets \citep{Mc95} and characterised by extreme kinematics (FWHM $>$ 1000 km s$^{-1}$; \citealt{Vo,VM1,Hu2}), and a low surface brightness region sometimes extending beyond the maximum extent of the radio structures, showing relatively quiescent kinematics (FWHM $\la$ 1000 km s$^{-1}$) and no clear relationship with the radio source \citep{Vo96,Vo,VM4,Sa}. In addition, \cite{Hu3} proposed that the quiescent gas of some HzRGs are in infall towards the central regions of the host galaxy, though it is not clear whether the infalling gas has an external origin, or is instead a back-flow of material after the passage of a large scale outflow.

The ionization of the Ly$\alpha$ haloes of HzRGs is not fully understood, but several different mechanisms have been proposed, including photoionisation by the central AGN \citep[see][]{Mc90,VM1,geach2009,kollmeier2010,Cai2017}, cooling radiation \citep[see][]{Haiman,steidel,fardal,yang2006,dijkstra2006,faucher2010,rosdahl2012}, shocks driven by radio jets \citep[see][]{emonts2005,VM2,Hu6}, or resonant scattering (for Ly$\alpha$) \citep[see][]{gould1996,VM2,dijkstra2006,loeb2009,Hu5,Cantalupo2014,yang2014}. 

Other structures that are also not fully understood are the extended Ly$\alpha$ absorbers that are associated with roughly half of all radio galaxies at $z > 2$ \citep{Vo,Ja,Wm}. They are at least as spatially extended as the Ly$\alpha$ haloes and appear to have column densities in the range $\sim10^{14-20}$ cm$^{-2}$ \citep{Ro95,Vo,Bi1,Ja,Wm,Bi2}, with properties suggestive of an expanding shell with covering factor of, or approaching, unity \citep[see][]{Bi1,Ja,Wm,Bi2}. It has been suggested that these giant absorbing structures are the product of a past feedback event in the galaxy \citep[e.g.][]{Bi3}, and that their properties are somehow related to the size or passage of the radio jets \citep{Vo}. The presence of this type of shell around some HzRGs may have important implications for the escape of Ly$\alpha$ and ionizing photons.

In this paper, we present previously unpublished VLT X-SHOOTER\footnote{We abbreviate X-SHOOTER to 'XSH' in some tables and figures in the interest of brevity.} spectroscopy of the $z = 2.92$ HzRG MRC 0943--242, and pool this data with archival spectroscopic data from other VLT instruments (MUSE and UVES), spectra from the AAT and Keck II, in order to conduct a detailed study of the spatially-resolved kinematics and ionization of this galaxy's large-scale Ly$\alpha$ emitting halo and its large-scale \ion{H}{I} absorber. \citet{Gu} presented a primarily morphological study of this object combining the MUSE dataset with multi-wavelength data (see Section 2 below); here we present a detailed study of the kinematic and ionization properties of the Ly$\alpha$ halo and the large-scale Ly$\alpha$ absorbing structure. 

The paper is organised as follows. In $\S$~\ref{prev} we describe the key previous results from the literature concerning this well studied HzRG. In $\S$~\ref{obs}, we introduce the selected observations and data reduction for MRC 0943--242. In $\S$~\ref{analysis}, we discuss our observational results. In $\S$~\ref{discussion}, we discuss the physical properties and several powering mechanisms that could be responsible for such extended \ion{Ly}{$\alpha$} emission and \ion{Ly}{$\alpha$} absorber. In $\S$~\ref{conclusions}, we give a brief summary concluding our results. 
A $\Omega _{\Lambda}$ = 0.713, $\Omega _{m}$ = 0.287 and $H_{\circ}$ = 69.3 km s$^{-1}$ Mpc$^{-1}$ \citep{wmap} cosmology is adopted in this paper so that 1\arcsec$\,$ corresponds to 7.94 kpc at the redshift of our target ($z = 2.92$).

\section{MRC 0943-242: previous results}
\label{prev}

MRC 0943-242 was first catalogued in the Molonglo Reference Catalogue \citep{large1981}, and selected as a possible HzRG on the basis of its ultra steep radio spectrum by \citet{Ro94}. Optical imaging and spectroscopy confirmed the high redshift of this galaxy ($z = 2.92$; \citealt{Ro95}). 

The radio continuum of MRC 0943--242 has a simple double-morphology (Fanaroff-Riley Class II), consisting of two bright hotspots separated by 3.9\arcsec,  with no core detected, at 1.4, 4.7 and 8.2 GHz \citep{Ro95,carilli1997}. Its spectral index between 1.5 GHz and 30 GHz is $\alpha$ = -- 1.44 \citep{emonts2011}. With a 1.4 GHz radio power of 10$^{35.4}$ erg s$^{-1}$ Hz$^{-1}$ \citep{Bre2000}, this is among the most radio-luminous galaxies known. 

Rest frame ultraviolet and optical HST images of MRC 0943--242 show a bright, elongated and curved morphology, with a close alignment between the major axis of this emission and that of the radio source \citep{Pe1999,Pe2001}. Using Keck II spectropolarimetry, \citet{Ve2001} found the extended UV continuum emission along the radio/optical axis to be significantly polarized ($P_\% = 6.6 \pm 0.9 \%$ at $\sim$1250 -- 1400 \AA) with the electric field vector approximately perpendicular to this axis, indicating a substantial contribution from scattered AGN continuum (22 -- 66 $\%$), but still allowing a potentially significant contribution from young stars (14 -- 53 $\%$). Along the radio/optical axis, \citet{Ve2001} also estimated that nebular continuum emission contributes around 20 $\%$ of the UV continuum. Ground-based K-band imaging shows a rounded and more centrally concentrated morphology than seen in the optical images \citep{vanBre98}. Moreover, studying the contribution of the host galaxy stellar emission at rest-frame H band, \citet{seymour2007} estimated the stellar luminosity of the radio galaxy which implied stellar mass of log(M$_{\star}$/M$_{\odot}$) = 11.22$^{+0.15} _{-0.07}$. This HzRG also appears to be located in a proto-cluster \citep{venemans2007}. 

Spectroscopic studies have revealed that MRC 0943--242 is embedded within a giant Ly$\alpha$ halo \citep{Ro95,Ro97,Bi1,VM1,Ja,venemans2007,Gu}. Using Keck II spectroscopy, \cite{VM1}  found that the Ly$\alpha$ halo is metal-enriched and has a high surface brightness, kinematically perturbed region (FWHM $\ga$ 1000 km s$^{-1}$) within the spatial extent of the radio structure, and a giant low surface brightness region with quieter kinematics (FWHM $\la$ 600 km s$^{-1}$) surrounding the entire object. \citet{VM1} and \citet{Hu2} argued that the close spatial association between the kinematically perturbed gas and the radio structures suggests that jet-gas interactions are responsible for the kinematic perturbation in the high surface brightness regions. Furthermore, \citet{Hu2} found that the kinematically more perturbed gas has a lower level of ionization than the quiescent gas, possibly due to the impact of shocks on the former. 

\citet{Ro95} detected a strong and spatially extended absorption feature in the profile of the Ly$\alpha$ line. By fitting the Ly$\alpha$ velocity profile, \citet{Ro95}, \citet{Bi1}, \citet{Ja} and \citet{Gu} have estimated the \ion{H}{I} column density of the absorbing gas to be log N(\ion{H}{I}/cm$^{-2}$) $\sim$ 19. It has been suggested that this absorbing structure is a large scale shell that might surround the radio galaxy and its Ly$\alpha$ halo \citep{Bi2,Gu}. Gas mass estimates for this absorbing structure range from $\gtrsim$ $10^{9}$ M$_{\odot}$ to $10^{12}$ M$_{\odot}$ \citep[see][]{Bi2,Gu}. 

In addition, \citet{Bi1} detected the \ion{C}{IV} doublet in absorption, with a column density log N(\ion{C}{IV}/cm$^{-2}$) = 14.5 $\pm$ 0.1 and a redshift close to that of the main Ly$\alpha$ absorber. Based on the ratio of column densities derived from the Ly$\alpha$ and \ion{C}{IV} absorption features, \citet{Bi1} argued that the absorbing structure has a low metallicity ($Z$ = 0.01Z$_{\sun}$) and is not co-spatial with the ionized gas responsible for the detected emission lines; the authors suggest that the absorbing gas represents material expelled from the HzRG during an earlier phase of starburst activity \citep[see][]{Bi3,Bi2}. 

\citet{Gu} presented a study of MRC 0943--242, combining ALMA sub-millimetre observations and MUSE IFU spectroscopy to perform a multiwavelength morphological study of the AGN, starburst and molecular gas components of the galaxy. They report a highly complex morphology, with a reservoir of molecular gas offset by $\sim$ 90 kpc from the AGN, and identify a linear feature that emits Ly$\alpha$, CO lines and dust continuum which they suggest may be due to an accretion flow onto the radio galaxy. Assuming the main extended absorber surrounds the HzRG and has a roughly spherical shape, \citet{Gu} estimated a total gas mass of the main Ly$\alpha$ absorber of M(\ion{H}{I}) $\gtrsim 3.8 \times 10^{9}$ M$_{\sun}$. 

In summary, the observational properties of MRC 0943--242 make it an excellent target for a detailed case study into various processes that are expected to play an important role in the evolution of massive galaxies, particularly the interplay between the "ingredients" of galaxy formation, such as AGN, gas, stars, feedback, and gas accretion. 

\section{Observations}
\label{obs}

This study makes use of deep spectroscopic observations (proprietary and archival) from several different telescopes and instruments (see Table \ref{observations}). 

\subsection{X-SHOOTER Long-Slit and IFU Spectra}

Here we present previously unpublished X-SHOOTER \citep[see][]{vernet} intermediate-resolution \'echelle spectroscopic observations of MRC 0943--242, obtained at the VLT UT3 on 2009 March 18 and May 3 -- 5, during commissioning of the instrument under the program 60.A-9022(C). The wavelength range of X-SHOOTER (3000 \AA$\,$ -- 25000 \AA) provides a continuous rest-frame wavelength range of 765 \AA$\,$ -- 6378 \AA$\,$ at $z = 2.92$, within which are expected to lie a multitude of diagnostically important emission or absorption lines. The pixel scale is $\sim$0.15\arcsec$\,$ for UVB and VIS arms, and $\sim$0.20\arcsec$\,$ for the NIR arm. 

The X-SHOOTER observations were taken in two different modes: long-slit and IFU. In long-slit mode, the slit widths were 1.0\arcsec (UVB), 0.9\arcsec (VIS), and 0.9\arcsec (NIR). In IFU mode, the 4\arcsec $\times$ 1.8\arcsec field of view was reformatted into 12\arcsec $\times$ 0.6\arcsec pseudo slits. 

The integration time using the slit mode was split into 6$\times$1500s exposures for the UVB and VIS arms, and 6$\times$500s exposures for the NIR arm, with the slit oriented at a position angle of 55$^{\circ}$. The slit observations were made in nodding mode. In IFU mode, the integration time was split into 1$\times$1700s for UVB arm, 1$\times$1700s for VIS arm, and 3$\times$600s for NIR arm, with the position angle identical to that used for the slit observations. The IFU observations were made in offset mode. 

The data were processed using ESO's X-SHOOTER reduction pipeline, which performs bias/dark subtraction, background subtraction, flat-fielding, order tracing and merging, wavelength calibration, and finally flux calibration, for which the spectrophotometric standard star EG274 was used. The spectral resolution (FWHM) measured from the sky-lines was $\sim$ 39 km s$^{-1}$  (UVB), $\sim$ 32 km s$^{-1}$  (VIS), and $\sim$ 51 km s$^{-1}$  (NIR) on slit mode, and $\sim$ 29 km s$^{-1}$  (UVB), $\sim$ 22 km s$^{-1}$  (VIS), and $\sim$ 39 km s$^{-1}$  (NIR) on IFU mode. 

The data were corrected for Galactic extinction using the {\scshape noao iraf} task {\scshape deredden}, assuming $E(B-V)$ = 0.0512 ($A_V$ = 0.1587) and the empirical selective extinction function of \citet{cardelli89}.

\subsection{VLT MUSE IFU Spectrum}

We also make use of IFU spectroscopy which was obtained using the Multi Unit Spectroscopic Explorer (MUSE, \citealt{bacon2010}) at the VLT UT4 on 2014 February 21, during the first commissioning run of the instrument \citep[see][]{bacon2014} under the program 60.A-9100(A). Wide Field Mode was used, resulting in a field of view of $1' \times 1'$ at 0.2\arcsec$\,$ spatial sampling. The wavelength range is 4650 \AA$\,$ -- 9300 \AA$\,$ and the mean spectral resolution is $\sim$ 100 km s$^{-1}$  (FWHM). The target was observed for a total of 3600s, which was split into 3$\times$1200s exposures oriented at the position angles 45$^{\circ}$, 135$^{\circ}$ and 225$^{\circ}$. Full details of the MUSE observations, their reduction and some analysis has been previously published by \citet{Gu}. 

\subsection{KECK II LRIS Long Slit Spectrum}

Additional spectral information comes from a Low Resolution Imaging Spectrometer (hereafter LRIS) spectrum taken at the Keck II 10 m telescope on 1997 December 27 \citep[see][]{Ve2001} under the program C56L. The observation was done in polarisation mode with the spectrum covering a wavelength range of  $\sim$ 3900 -- 9000 \AA, and a spectral resolution of $\sim$ 493 km s$^{-1}$  (FWHM). The 1\arcsec$\,$ slit was oriented at a position angle of 73$^{\circ}$, i.e. along the radio axis. This spectrum has been presented and discussed in several previous publications \citep{Ve2001,VM4,VM1,Hu2,Hu3}. See \citep{Ve2001} for full details of the observation and reduction of this data. 

\subsection{VLT UVES Archival Spectrum}

To complement the above observations, we have also made use of VLT Ultraviolet and Visual Echelle Spectrograph (hereafter UVES) observations on the night of 2001 December 8 -- 9 of the radio galaxy, previously published by \citet{Ja}. The spectrum comprises only the red arm and is centred on 5200 \AA$\,$ so as to include Ly$\alpha$ and \ion{C}{IV}, and the spectral resolution is $\sim$ 7 km s$^{-1}$  (FWHM). The raw spectra taken under the program 68.B-0086(A), were obtained from the ESO VLT/UVES archive and were reduced using the UVES pipeline in which the data was automatically bias-subtracted, flat-fielded, wavelength calibrated using Th--Ar arc lamp spectra, and flux calibrated using the spectrophotometric standard star LTT3864. The spectrum was corrected for Galactic extinction with {\scshape iraf's deredden} task, using  $E(B-V)$ = 0.0512 and the extinction function of \citet{cardelli89}.

\subsection{AAT RGO Archival Spectrum}

In addition, a spectrum of MRC 0943--242 taken using the RGO spectrograph of the Anglo Australian Telescope (hereafter AAT) is used, which was previously published by \citet{Ro95} and \citet{Bi1}. The spectrum covers the spectral regions around Ly$\alpha$, \ion{C}{IV} and \ion{He}{II} at a resolution of $\sim$ 91 km s$^{-1}$  (FWHM) with a slit PA of 74$^{\circ}$, i.e. aligned with the radio axis. The raw spectra were obtained from the AAT archive and were bias subtracted, flat-fielded, and then wavelength calibrated using Cu-Ar arc lamp spectra. In addition, flux calibration was done with the spectrophotometric standard stars FEIGE110 and LTT3864. Finally, the spectrum was corrected for Galactic extinction using {\scshape iraf's deredden}, assuming $E(B-V)$ = 0.0512 ($A_V$ = 0.1587) and the extinction curve of \citet{cardelli89}.

\begin{table*}
	\centering
	\caption{Long slit and IFU spectroscopic observations. (1) Instruments used in the observation. (2) Period during which the observation was made. (3) Spectral Resolution of the instrument. (4) Aperture of the spectrum extraction. (5) Slit width. (6) Total observation time. (7) Spatial sampling. (8) Position Angle.}
	\label{observations}
	\begin{tabular}{lcccccccr} 
		\hline
		\hline
		Instrument & Date & Resolution & Aperture & Slit width & Exp. Time & Scale/pixel & P.A.\\
				&		&	(km s$^{-1}$ )	&  (\arcsec) & (\arcsec) & (sec)   &  (\arcsec)  & ($^{o}$) \\
				(1)			 &	(2)	&	(3)		 &	 (4)	&    (5)	 &   (6)     &	(7)		 &	(8) \\
		\hline
		 XSH SLIT 	 &  March 2009    &	 39, 32, 51    & 0.75,0.8 & 1.0,0.9  & 9000 & 0.16, 0.21 & $55$ \\
		 \hline
		 XSH IFU 	 &  March 2009    &	 29, 22, 39    & 0.8 & --  & 1700 & 0.16, 0.21 & $55$\\
		 \hline
		 MUSE IFU 	 &  February 2014    &	 100    & 0.8 & --  & 3600 & 0.20 & $45,135,225$\\
		 \hline
		 UVES 	 &  December 2001    &	 7    & 1.0 & 1.2  & 10800 & 0.18 & $74$\\
		 \hline
		 AAT 	 &  April 1993, March 1995    &	 91    & 0.79 & 1.6  & 25000 & 0.79 & $74$\\
		 \hline
		 LRIS 	 &  December 1997    &	 493    & 0.856 & 1.0  & 13800 & 0.214 & $73$\\
		 \hline
		 \hline

	\end{tabular}
\end{table*}

\section{Data Analysis}
\label{analysis}

\subsection{Emission and Absorption Line Fitting}
\label{fitting}

\subsubsection{Fitting routine}
We created a Python routine to fit the emission and absorption line parameters, with Gaussian and Voigt profiles being used to model the emission and absorption lines, respectively. The routine minimizes the sum of the squares of the difference between the model and data using the LMFIT algorithm \citep{lmfit}. Parameters for the best fits are shown in Tables \ref{instru}, \ref{instru02}, \ref{instru03}, and also Tables \ref{instru_ap}, \ref{instru02_ap} and \ref{instru03_ap}.

\subsubsection{Ly$\alpha$ and the degeneracy in \textit{N} and \textit{b}}
\begin{figure*}
	\textsc{High N(\ion{H}{I}) solution} \hspace{2cm} \textsc{Low N(\ion{H}{I}) solution}
	\subfloat[XSHOOTER SLIT]{
		\includegraphics[width=\columnwidth,height=5.5cm,keepaspectratio]{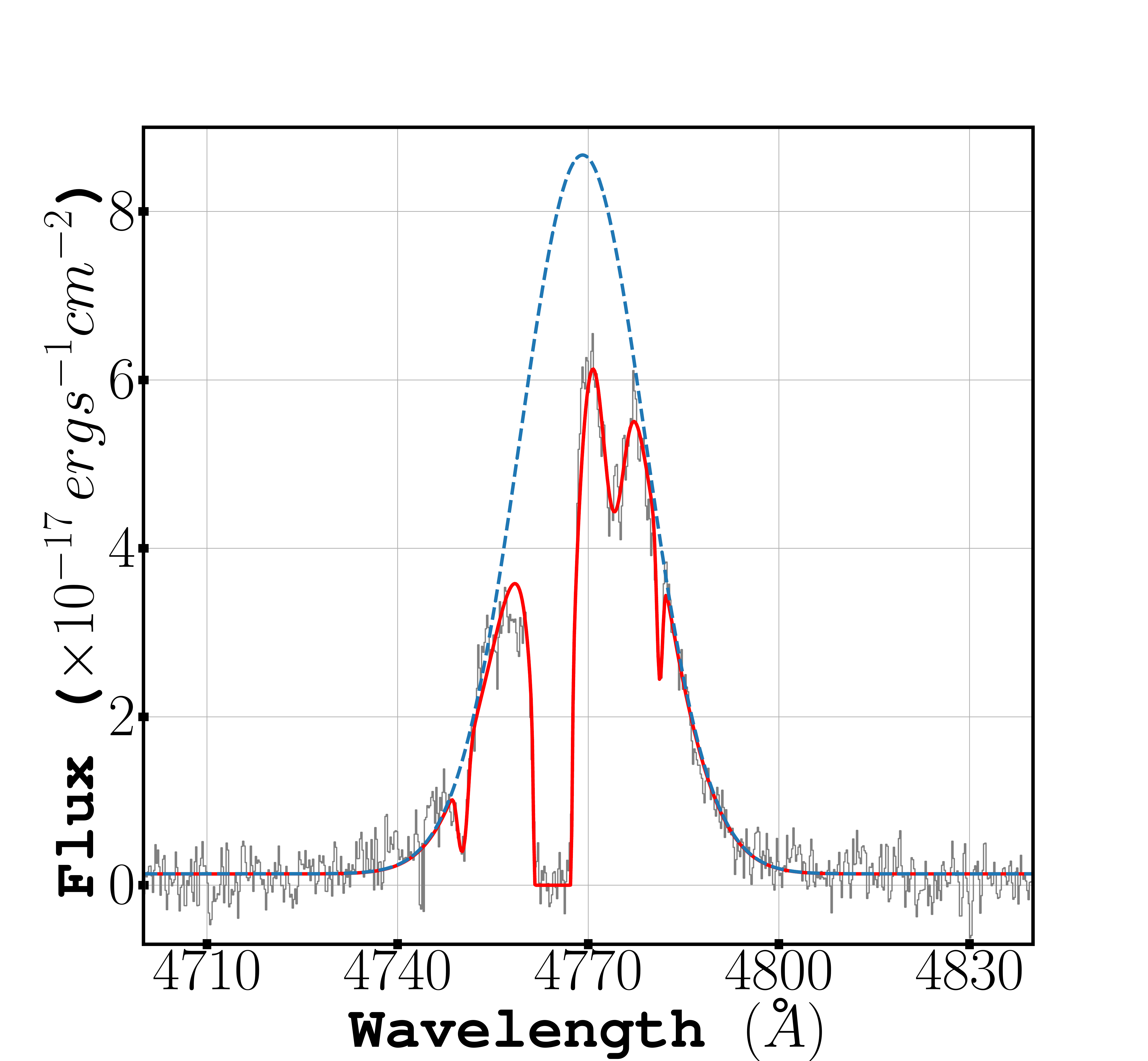}
		\includegraphics[width=\columnwidth,height=5.5cm,keepaspectratio]{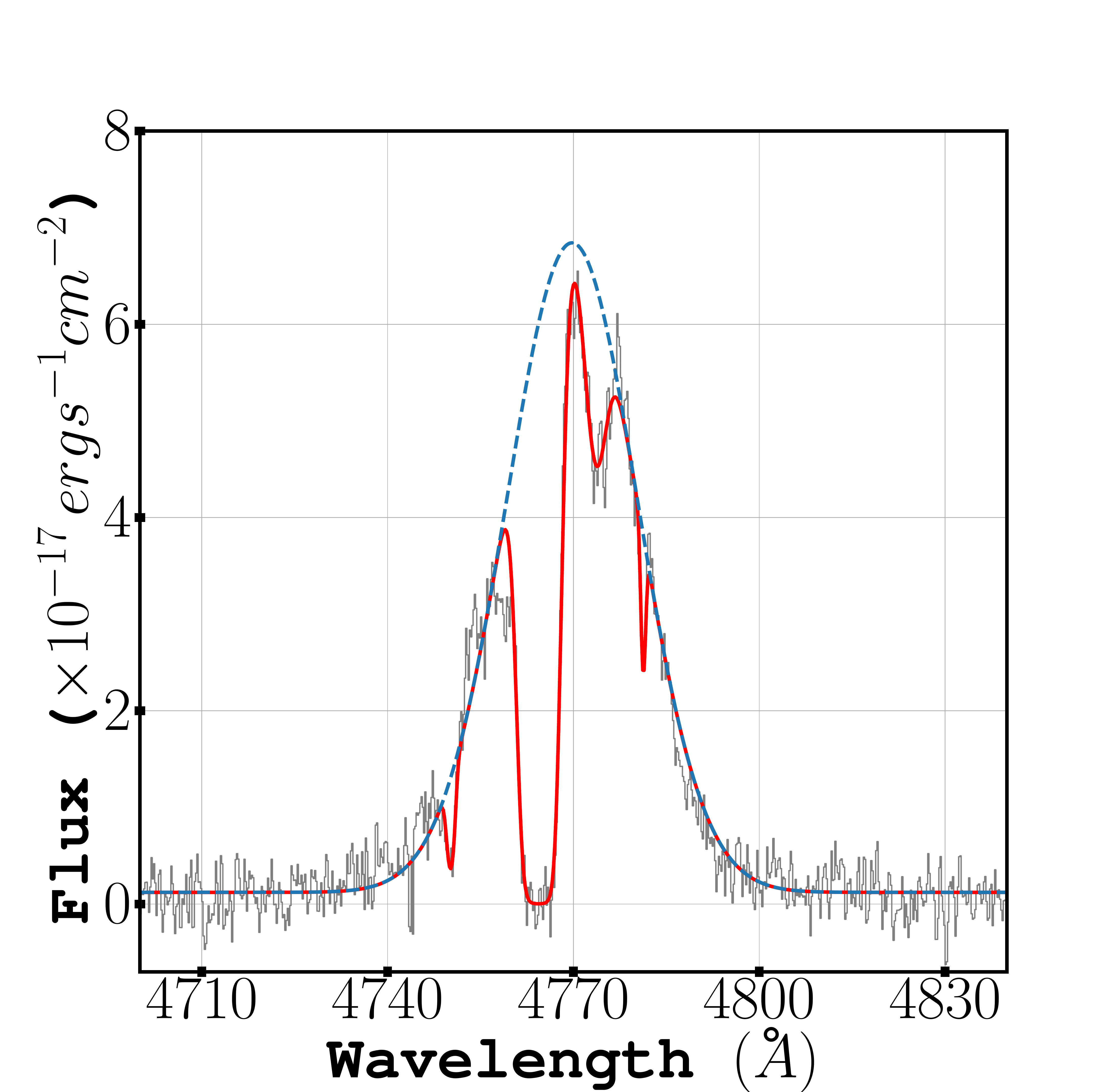}
		\label{ly-xsh}}
	\caption{Ly$\alpha$ profile of MRC 0943--242 extracted from the X-SHOOTER long-slit, with the Gaussian emission component (dashed blue line) plus absorption model overlaid (red line). The left and right columns show the high and low column density best-fitting, respectively. See other instruments in Fig. \ref{ly_ap}.}
	\label{ly}
\end{figure*}

The Ly$\alpha$ profile was parametrized assuming that the underlying emission line is a single Gaussian, and adopting Voigt profiles for the superimposed \ion{H}{I} absorption features. Figure \ref{ly} shows the one-dimensional spectra of the Ly$\alpha$ emission-line from the X-SHOOTER observations of MRC 0943--242. In addition, Figure \ref{ly_ap} shows the Ly$\alpha$ profile from the other telescopes/instruments.

When fitting absorption lines there can be a strong degeneracy between column density (\textit{N}) and the Doppler width (\textit{b}), and we find this to be the case with the main \ion{H}{I} absorption feature seen in the Ly$\alpha$ profile of MRC 0943--242. We find two widely separated `best-fitting' to the column density of this absorber, one at log N(\ion{H}{I}/cm$^{-2}$) = 15.20 with $b$ $\sim$ 153 km s$^{-1}$ ($\tilde{\chi}^2_{\nu}$ = 0.08), and another at log N(\ion{H}{I}/cm$^{-2}$) = 19.63 with $b$ $\sim$ 52 km s$^{-1}$ ($\tilde{\chi}^2_{\nu}$ = 0.07). 

\begin{figure*}
	\includegraphics[width=15cm,height=15cm]{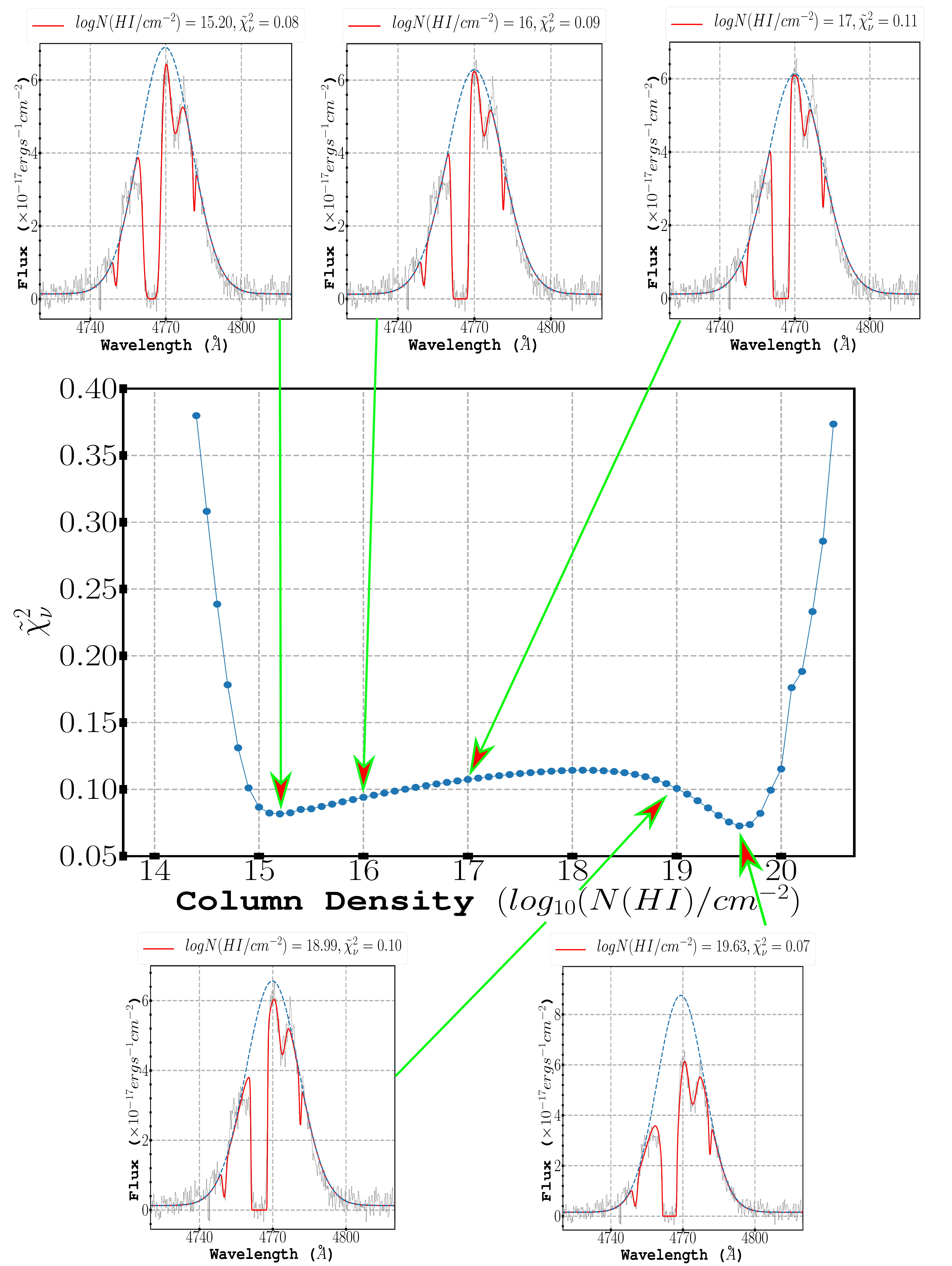}
	\caption{Plot of column density ($N(\ion{H}{I})$) \textit{versus} reduced chi-square ($\tilde{\chi}^2_{\nu}$), showing the presence of two widely-spaced `best-fitting', between which lie a broad range of inferior, but nonetheless reasonable fits. Data points other than the two `best-fitting' were produced by running our fitting code with N(\ion{H}{I}) fixed to specific values. Around the main plot we show selected fits to the X-SHOOTER long-slit Ly$\alpha$ profile, to further illustrate how the fits evolve as a function of $N(\ion{H}{I})$.}
	\label{degeneracy}
\end{figure*}

We illustrate this degeneracy in Figure \ref{degeneracy}, where we show column density ($N(\ion{H}{I})$) \textit{versus} reduced chi-square ($\tilde{\chi}^2_{\nu}$). With the exception of the two `best-fitting', the data points in this Figure were obtained by running our fitting code with N(\ion{H}{I}) fixed at specific values, but with all other parameters left free to vary. In addition to the presence of the two `best-fitting', Figure \ref{degeneracy} also illustrates the presence of a broad range of intermediate N(\ion{H}{I}) values where the fits are still reasonably good. 

Throughout this paper, we will ensure that the degeneracy described above is fully taken into account when we derive properties of the extended \ion{H}{I} absorber. Where we make use of the flux of the Ly$\alpha$ emission line (e.g., Sect. 5.2), we adopt the absorption-corrected flux obtained from our log N(\ion{H}{I}/cm$^{-2}$) = 19.63 fit to the line profile (20.27 $\pm$ 0.38 $\times10^{-16}$ erg cm$^{-2}$ s$^{-1}$). Using instead the flux obtained from our log N(\ion{H}{I}/cm$^{-2}$) = 15.20 fit (17.64 $\pm$ 0.38 $\times10^{-16}$ erg cm$^{-2}$ s$^{-1}$) does not have any significant impact on our conclusions. 

\subsubsection{C$\,${\scriptsize IV}}
\begin{figure*}
\subfloat[XSHOOTER SLIT]{
\includegraphics[width=\columnwidth,height=5.5cm,keepaspectratio]{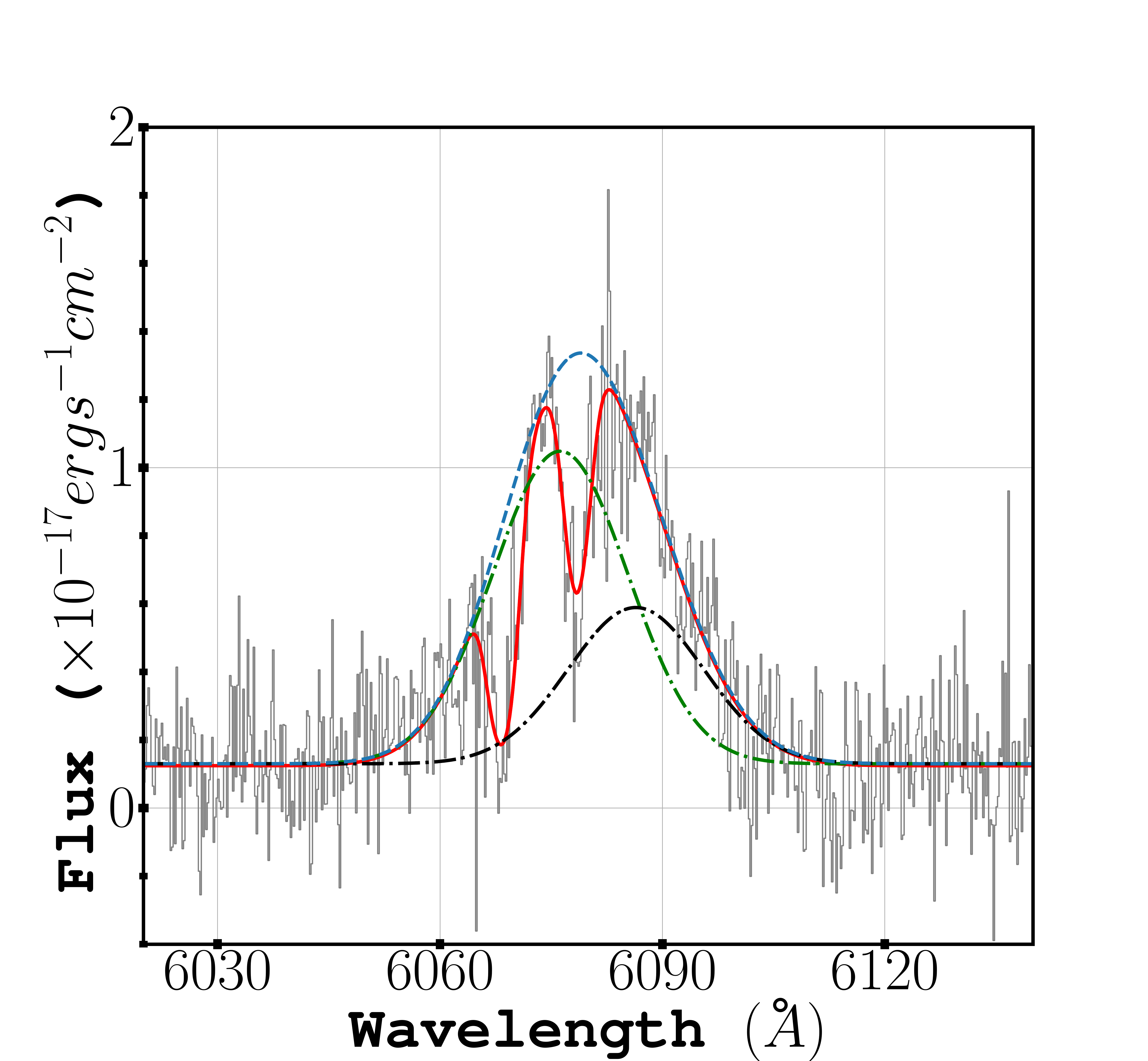}
\label{xsh}}
\quad
\subfloat[XSHOOTER IFU]{
\includegraphics[width=\columnwidth,height=5.5cm,keepaspectratio]{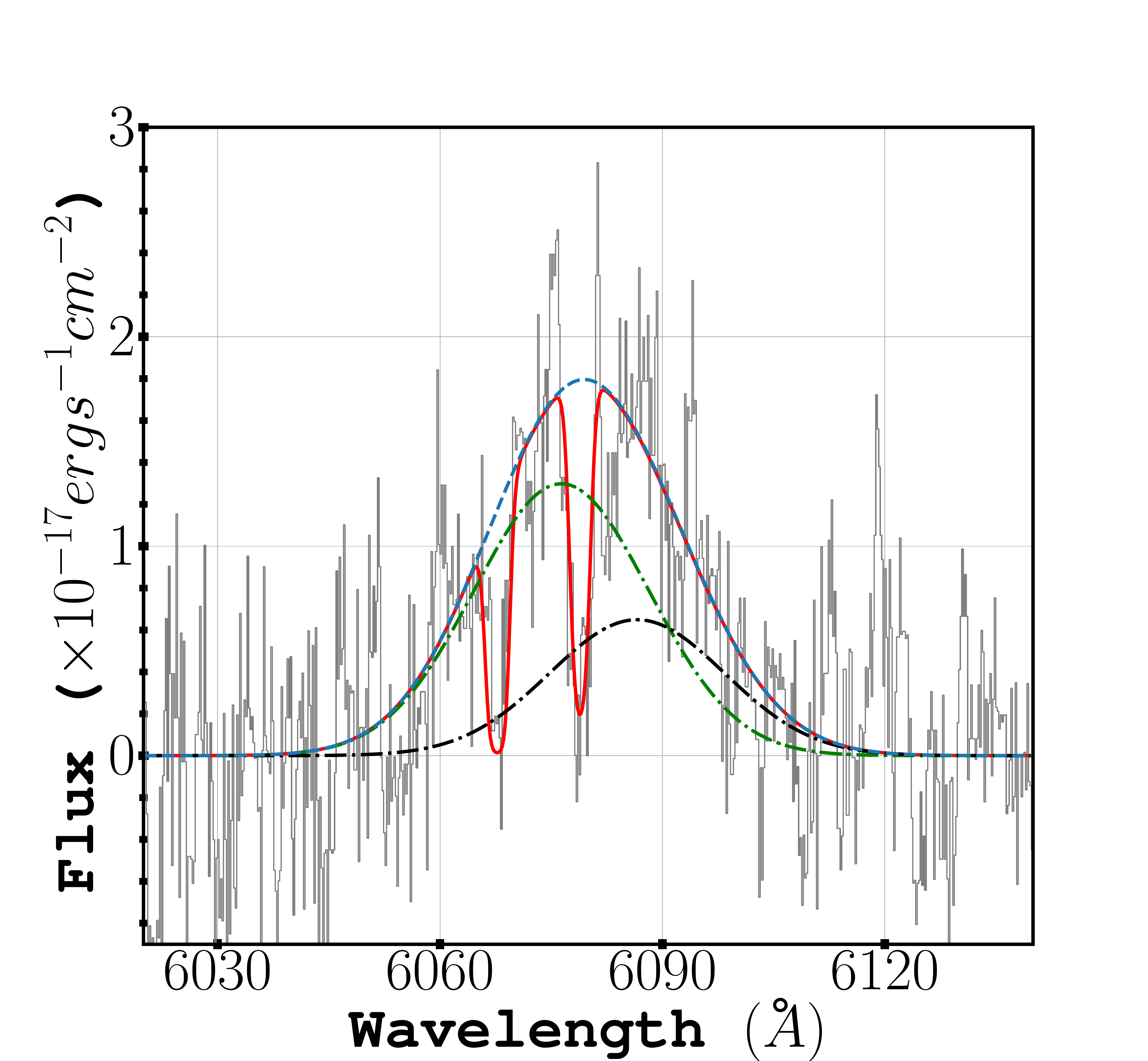}
\label{ifu}}
\caption{The \ion{C}{IV} profile of MRC 0943--242, with the Gaussian emission component (dashed blue line) and emission plus absorption model overlaid (red line). The two individual doublet components are also shown. See the \ion{C}{IV} profile from other telescopes/ instruments in Fig. \ref{civ-ap}.}
\label{civ}
\end{figure*}

The methodology used to fit the \ion{C}{IV} profile is similar to that used for Ly$\alpha$. Because \ion{C}{IV} is a doublet, two gaussians were used for the emission and two Voigt profiles for the absorption, with the doublet's (rest-frame) wavelength separation set to its theoretical value (see Fig. \ref{civ} and Fig. \ref{civ-ap}). In the case of the emission components, we set the flux ratio \ion{C}{IV} $\lambda$1548.2 / \ion{C}{IV} $\lambda$1550.8 to 0.5 and constrained both lines to have equal FWHM and redshift (\textit{z}) . For the absorption lines, we constrained both lines to have the same values for \textit{z}, \textit{b} and \textit{N}. 

\subsubsection{Other lines}
Other emission lines were fitted with a single Gaussian component. Emission doublets were specified to have equal FWHM, fixed wavelength separation and a fixed flux ratio. For instance, we set the flux ratio \ion{N}{V} $\lambda \lambda$1239,1243 to 1.1, \ion{C}{III]} $\lambda \lambda$1907,1909 to 0.66, \ion{[Ne}{IV]} $\lambda \lambda$2422,2424 to 1.5, \ion{[O}{II]}  $\lambda \lambda$3726,3729 to 1.4 and \ion{[O}{III]} $\lambda \lambda$4959,5007 to 0.35. In order to obtain a fiducial systemic velocity we use the non-resonant \ion{He}{II} $\lambda$1640 emission-line. 

\subsection{Kinematic maps}
\label{maps}
From the MUSE datacube, we have extracted the spatially resolved kinematic properties of the UV emission and absorption lines using the fitting procedure outlined in $\S$\ref{fitting}. Due to the relatively low spectral resolution of the MUSE data, we only included a single absorption line in our fit to Ly$\alpha$, which represents the main absorber. 

In Figure \ref{kin} we show the maps of the FWHM, and the velocity offset relative to the \ion{He}{II} emission at the position of the UV continuum peak. Figures \ref{ly-abs} and \ref{civ-abs} show the velocity offset for the extended absorber in Ly$\alpha$ and \ion{C}{IV}, respectively. To facilitate consistency checks, we also extracted the kinematic properties of the Ly$\alpha$, \ion{He}{II} and/or \ion{C}{IV} from the X-SHOOTER and Keck II long slit spectra (see Figures \ref{1dim_kinXSH} and \ref{1dim_kinKECK}).

\subsection{Pseudo-Narrowband images and line ratio maps}
\label{pseudoNB}
\begin{figure*}
\subfloat[Ly$\alpha\,$ $\lambda$1216]{
\includegraphics[width=7.0cm,height=4.75cm,keepaspectratio]{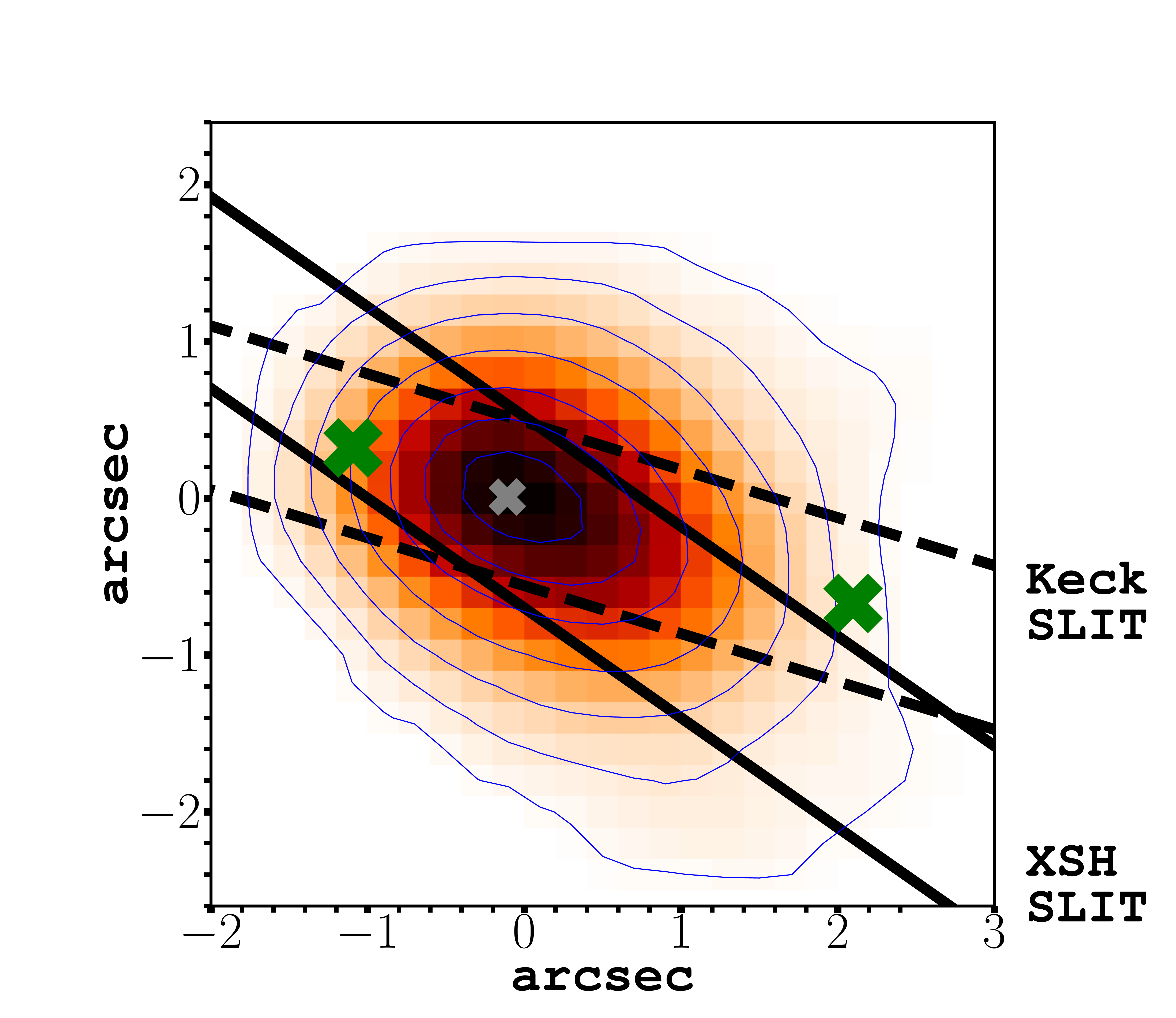}
\includegraphics[width=7.0cm,height=4.75cm,keepaspectratio]{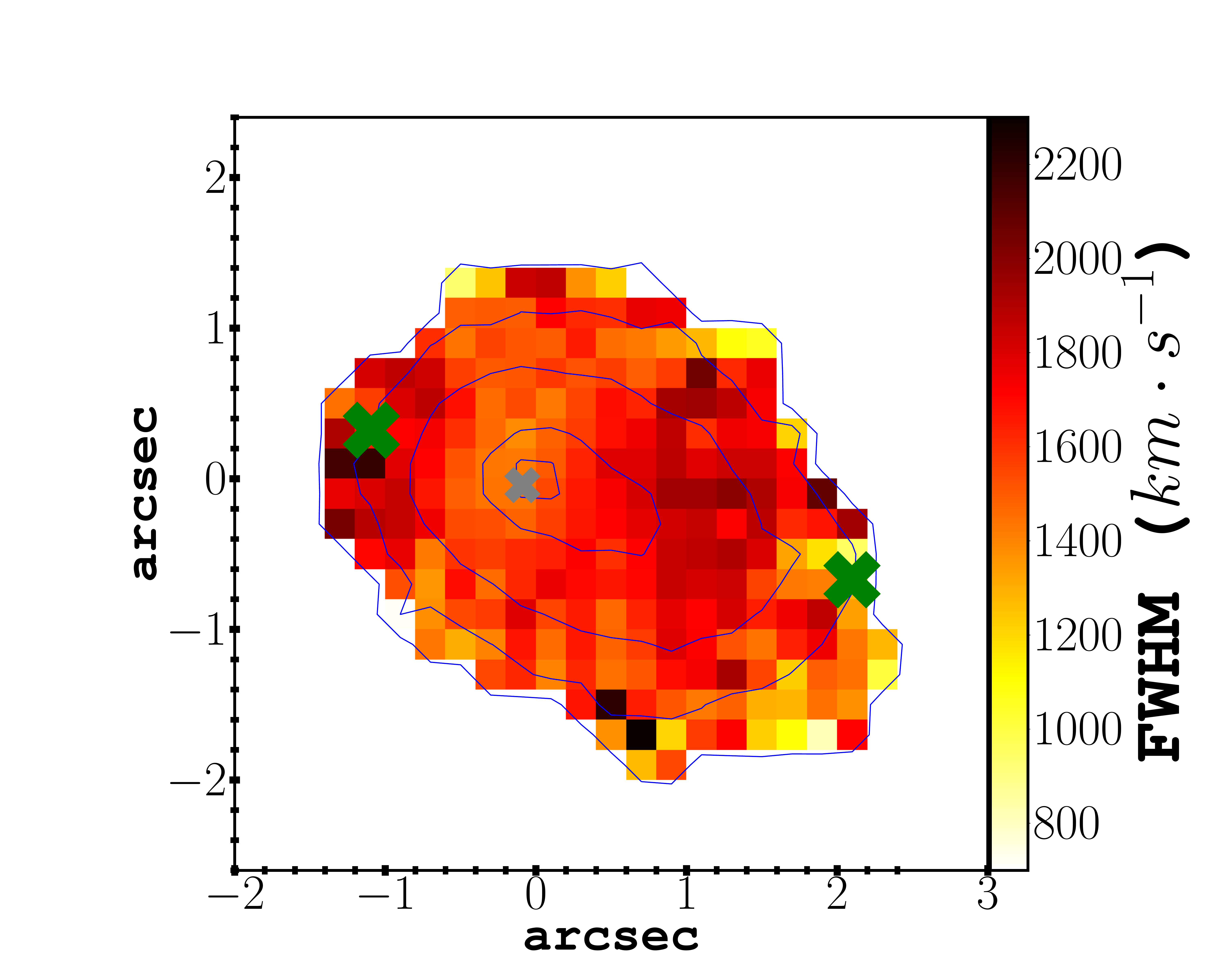}
\includegraphics[width=7.0cm,height=4.75cm,keepaspectratio]{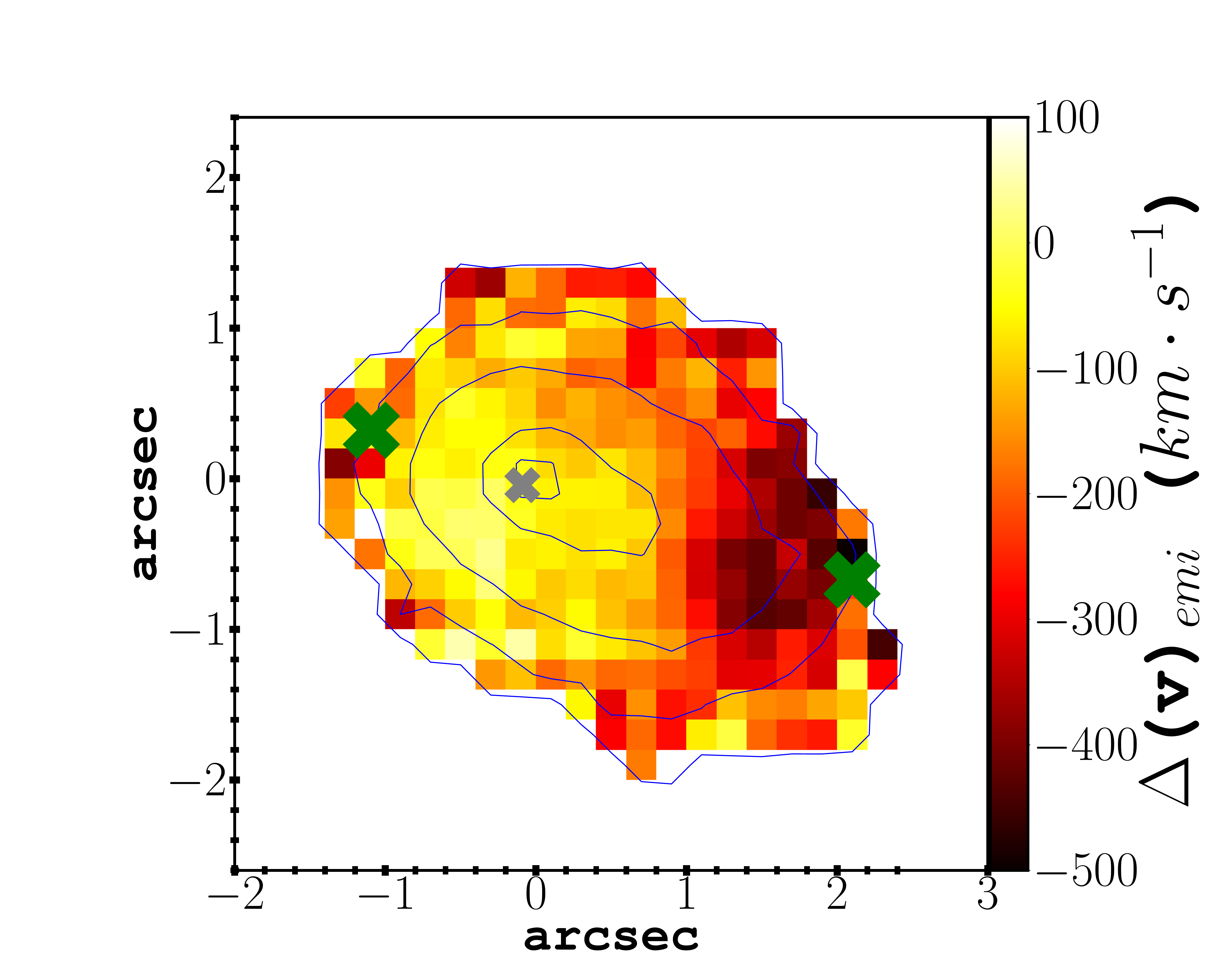}
\label{kin-ly}}
\quad
\subfloat[\ion{C}{IV} $\lambda\lambda$1548,1550]{
\includegraphics[width=7.0cm,height=4.75cm,keepaspectratio]{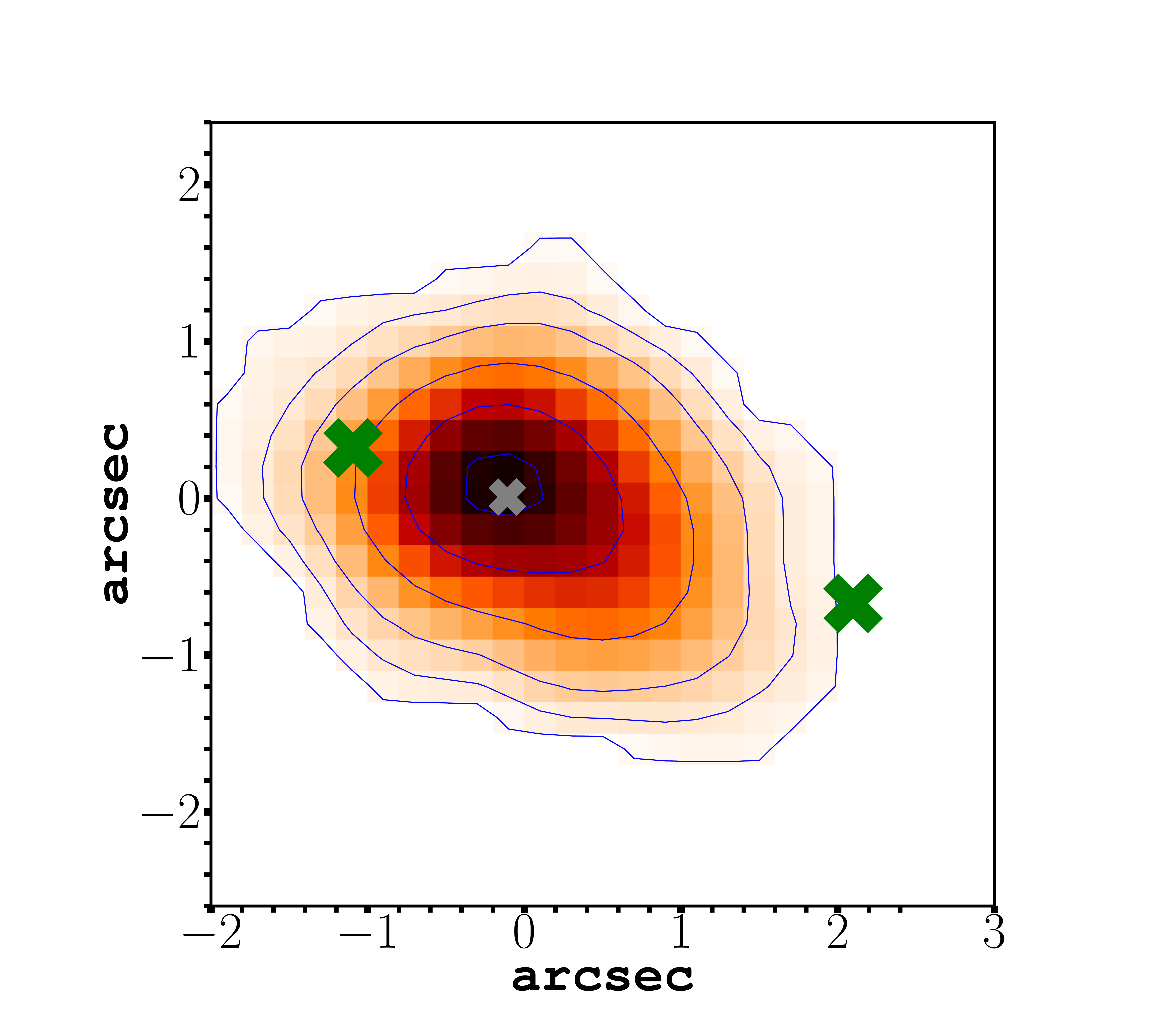}
\includegraphics[width=7.0cm,height=4.75cm,keepaspectratio]{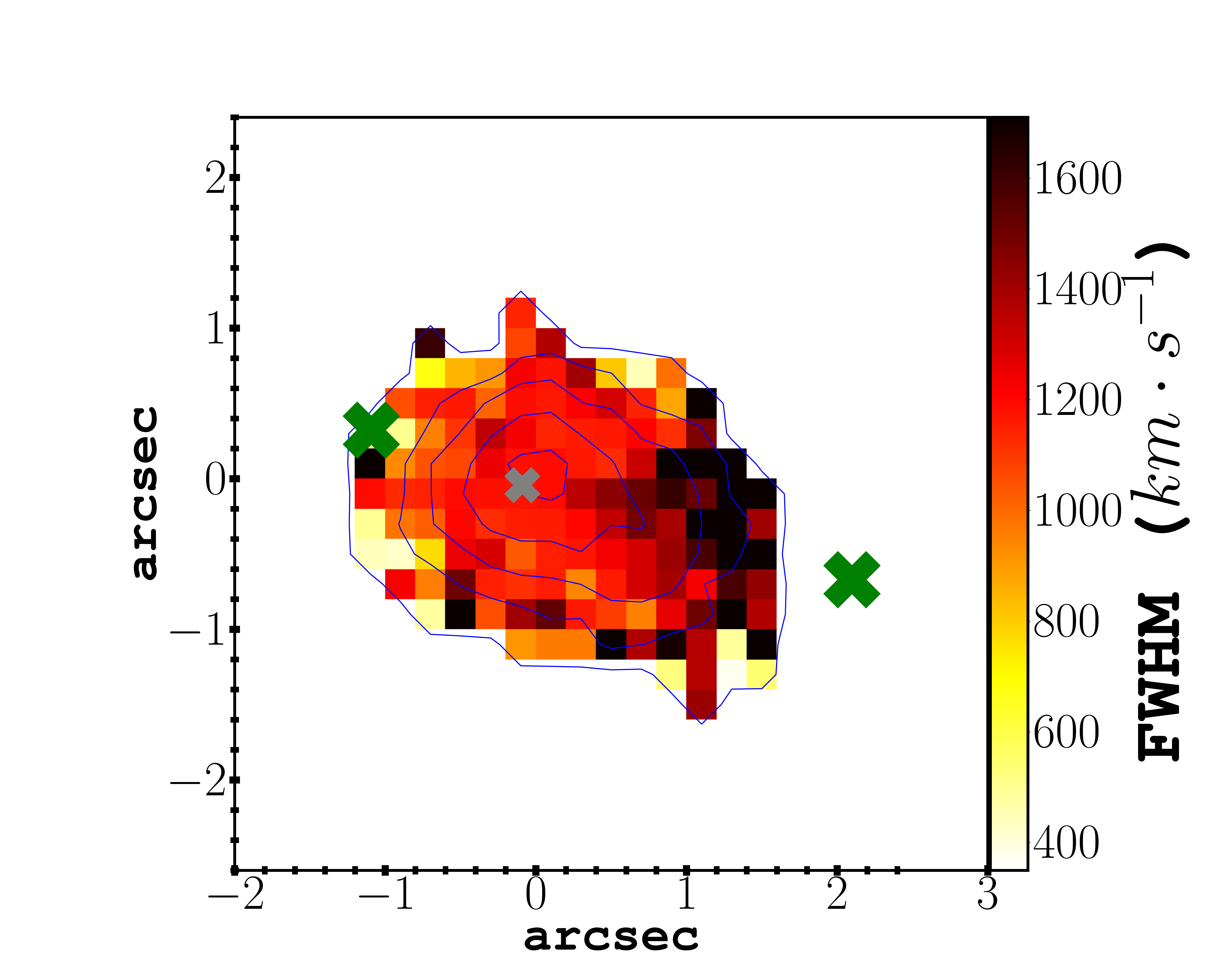}
\includegraphics[width=7.0cm,height=4.75cm,keepaspectratio]{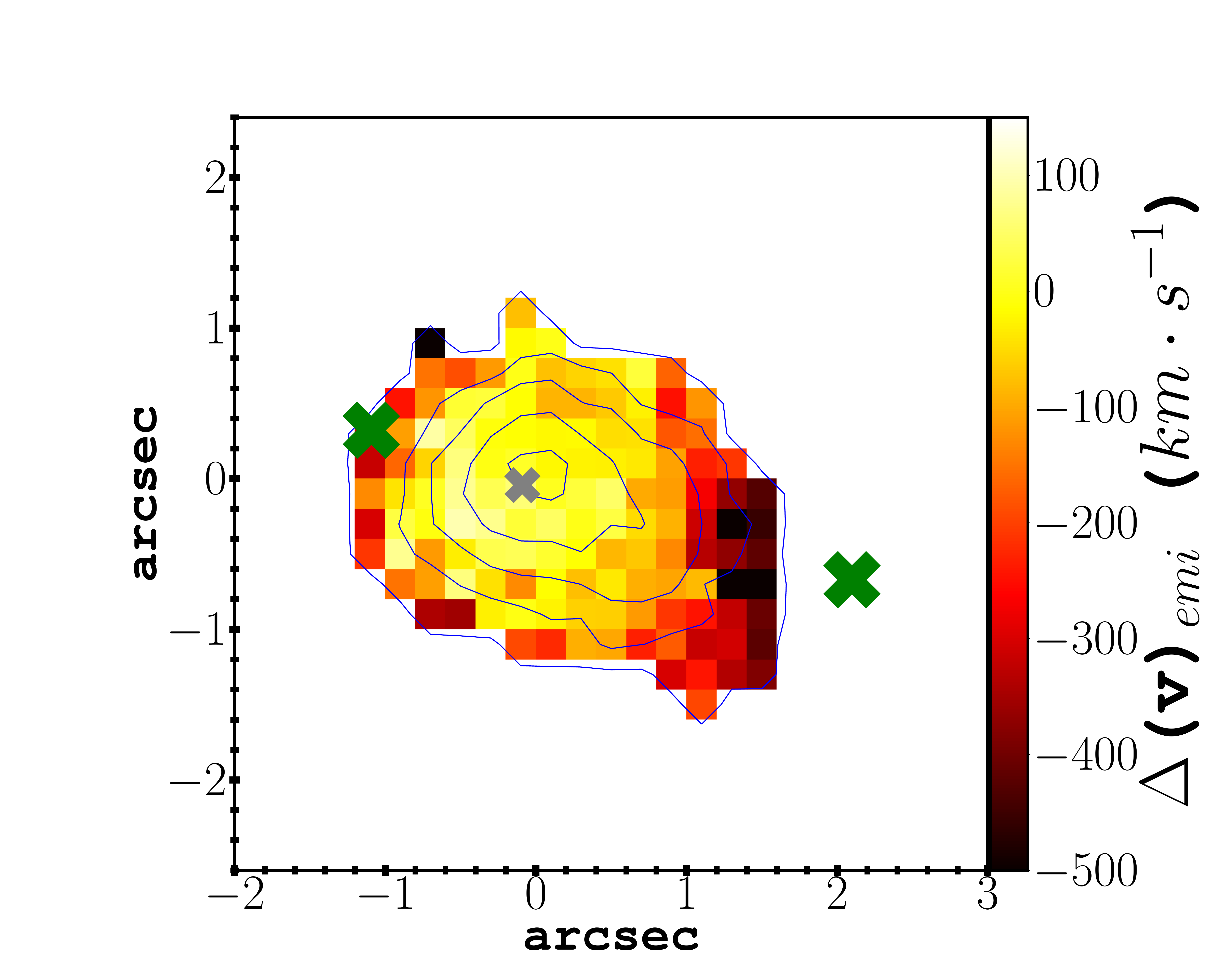}
\label{kin-civ}}
\quad
\subfloat[\ion{He}{II} $\lambda$1640]{
\includegraphics[width=7.0cm,height=4.75cm,keepaspectratio]{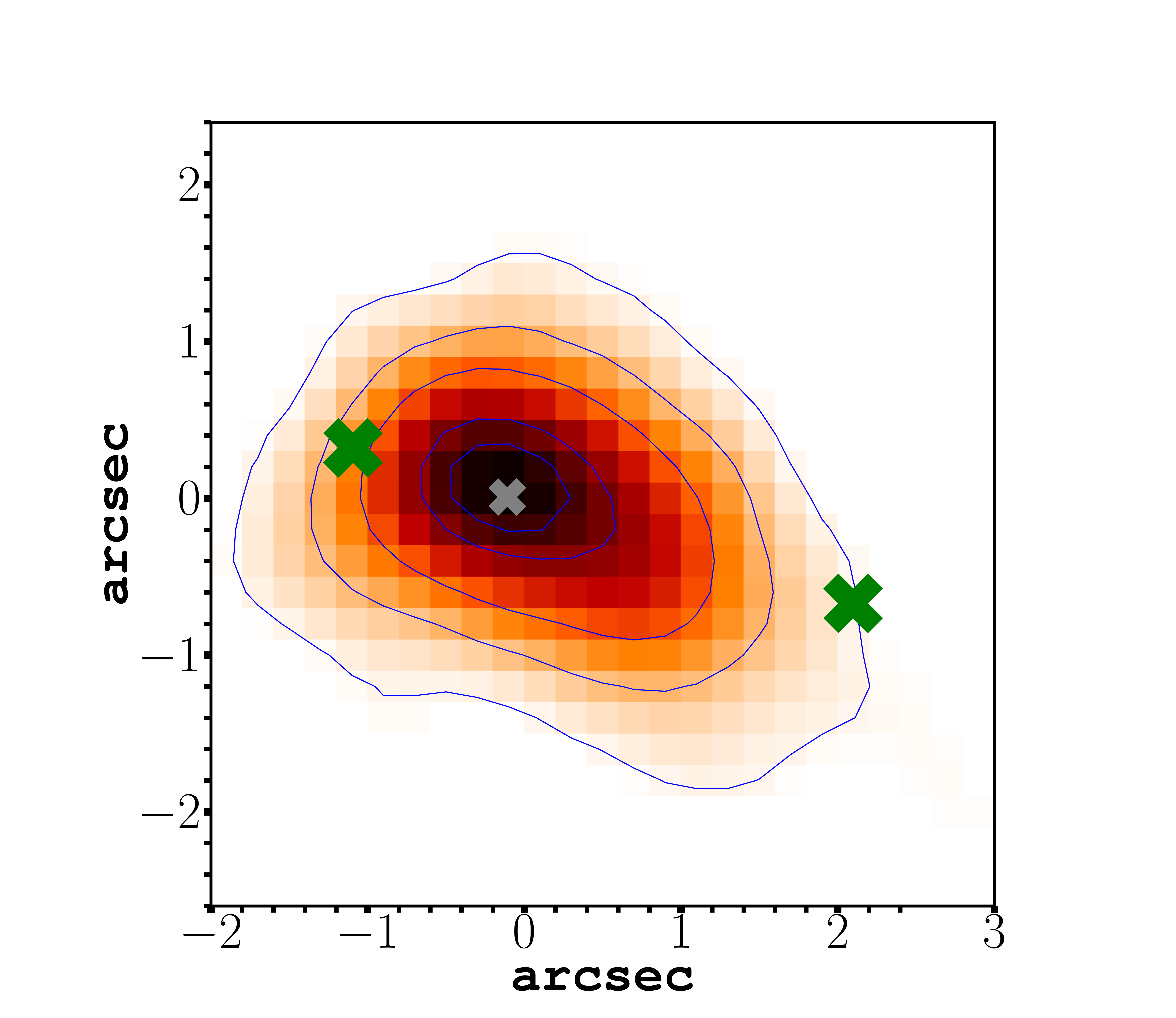}
\includegraphics[width=7.0cm,height=4.75cm,keepaspectratio]{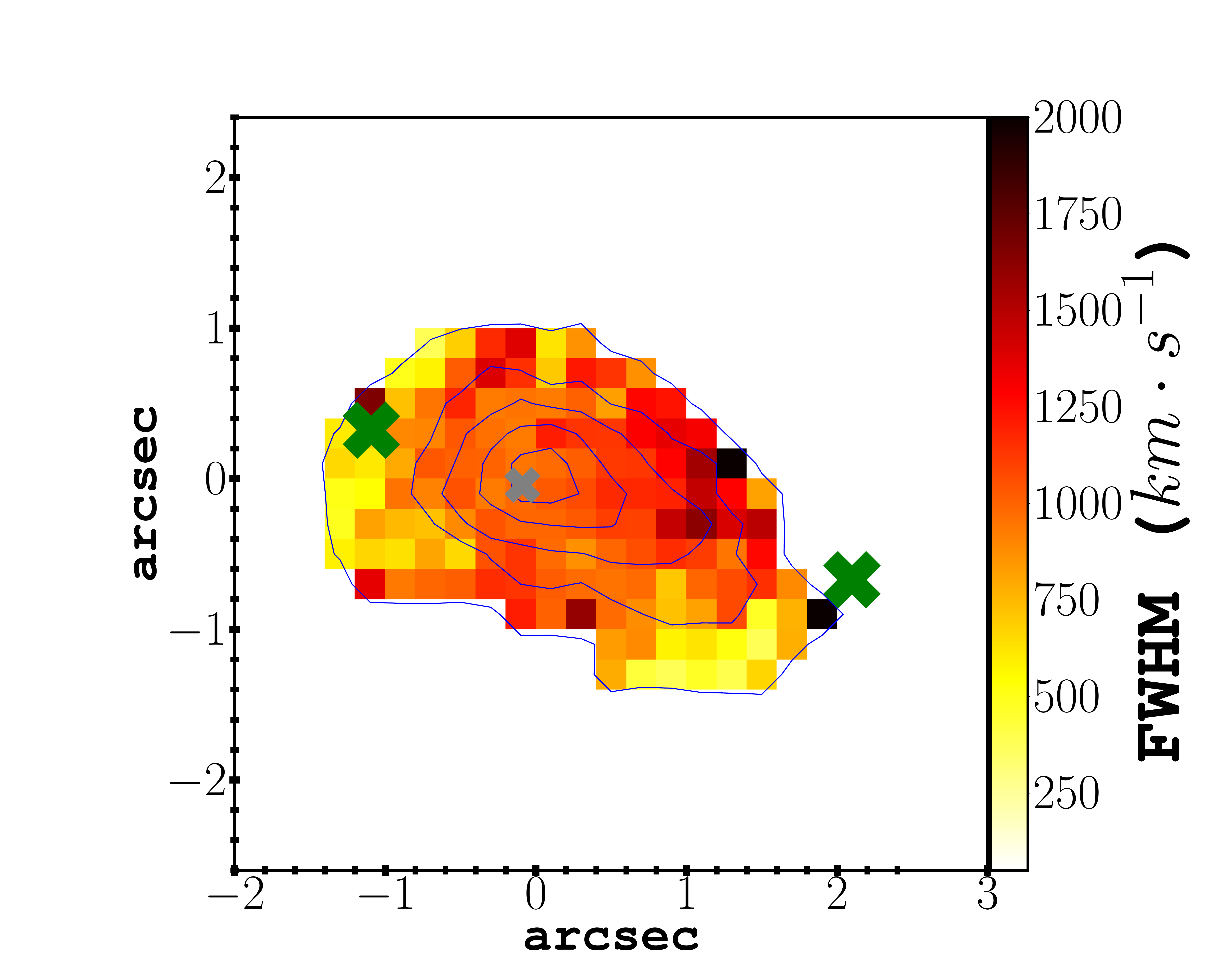}
\includegraphics[width=7.0cm,height=4.75cm,keepaspectratio]{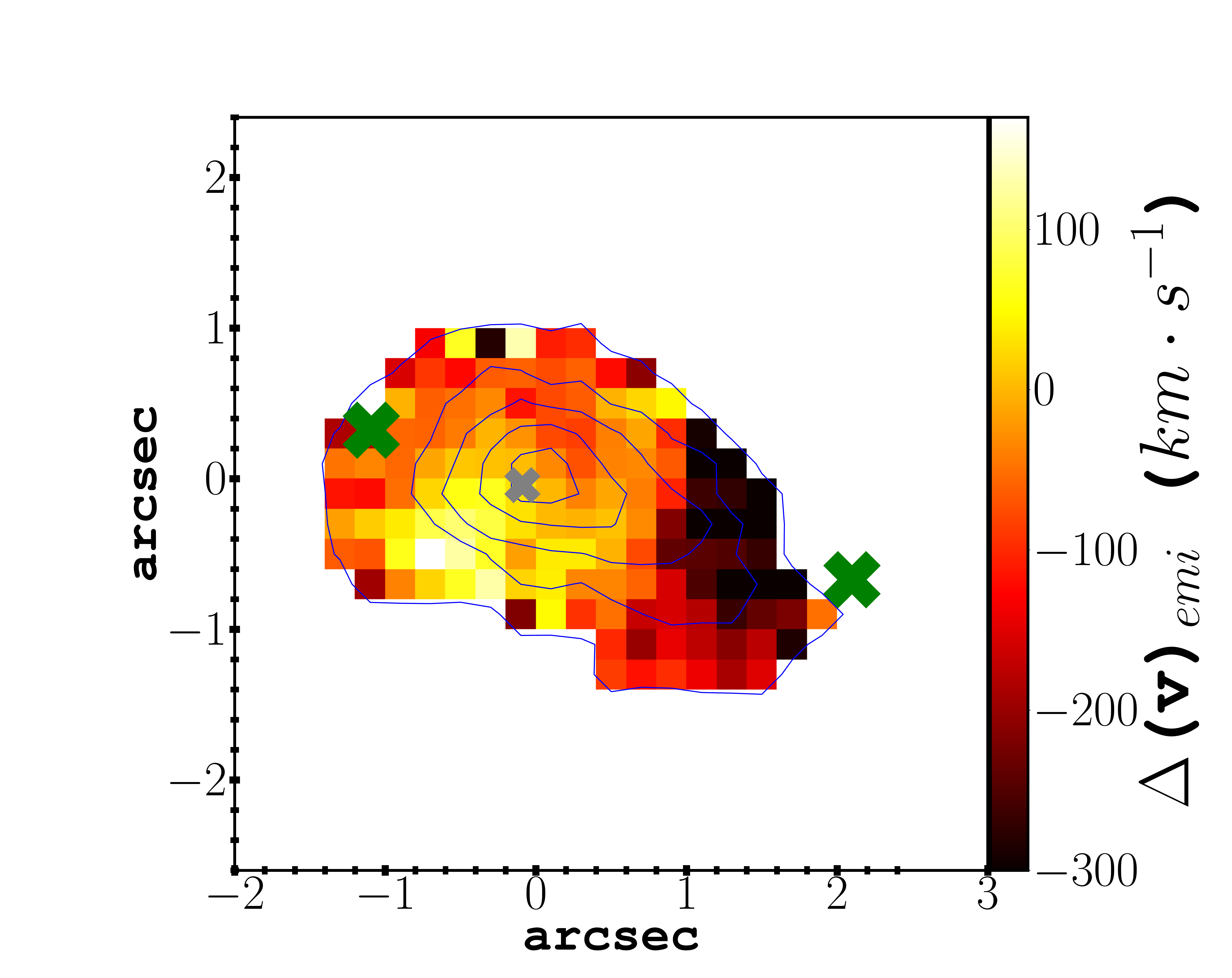}
\label{kin-he2}}
\quad
\subfloat[\ion{C}{III]} $\lambda\lambda$1907,1909]{
\includegraphics[width=7.0cm,height=4.75cm,keepaspectratio]{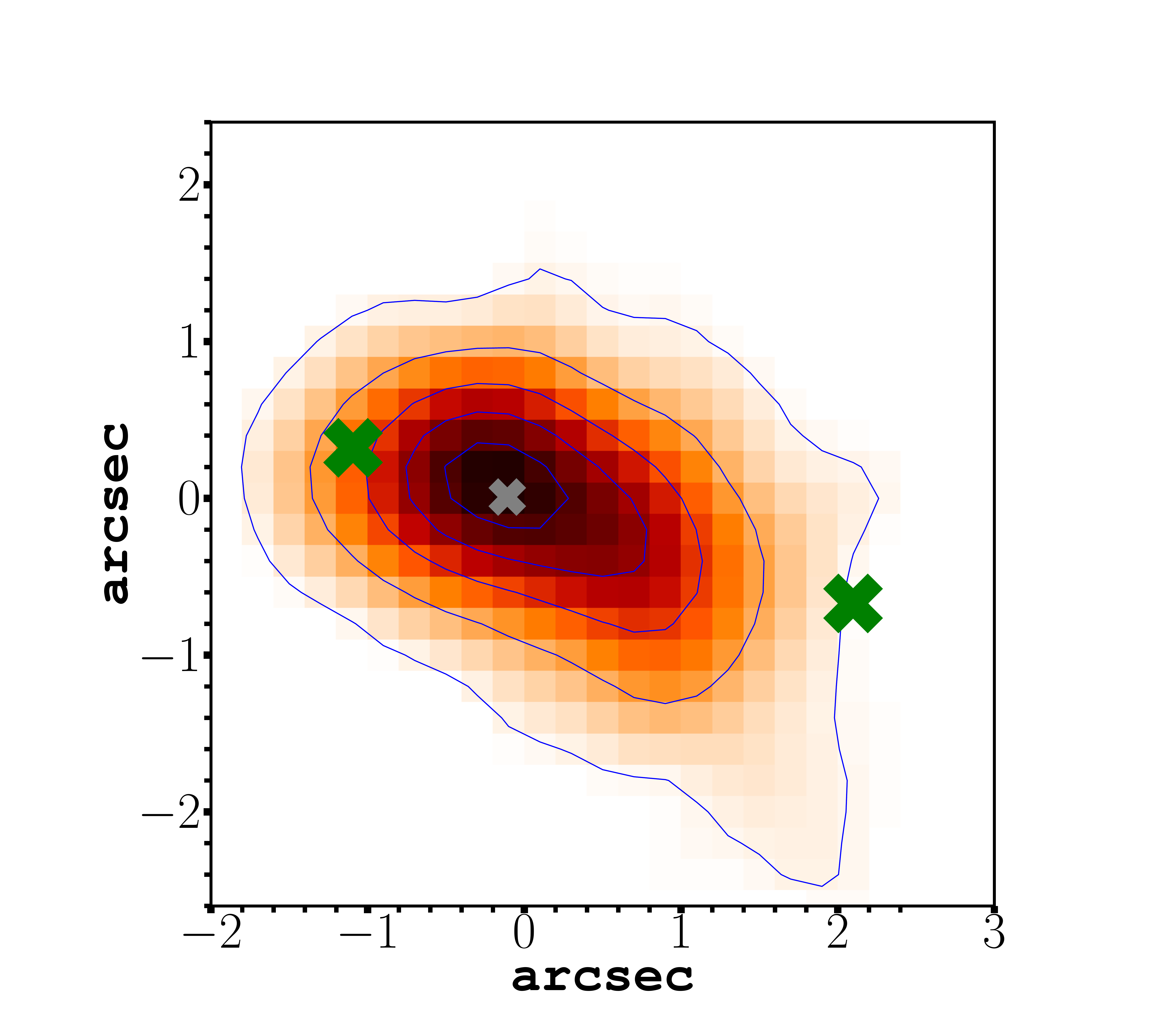}
\includegraphics[width=7.0cm,height=4.75cm,keepaspectratio]{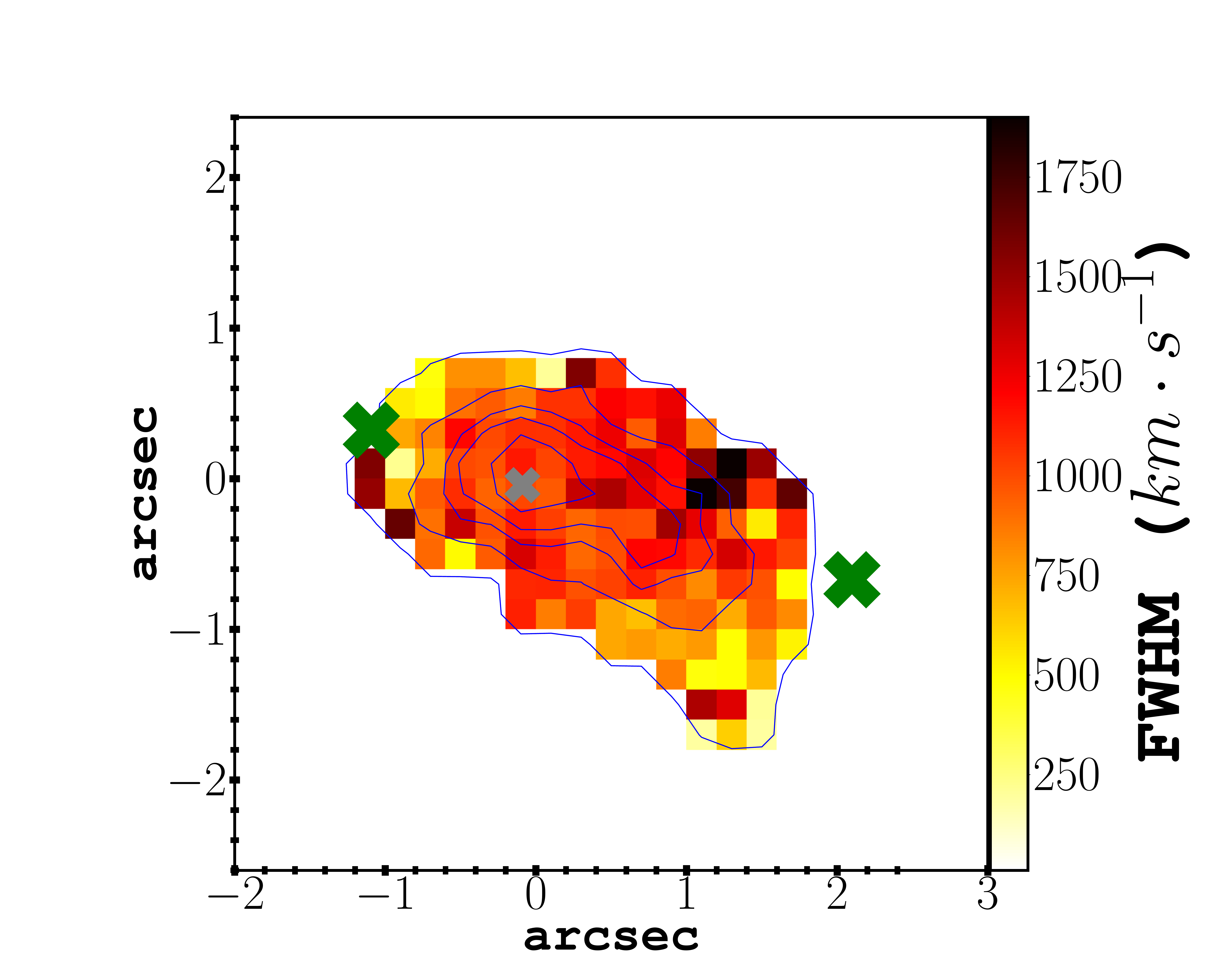}
\includegraphics[width=7.0cm,height=4.75cm,keepaspectratio]{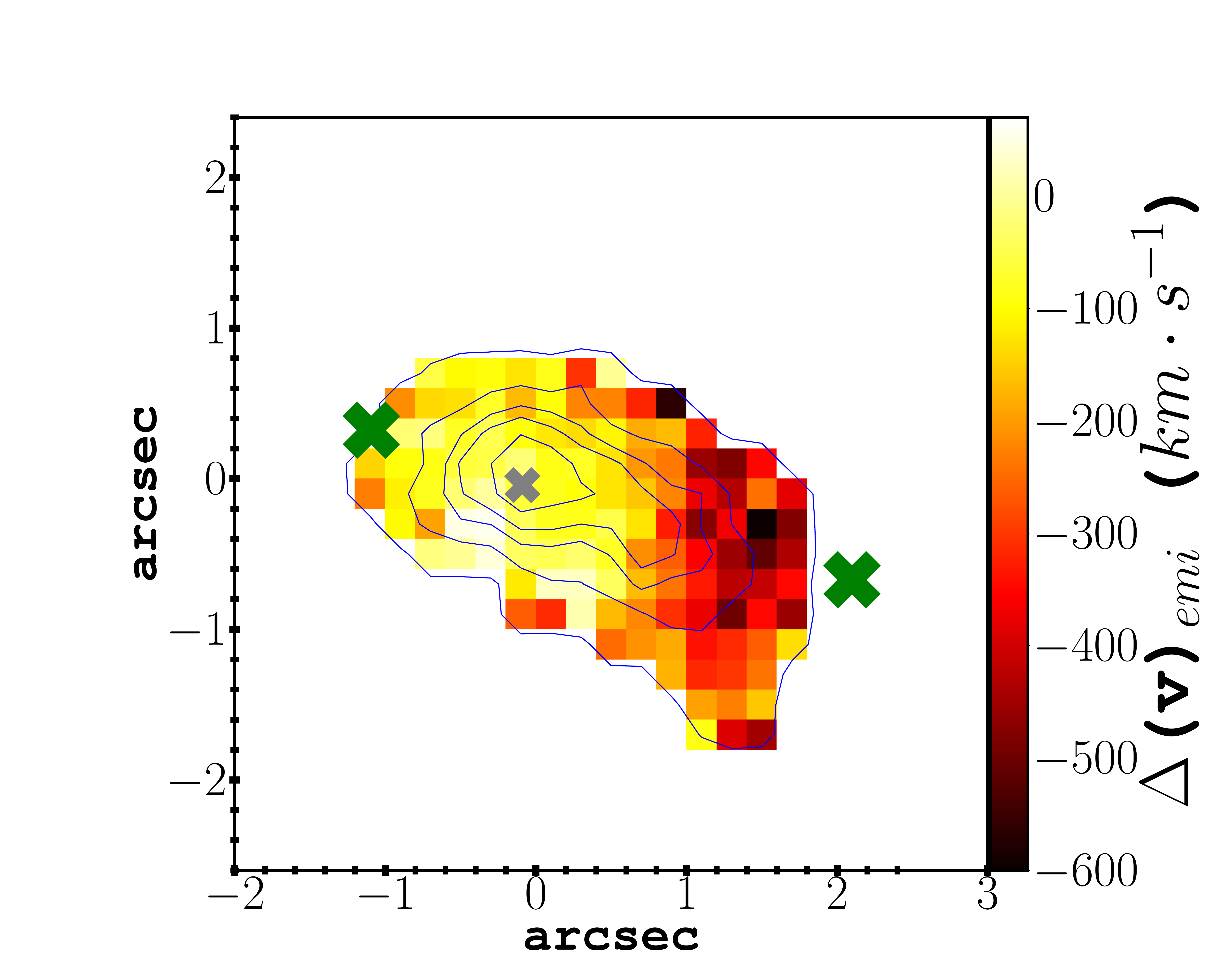}
\label{kin-ciii}}

\caption{From left to right, we show the pseudo-narrowband images, the FWHM map and the velocity map. The pseudo-narrowband image of Ly$\alpha$ is overlaid with the position of the X-SHOOTER slit (solid lines) and KECK slit (dashed lines). Contour levels: Ly$\alpha$ - (0.3,1.3,3.8,12.5,16.8)$\times 10^{-16}$, \ion{C}{IV} - (0.5,1.0,2.0,3.5,4.3)$\times 10^{-16}$, \ion{He}{II} - (0.1,0.8,1.0,1.5,2.0)$\times 10^{-16}$ and \ion{C}{III]} - (0.08,0.5,0.8,1.0,1.3)$\times 10^{-16}$ erg cm$^{-2}$ s$^{-1}$ arcsec$^{-2}$. The green "X" represents the positions of the radio hotspots. The coordinate (0,0) correponds to the assumed position of the AGN which is marked with a grey "x". The scale of 5\arcsec $\times$ 5\arcsec on the axes corresponds to the physical scale of 40 kpc $\times$ 40 kpc.}
\label{kin}
\end{figure*}

In order to study the emission line morphology, we have produced pseudo-narrowband (hereafter, pseudo-NB) images from the MUSE datacube. These images were created by summing pixels along the dispersion axis within spectral windows that include all of the detected line flux: $\lambda _{obs}$ = 4705.55 -- 4718.05, 4722.68 -- 4726.84, 4734.71 -- 4740.72, 4742.64 -- 4754.61 \AA$\,$ for Ly$\alpha$ aiming to exclude the absorption troughs of the emission-line, $\lambda _{obs}$ = 5745.74 -- 5812.24 for \ion{C}{IV}, $\lambda _{obs}$ = 6037.49 -- 6094 for \ion{He}{II} and  $\lambda _{obs}$ = 6875.09 -- 6931.81 for \ion{C}{III]}. An image of the adjacent continuum was then subtracted from each, giving a pure emission-line image. The resulting continuum-subtracted, pseudo-NB images are shown on the left of Fig. \ref{kin}. 

\begin{figure*}
	
	\includegraphics[width=5.8cm,height=5.7cm,keepaspectratio]{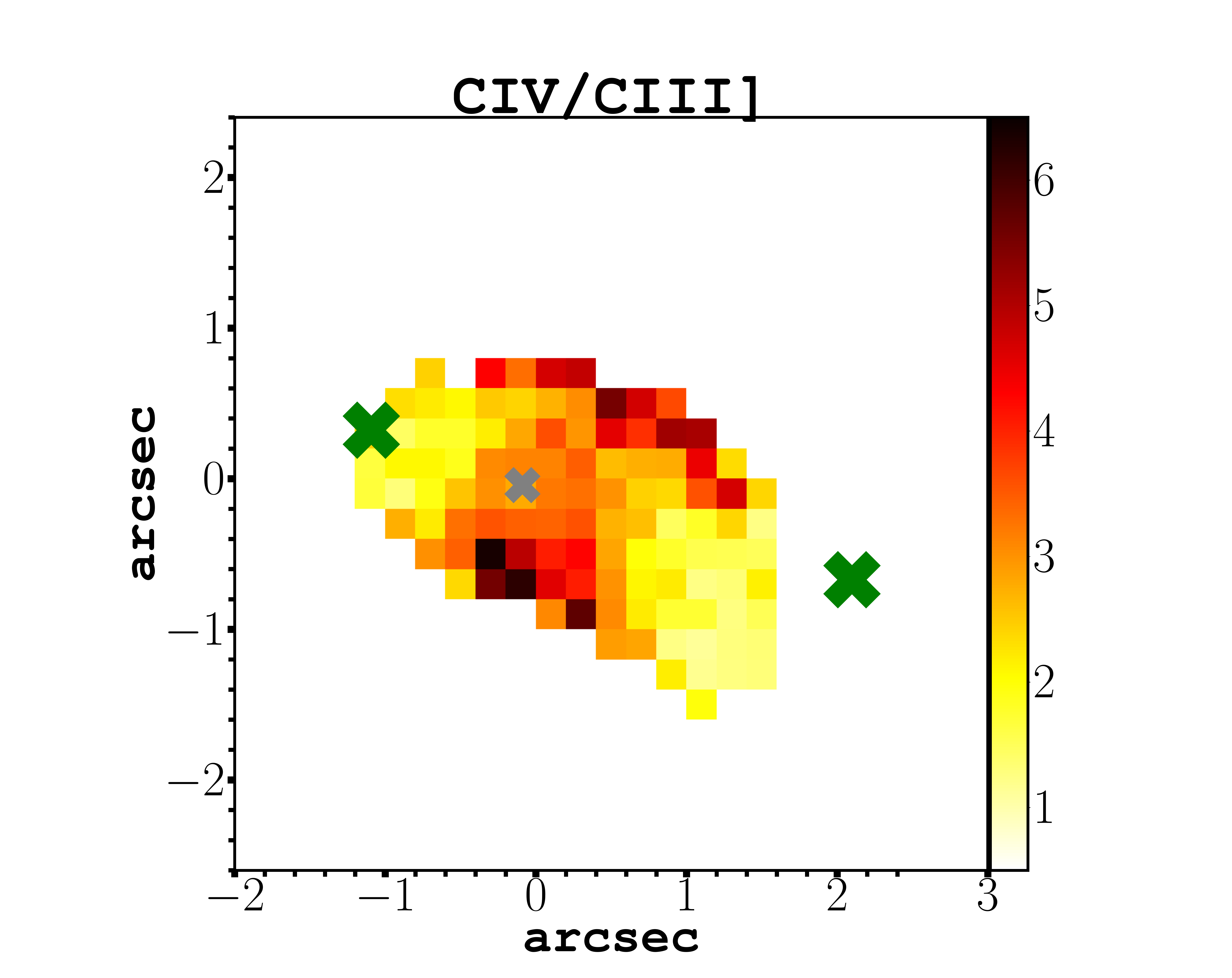}
	\includegraphics[width=5.8cm,height=5.7cm,keepaspectratio]{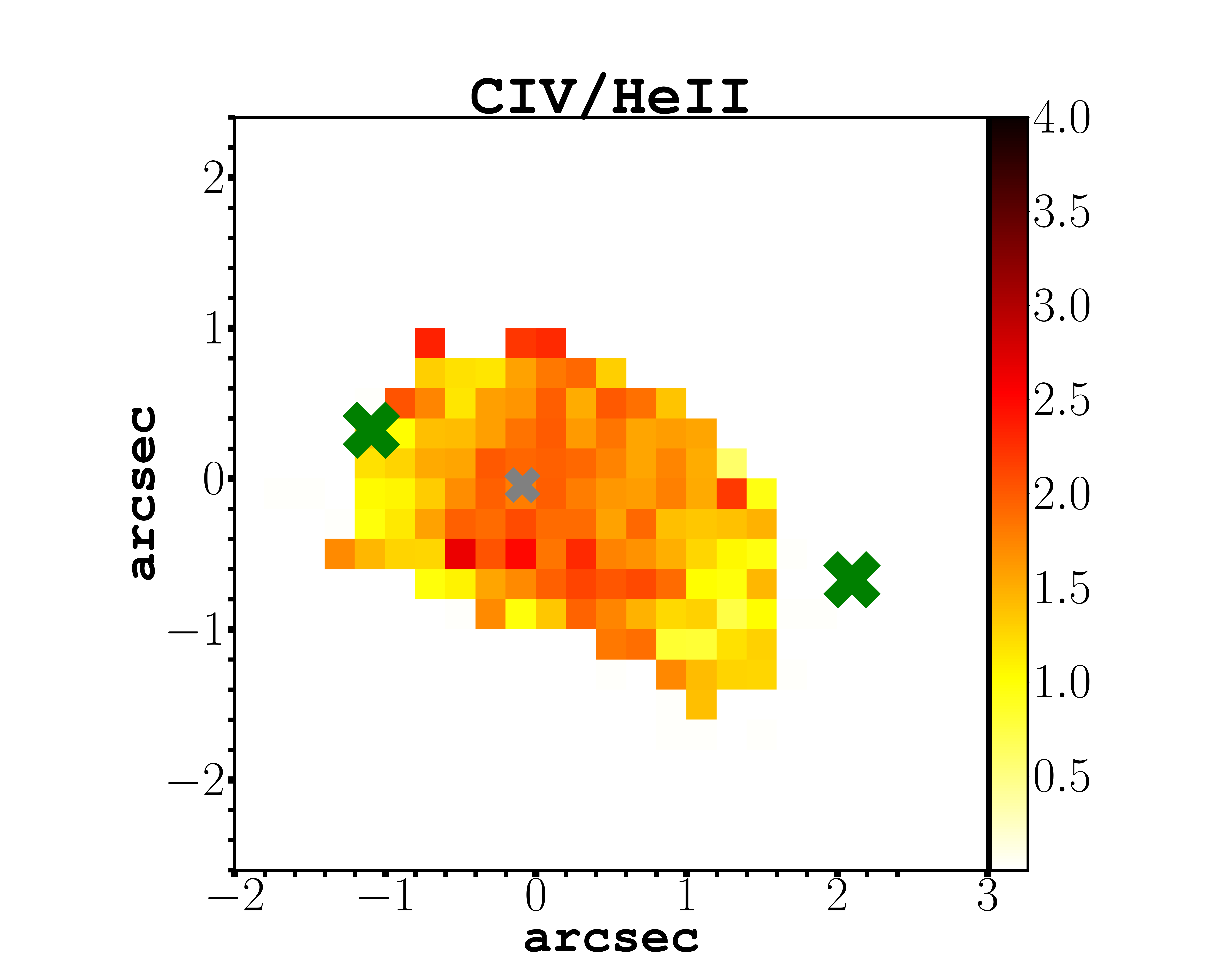}
	\includegraphics[width=5.8cm,height=5.7cm,keepaspectratio]{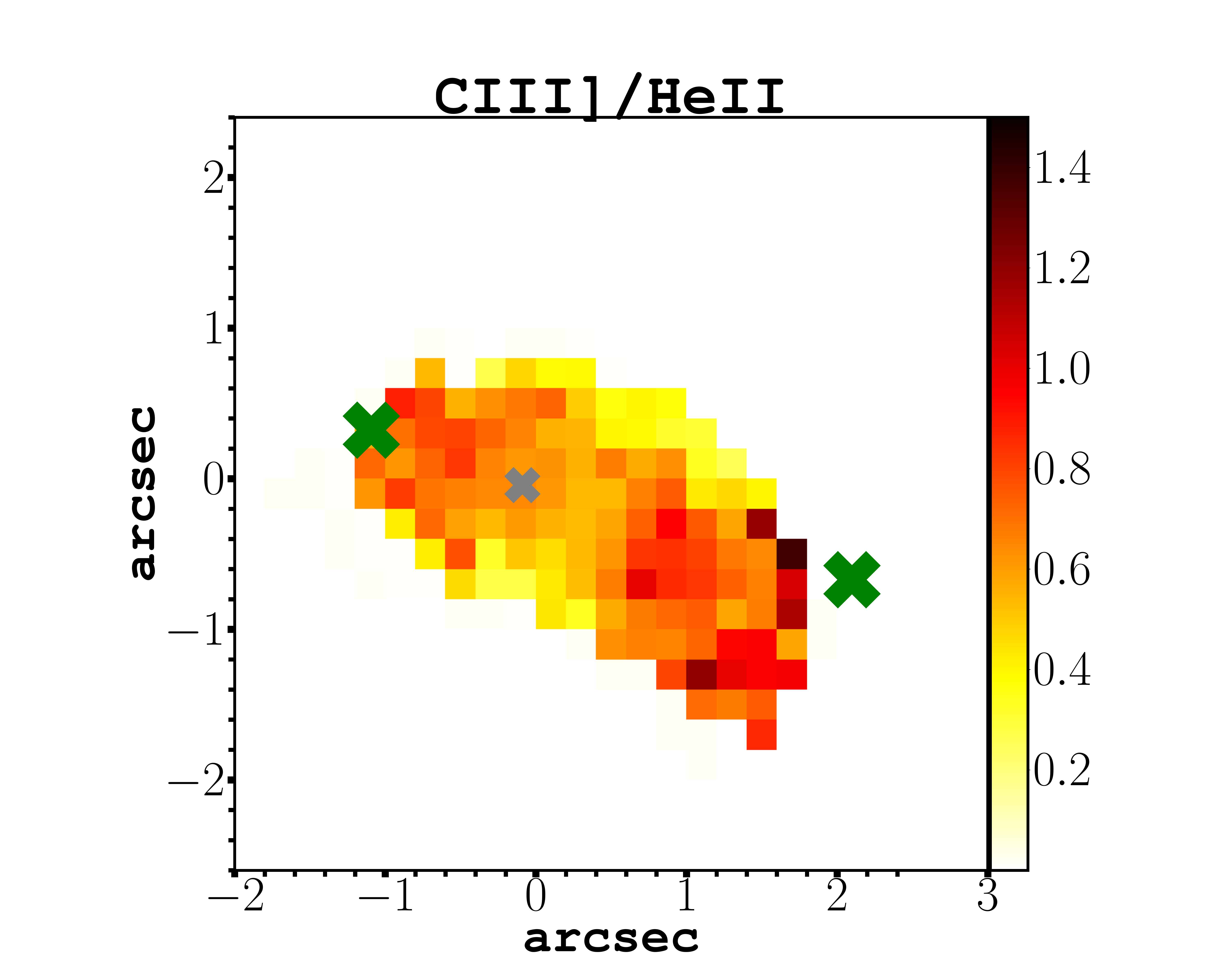}
	\quad
	\includegraphics[width=5.8cm,height=5.7cm,keepaspectratio]{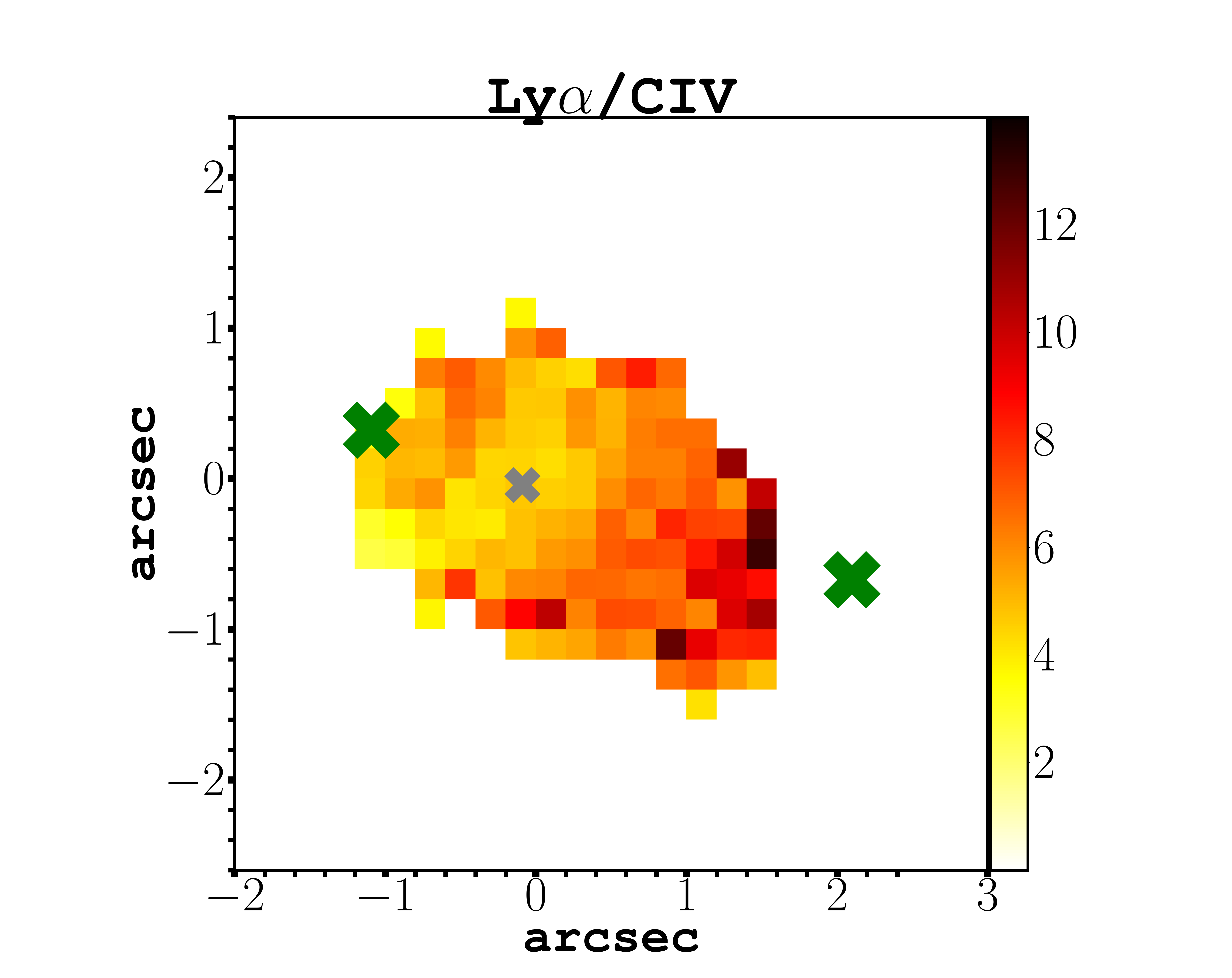}
	\includegraphics[width=5.8cm,height=5.7cm,keepaspectratio]{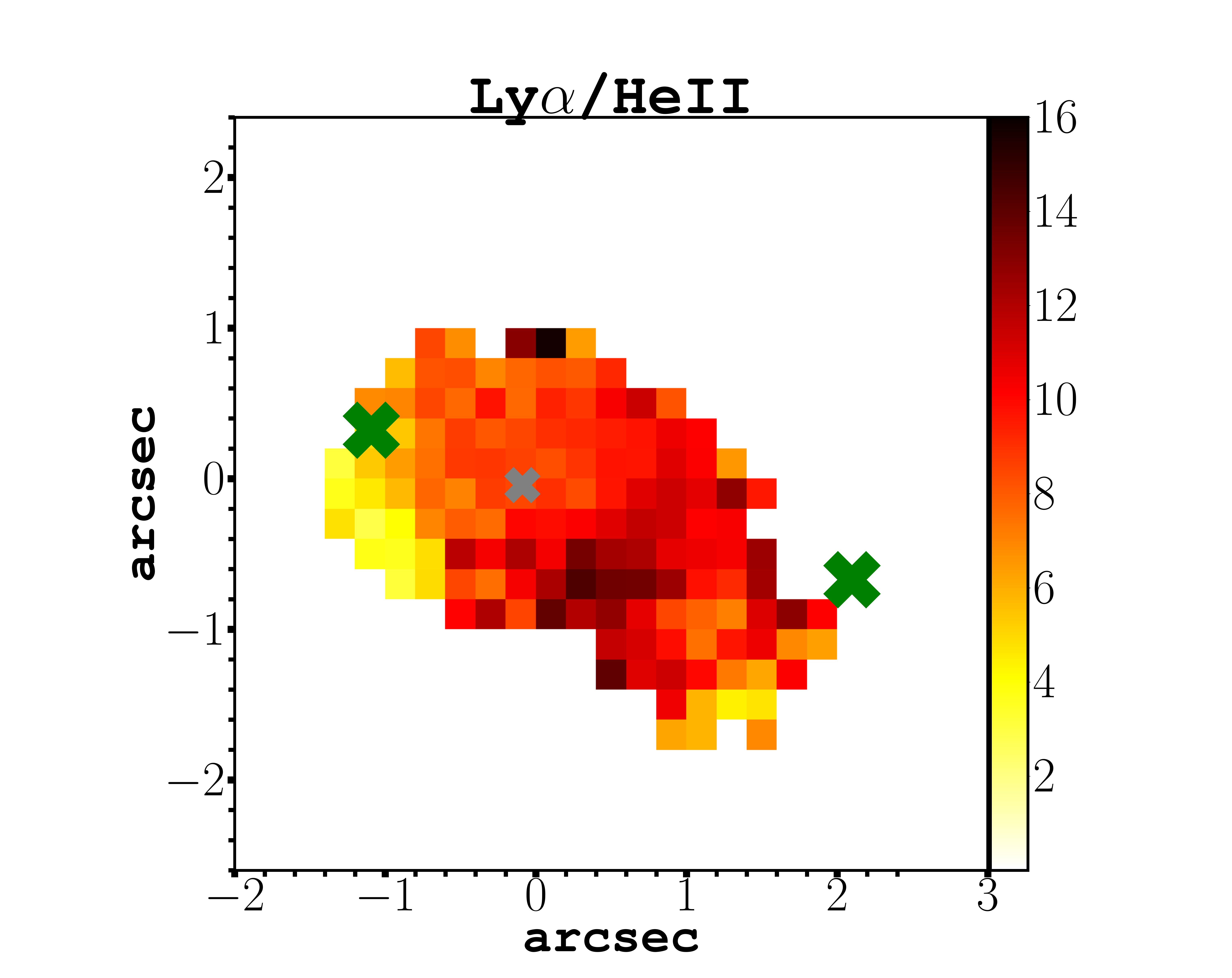}
	\includegraphics[width=5.8cm,height=5.7cm,keepaspectratio]{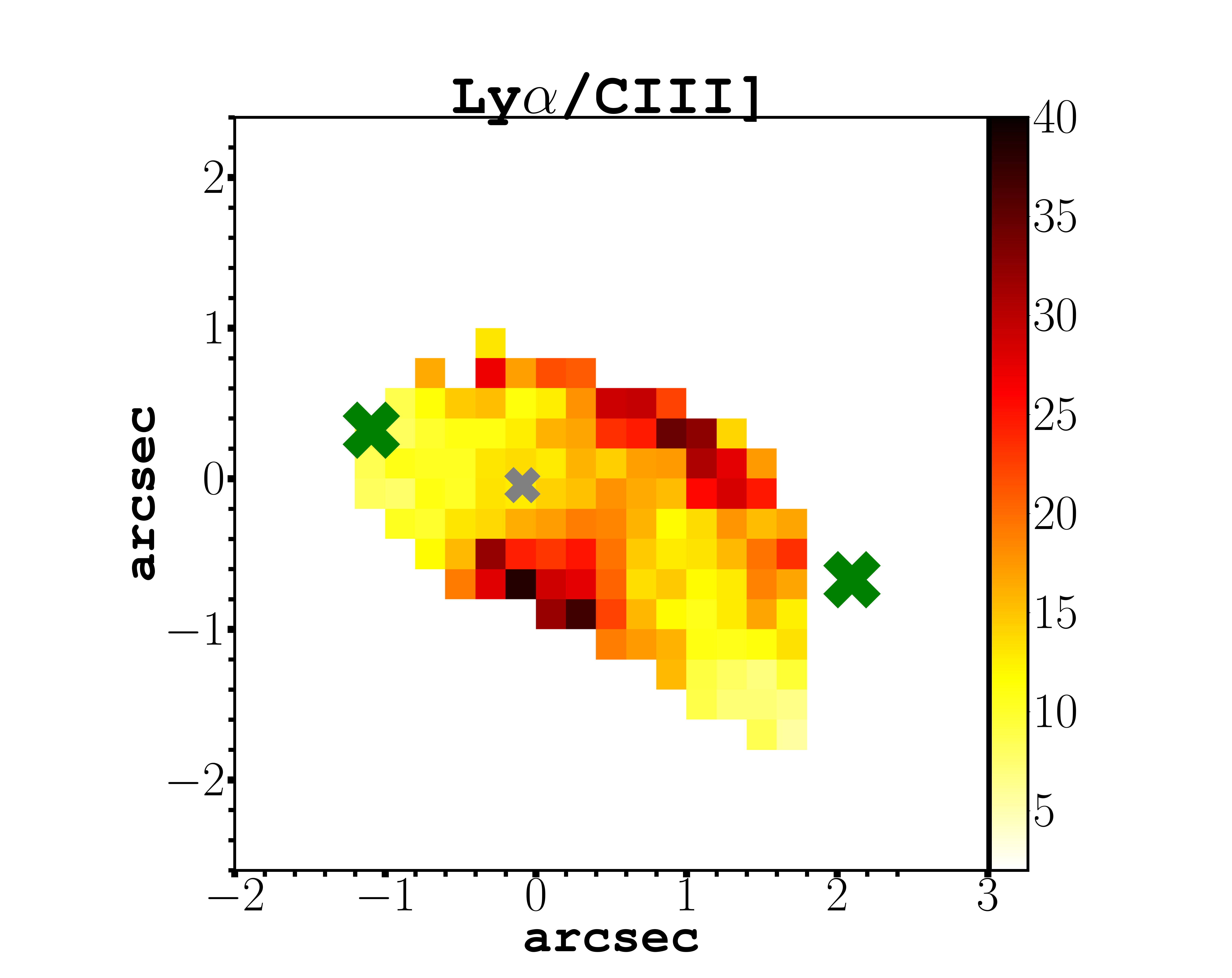}
	
	\caption{Flux ratio maps of the MRC 0943--242. See text for a detailed description of the flux ratio maps. The most remarkable feature in this maps is that the gas close to the locations of the hotspots appear clearly differentiated in several maps, such as \ion{C}{IV}/\ion{C}{III]}, \ion{C}{IV}/\ion{He}{II}, \ion{C}{III]}/\ion{He}{II} and Ly$\alpha$/\ion{C}{III]}. The green "X" represent the positions of the radio hotspots and the grey "x" the position of the AGN.}
	\label{lnr}
\end{figure*}

Our flux ratio maps use the flux maps produced by our line fitting procedure ($\S$\ref{fitting} and $\S$\ref{maps}), and we have simply divided one line flux map by another (see Fig. \ref{lnr}).

\begin{table*}
	\centering
	\caption{Measurements of the rest-frame UV and optical emission lines obtained with the fitting routine. Ly$\alpha _{15}$ and Ly$\alpha _{19}$ correspond to the lower (log N(\ion{H}{I}/cm$^{-2}$) $\sim$ 15.20) and higher (log N(\ion{H}{I}/cm$^{-2}$) $\sim$ 19.63) column density results, respectively. See Table \ref{instru_ap} for other telecopes/instruments.}
	\label{instru}
	\begin{tabular}{lccccr} 
		\hline
		Line & $\lambda _{rest}$ & $\lambda _{obs}$ & Line Flux & FWHM  &  $\Delta$v \\
		& \AA		      &		\AA		     & ($\times10^{-16}$ erg cm$^{-2}$ s$^{-1}$) & (km s$^{-1}$)  &  (km s$^{-1}$) \\
		\hline	
		\multicolumn{5}{c}{XSHOOTER SLIT} \\
		\hline
		\ion{O}{VI}+\ion{C}{II} & 1031.9,1037.6 & 4050.0 $\pm$ 0.7, 4072.4 $\pm$ 0.7 & 1.57 $\pm$ 0.24  &  869 $\pm$ 100  & 6 $\pm$ 48 \\
		Ly$\alpha _{15}$ & 1215.7 & 4769.8 $\pm$ 0.1 & 17.64 $\pm$ 0.29  &  1557 $\pm$ 18  & -84 $\pm$ 6 \\
		Ly$\alpha _{19}$ & 1215.7 & 4769.1 $\pm$ 0.1 & 20.27 $\pm$ 0.38  &  1458 $\pm$ 16  & -126 $\pm$ 5 \\
		\ion{N}{V}	  & 1238.8, 1242.8	&   4858.7 $\pm$ 1.0, 4874.3 $\pm$ 1.0  &  0.62 $\pm$ 0.11  &  883 $\pm$ 130 & -204 $\pm$ 64 \\
		\ion{C}{IV} & 1548.2,1550.8 & 6076.2 $\pm$ 0.5, 6086.4 $\pm$ 0.3 & 3.02 $\pm$ 0.20 & 1045 $\pm$ 53  &  -1 $\pm$ 23 \\
		\ion{He}{II} & 1640.4 & 6436.5 $\pm$ 0.26 & 2.05 $\pm$ 0.08 & 1018 $\pm$ 32 & 0 $\pm$ 12 \\
		\ion{C}{III]}  & 1906.7, 1908.7	& 7481.9 $\pm$ 0.6,  7489.9 $\pm$ 0.6 & 1.11 $\pm$ 0.07 & 1087 $\pm$ 59 & -49 $\pm$ 22 \\
		\ion{C}{II]}  & 2325.4, 2326.9	& 9123.2 $\pm$ 2.0, 9129.2 $\pm$ 2.0 & 0.55 $\pm$ 0.09 & 1298 $\pm$ 181 & -105 $\pm$ 64 \\
		\ion{[Ne}{IV]}  & 2421.8, 2424.4	& 9502.6 $\pm$ 0.9, 9512.9 $\pm$ 0.9 & 0.37 $\pm$ 0.10 & 386 $\pm$ 87 & -69 $\pm$ 29 \\
		\ion{Mg}{II}  & 2795.5, 2802.7 & 10974.0 $\pm$ 1.4, 11002.2 $\pm$ 1.4 & 0.97 $\pm$ 0.14 & 565 $\pm$ 66 & 67 $\pm$ 38 \\
		\ion{[Ne}{V]}  & 3425.9	& 13437.0 $\pm$ 1.2 & 1.27 $\pm$ 0.11 & 951 $\pm$ 64 & -188 $\pm$ 27 \\
		\ion{[O}{II]}  & 3726.0, 3728.8 & 14617.9 $\pm$ 0.8,  14628.8 $\pm$ 0.8 & 4.83 $\pm$ 0.18 & 1238 $\pm$ 39 & -113 $\pm$ 16 \\
		\ion{[Ne}{III]}  & 3868.7	& 15180.2 $\pm$ 0.8 & 1.79 $\pm$ 0.10 & 992 $\pm$ 45 & -67 $\pm$ 17 \\
		\ion{H}{$\gamma$}  & 4340.4	& 17032.8 $\pm$ 2.1 & 0.84 $\pm$ 0.12 & 844 $\pm$ 91 & -34 $\pm$ 37 \\
		\ion{[O}{III]}  & 4363.2	& 17118.4 $\pm$ 0.4 & 0.47 $\pm$ 0.04 & 327 $\pm$ 18 & -101 $\pm$ 8 \\
		\ion{[O}{III]}  & 4958.9	& 19459.4 $\pm$ 0.4 & 5.18 $\pm$ 0.12 & 1002 $\pm$ 17 & -42 $\pm$ 6 \\
		\ion{[O}{III]}  & 5006.8	& 19647.5 $\pm$ 0.4 & 14.97 $\pm$ 0.33 & 1002 $\pm$ 17 & -42 $\pm$ 6 \\
		\hline
		\multicolumn{5}{c}{XSHOOTER IFU} \\
		\hline
		Ly$\alpha _{15}$ & 1215.7 & 4768.2 $\pm$ 0.3 & 18.14 $\pm$ 0.67  &  1778 $\pm$ 45  & -189 $\pm$ 18 \\
		Ly$\alpha _{19}$ & 1215.7 & 4767.2 $\pm$ 0.3 & 20.81 $\pm$ 0.99  &  1674 $\pm$ 42  & -250 $\pm$ 18 \\
		\ion{C}{IV} & 1548.2,1550.8 & 6076.4 $\pm$ 0.5, 6086.6 $\pm$ 5.8 & 5.78 $\pm$ 0.32 & 1372 $\pm$ 61  &  -64 $\pm$ 25 \\
		\ion{He}{II} & 1640.4 & 6438.0 $\pm$ 0.5 & 1.30 $\pm$ 0.11 & 773 $\pm$ 51 & 0 $\pm$ 21 \\
		\ion{C}{III]}  & 1906.7, 1908.7	& 7478.6 $\pm$ 0.6,  7486.7 $\pm$ 0.6 & 1.14 $\pm$ 0.08 & 1050 $\pm$ 66 & -252 $\pm$ 24 \\
		\ion{[O}{III]}  & 4958.9	& 19461.2 $\pm$ 0.5 & 8.15 $\pm$ 0.16 & 1030 $\pm$ 19 & -88 $\pm$ 8 \\
		\ion{[O}{III]}  & 5006.8	& 19649.3 $\pm$ 0.5 & 23.56 $\pm$ 0.16 & 1030 $\pm$ 19 & -88 $\pm$ 8 \\
		\hline
		\multicolumn{5}{c}{MUSE IFU} \\
		\hline
		Ly$\alpha _{15}$ & 1215.7 & 4769.4 $\pm$ 0.1 & 18.28 $\pm$ 0.50  &  1572 $\pm$ 19  & $-$86 $\pm$ 6 \\
		Ly$\alpha _{19}$ & 1215.7 & 4769.0 $\pm$ 0.1 & 20.67 $\pm$ 1.00  &  1532 $\pm$ 25  & $-$92 $\pm$ 7 \\
		\ion{N}{V}	  & 1238.8, 1242.8	&   4860.4 $\pm$ 0.3, 4876.1 $\pm$ 0.3  &  0.70 $\pm$ 0.04  &  746 $\pm$ 34 & $-$52 $\pm$ 18 \\
		\ion{Si}{IV}  & 1402.8  & 5501.0 $\pm$ 1.2 & 0.46 $\pm$ 0.05 & 2244 $\pm$ 182 & $-$207 $\pm$ 66  \\
		\ion{N}{IV]}  & 1483.3, 1486.5	& 5825.0 $\pm$ 2.8, 5837.5 $\pm$ 2.8  &  0.29 $\pm$  0.09 &  1785 $\pm$ 321 & 220 $\pm$ 142 \\
		\ion{C}{IV} & 1548.2,1550.8 & 6075.6 $\pm$ 0.2, 6085.8 $\pm$ 0.1 & 2.88 $\pm$ 0.06 & 1120 $\pm$ 16  &  12 $\pm$ 8 \\
		\ion{He}{II} & 1640.4 & 6435.6 $\pm$ 0.2 & 1.91 $\pm$ 0.05 & 1034 $\pm$ 19 & 0 $\pm$ 7 \\
		\ion{O}{III]}  & 1660.8, 1666.1	& 6512.6 $\pm$ 0.9,  6535.7 $\pm$ 0.9 & 0.33 $\pm$ 0.04 & 921 $\pm$ 88 & -211 $\pm$ 40 \\
		\ion{C}{III]}  & 1906.7, 1908.7	& 7480.6 $\pm$ 0.3,  7488.6 $\pm$ 0.3 & 1.21 $\pm$ 0.04 & 1008 $\pm$ 32 & -60 $\pm$ 12 \\
		\ion{C}{II]}  & 2325.4, 2326.9	& 9119.3 $\pm$ 1.1, 9125.3 $\pm$ 1.1 & 0.49 $\pm$ 0.04 & 1445 $\pm$ 110 & -193 $\pm$ 37 \\
		\hline
	\end{tabular}
\end{table*}

\begin{table*}
	\centering
	\caption{Best fit parameters for the Ly$\alpha$ absorption features, for different instruments. Column (1) gives the redshift for the Ly$\alpha$ emission Gaussian. Column (2) gives the redshift for each Ly$\alpha$ absorption. Column (3) gives the column density ($N_{\ion{H}{I}}$). Column (4) gives the Doppler width b. Column (5) gives the velocity shift of the main absorber with respect to \ion{He}{II} emission in the same spectrum. Note: The \ion{He}{II} emission line was outside the spectral range covered by the red arm of VLT UVES and thus we do not give the velocity shift for this instrument. See Table \ref{instru02_ap} for more results.}
	\label{instru02}
	\begin{tabular}{lcccr} 
		\hline
		Ly$\alpha$ emission redshift & Absorption redshift & Column Density & Doppler b Parameter &  $\Delta$v \\
		($z_{em}$)  		    &      ($z_{abs}$)	     &	  (cm$^{-2}$)      & (km s$^{-1}$) & (km s$^{-1})$  \\
		\hline	
		\multicolumn{5}{c}{XSHOOTER SLIT} \\
		\hline
		& 2.90746 $\pm$ 0.00009 & (9.17 $\pm$ 1.57)$\times10^{13}$ & 55 $\pm$ 8  &     \\
		2.92303 $\pm$ 0.00007 & 2.91917 $\pm$ 0.00002 & (4.27 $\pm$ 0.20)$\times10^{19}$ & 52 $\pm$ 1  &   -421 $\pm$ 1  \\
		& 2.92702 $\pm$ 0.00006 & (8.06 $\pm$ 0.70)$\times10^{13}$ & 142 $\pm$ 9  &     \\
		& 2.93301 $\pm$ 0.00004 & (2.30 $\pm$ 0.30)$\times10^{13}$ & 35 $\pm$ 5  &     \\
		\hline
		& 2.90747 $\pm$ 0.00311 & (8.09 $\pm$ 1.66)$\times10^{13}$ & 48 $\pm$ 9  &     \\
		2.92360 $\pm$ 0.00007 & 2.91915 $\pm$ 0.00001 & (1.52 $\pm$ 0.12)$\times10^{15}$ & 153 $\pm$ 4  &   -422 $\pm$ 2  \\
		& 2.92672 $\pm$ 0.00002 & (6.39 $\pm$ 0.73)$\times10^{13}$ & 142 $\pm$ 12  &     \\
		& 2.93304 $\pm$ 0.00001 & (1.95 $\pm$ 0.30)$\times10^{13}$ & 31 $\pm$ 5  &     \\
		\hline	
		\multicolumn{5}{c}{XSHOOTER IFU} \\
		\hline
		& 2.90643 $\pm$ 0.00003 & (3.15 $\pm$ 2.51)$\times10^{14}$ & 54 $\pm$ 17  &     \\
		2.92165 $\pm$ 0.00024 & 2.91777 $\pm$ 0.00006 & (3.46 $\pm$ 0.50)$\times10^{19}$ & 52 $\pm$ 2  &   -533 $\pm$ 4  \\
		& 2.92525 $\pm$ 0.00012 & (1.95 $\pm$ 0.63)$\times10^{13}$ & 40 $\pm$ 13  &     \\
		& 2.93178 $\pm$ 0.00010 & (1.84 $\pm$ 0.76)$\times10^{13}$ & 23 $\pm$ 10  &     \\
		\hline
		& 2.90643 $\pm$ 0.00015 & (3.16 $\pm$ 0.14)$\times10^{14}$ & 51 $\pm$ 19  &     \\
		2.92226 $\pm$ 0.00024 & 2.91778 $\pm$ 0.00003 & (1.86 $\pm$ 0.65)$\times10^{15}$ & 136 $\pm$ 12  &   -600 $\pm$ 6  \\
		& 2.92522 $\pm$ 0.00012 & (1.58 $\pm$ 0.58)$\times10^{13}$ & 34 $\pm$ 13  &     \\
		& 2.93178 $\pm$ 0.00010 & (1.66 $\pm$ 0.75)$\times10^{13}$ & 22 $\pm$ 10  &     \\
		\hline	
		\multicolumn{5}{c}{MUSE IFU} \\
		\hline
		& 2.90692 $\pm$ 0.00020 & (1.13 $\pm$ 0.23)$\times10^{14}$ & 67 $\pm$ 32  &     \\
		2.92291 $\pm$ 0.00010 & 2.91843 $\pm$ 0.00008 & (3.50 $\pm$ 0.39)$\times10^{19}$ & 50 $\pm$ 3  &   -434 $\pm$ 6  \\
		& 2.92714 $\pm$ 0.00018 & (2.76 $\pm$ 1.04)$\times10^{13}$ & 114 $\pm$ 40  &     \\
		& 2.93192 $\pm$ 0.00021 & (2.69 $\pm$ 0.83)$\times10^{13}$ & 92 $\pm$ 39  &     \\
		\hline
		& 2.90686 $\pm$ 0.00032 & (9.27 $\pm$ 3.44)$\times10^{13}$ & 60 $\pm$ 21  &     \\
		2.92324 $\pm$ 0.00008 & 2.91858 $\pm$ 0.00002 & (1.06 $\pm$ 0.07)$\times10^{15}$ & 175 $\pm$ 6  &   -442 $\pm$ 2  \\
		& 2.92666 $\pm$ 0.00020 & (1.77 $\pm$ 0.76)$\times10^{13}$ & 88 $\pm$ 51  &     \\
		& 2.93207 $\pm$ 0.00033 & (1.54 $\pm$ 0.67)$\times10^{13}$ & 55 $\pm$ 41  &     \\
		\hline	
	\end{tabular}
\end{table*}

\begin{table*}
	\centering
	\caption{Best fit parameters for the \ion{C}{IV} absorption features, for different instruments. Column (1) gives the redshift for the \ion{C}{IV} emission Gaussian. Column (2) gives the redshift for each \ion{C}{IV} absorption. Column (3) gives the column density (N$_{\ion{C}{IV}}$). Column (4) gives the Doppler width b. Column (5) gives the velocity shift of the main absorber with respect to \ion{He}{II} emission in the same spectrum. Note: The \ion{He}{II} emission line was outside the spectral range covered by the red arm of VLT UVES and thus we do not give the velocity shift for this instrument. See Table \ref{instru03_ap} for more results.}
	\label{instru03}
	\begin{tabular}{lcccr} 
		\hline
		CIV emission redshift & Absorption redshift & Column Density & Doppler Parameter &  $\Delta$v \\
		($z_{em}$)  		    &      ($z_{abs}$)	     &	  (cm$^{-2}$)      & (km s$^{-1}$) & (km s$^{-1}$)  \\
		\hline
		\multicolumn{5}{c}{XSHOOTER SLIT} \\
		\hline
		2.92466 $\pm$ 0.00030 & 2.91965 $\pm$ 0.00011 & (3.65 $\pm$ 0.53)$\times10^{14}$ & 114 $\pm$ 14  &  -384 $\pm$ 8   \\
		\hline
		\multicolumn{5}{c}{XSHOOTER IFU} \\
		\hline
		2.92479 $\pm$ 0.00033 & 2.91921 $\pm$ 0.00031 & (6.19 $\pm$ 1.27)$\times10^{14}$ & 62 $\pm$ 7  &  -491 $\pm$ 9   \\
		\hline
		\multicolumn{5}{c}{MUSE IFU} \\
		\hline
		2.92427 $\pm$ 0.00010 & 2.91927 $\pm$ 0.00004 & (3.14 $\pm$ 0.18)$\times10^{14}$ & 100 $\pm$ 7  &  -371 $\pm$ 3   \\
		\hline
		
	\end{tabular}
\end{table*}

\subsection{Ionization Models}
\label{model_grid}

\begin{figure*}
	\subfloat[$\alpha$ = --1.5]{
		\includegraphics[width=\columnwidth,height=6.8cm,keepaspectratio]{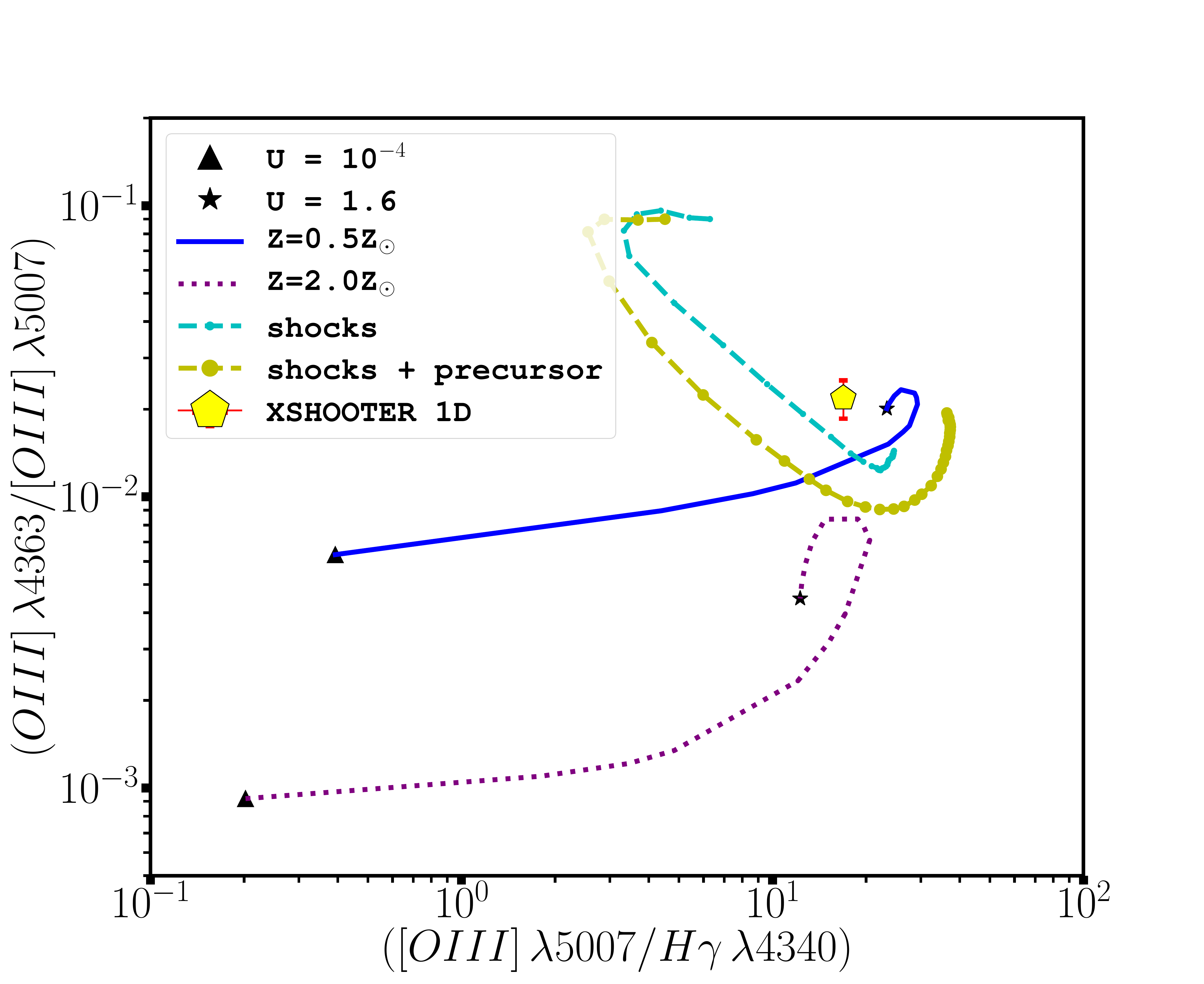}}
	\subfloat[$\alpha$ = --1.0]{
		\includegraphics[width=\columnwidth,height=6.8cm,keepaspectratio]{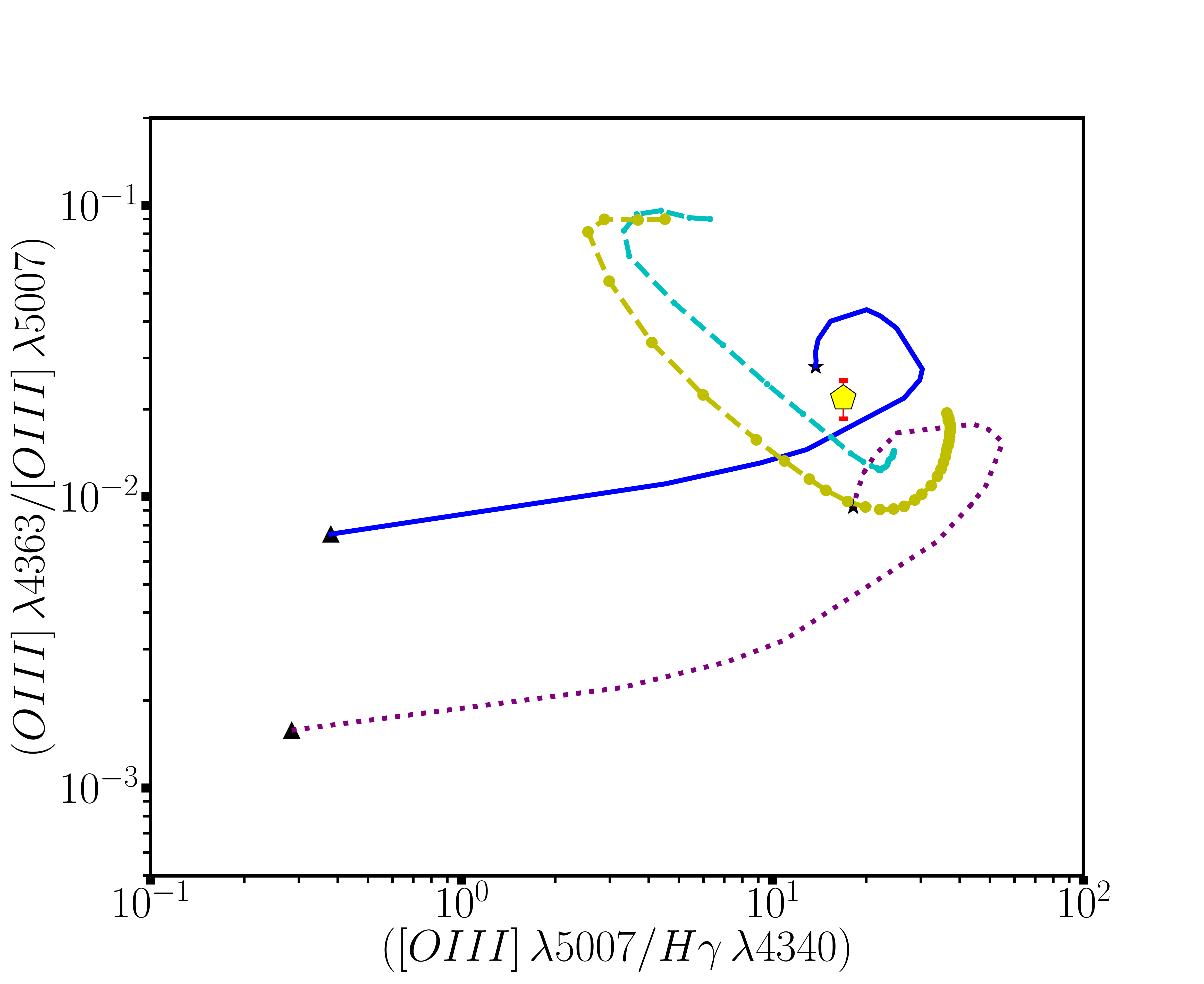}}
	
	\caption{Comparison of the observed emission line ratios using integrated spectra from the X-SHOOTER long-slit (yellow pentagon) with photoionization ($0.5Z_{\odot}$ sequence is represented by the solid blue line and the $2.0Z_{\odot}$ sequence by the purple dotted line), pure shock models (blue solid circles connected by a dashed line) and the composite shock + precursor models (large yellow solid circles connected by a dashed line). In the case of the photoionization models, we use ionizing continuum power law index $\alpha$=--1.5 (left side) or $\alpha$=--1.0 (right side). At the end of each sequence, a solid black triangle corresponds to the initial value of the ionization parameter ($U$ = 10$^{-4}$) and a solid black star that corresponds to the maximum value of the ionization parameter ($U$ = 1.6). The pure shock and the composite shock + precursor models are from \citet{allen2008}. Both shock model sequences are characterized by hydrogen density 100 cm$^{-3}$, magnetic field 100 $\mu$ G and velocity covering the range $v_{s}$ = 100 up to 1000 km s$^{-1}$.}
	\label{models05}
\end{figure*}

\begin{figure*}
	\subfloat[$\alpha$ = --1.5]{
		\includegraphics[width=\columnwidth,height=6.8cm,keepaspectratio]{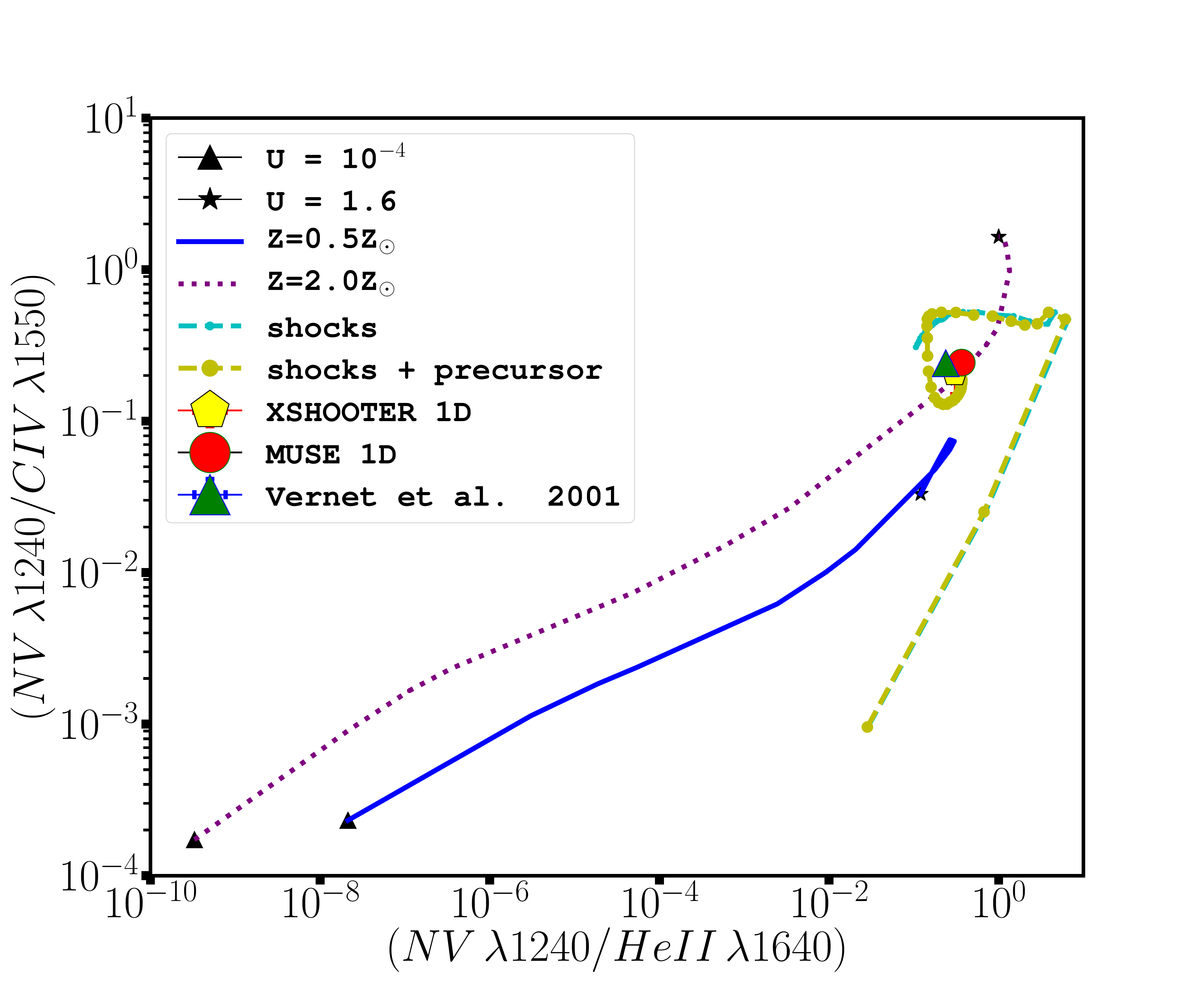}}
	\subfloat[$\alpha$ = --1.0]{
		\includegraphics[width=\columnwidth,height=6.8cm,keepaspectratio]{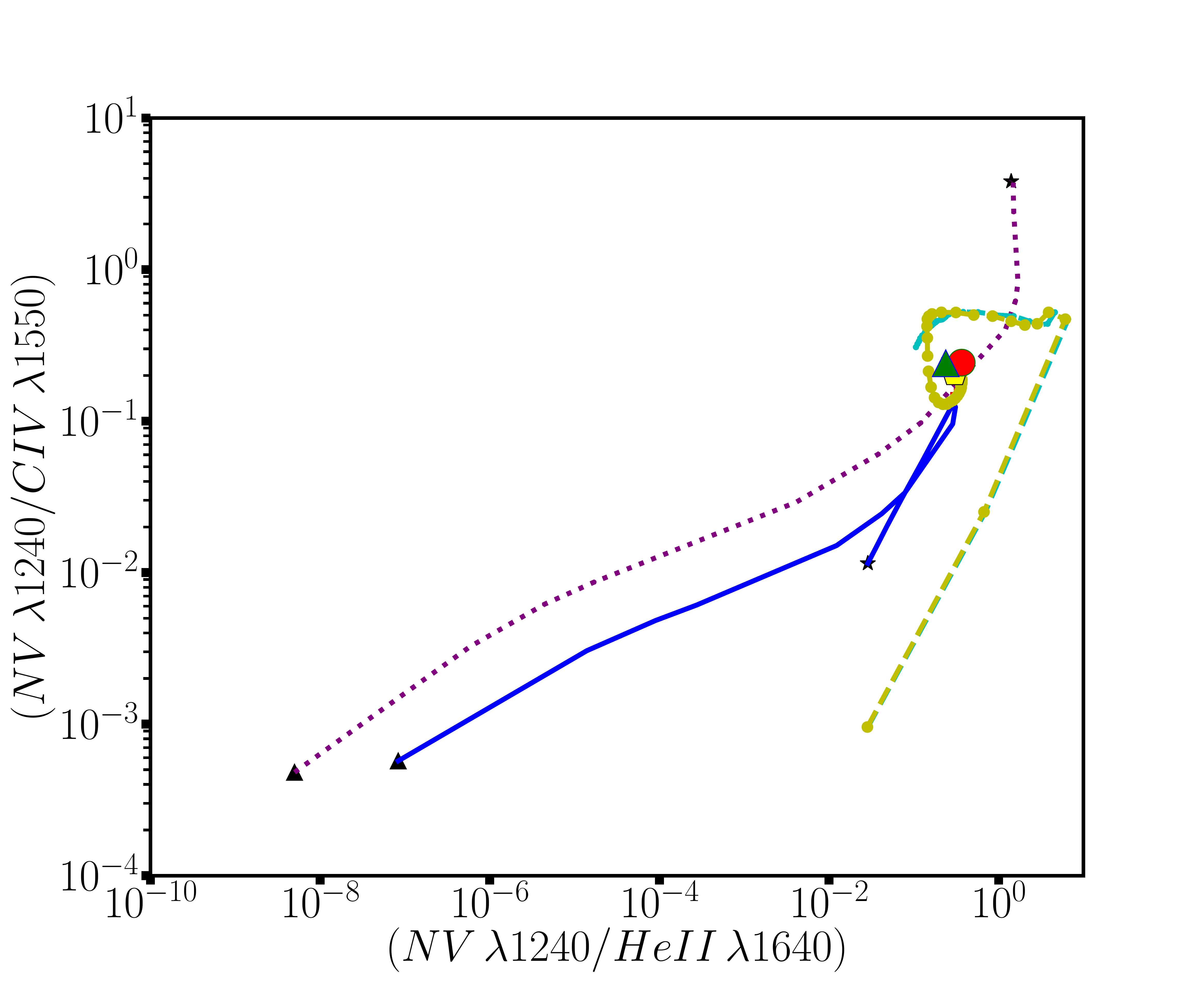}}
	\quad
	\subfloat[$\alpha$ = --1.5]{
		\includegraphics[width=\columnwidth,height=6.8cm,keepaspectratio]{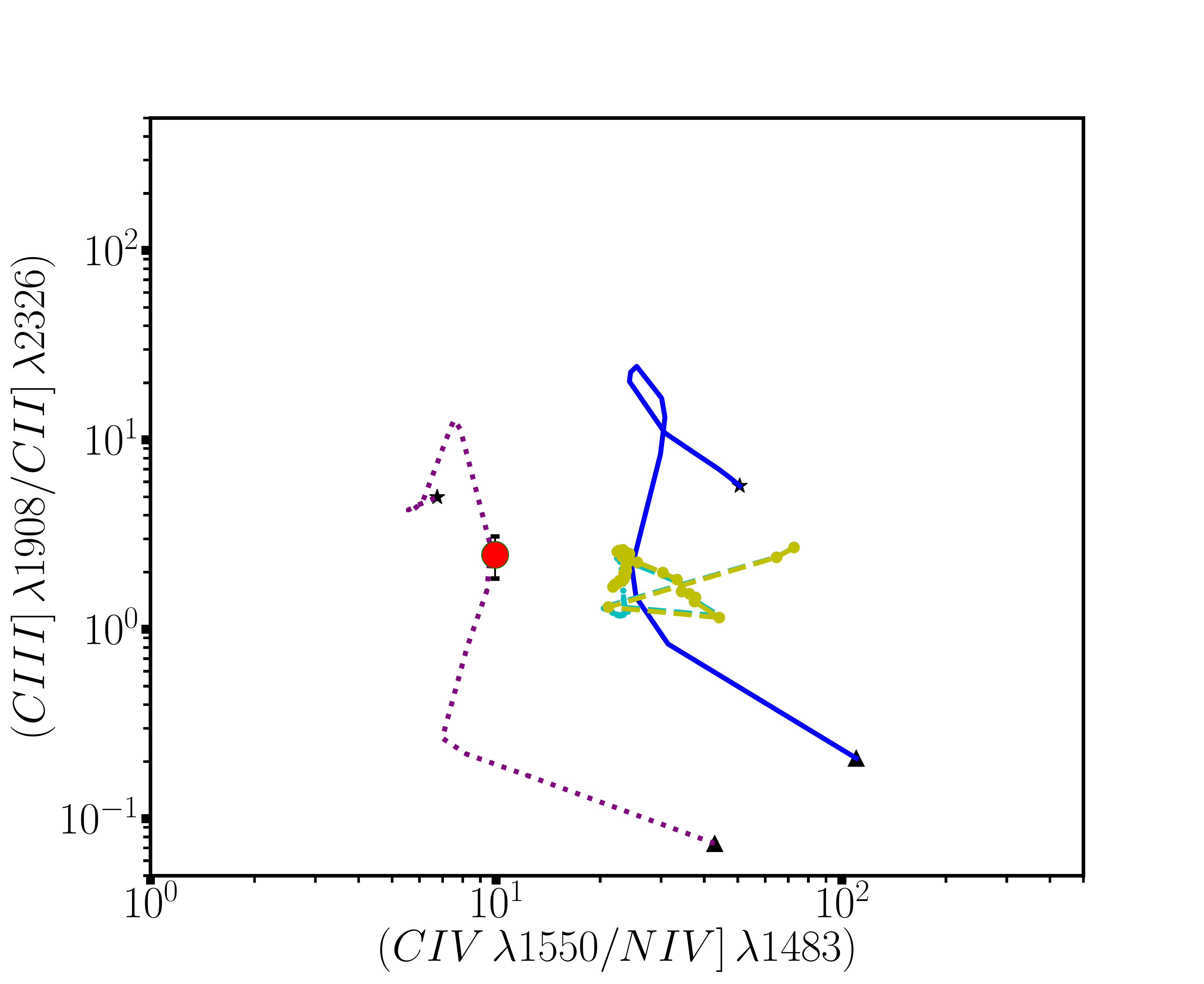}}
	\subfloat[$\alpha$ = --1.0]{
		\includegraphics[width=\columnwidth,height=6.8cm,keepaspectratio]{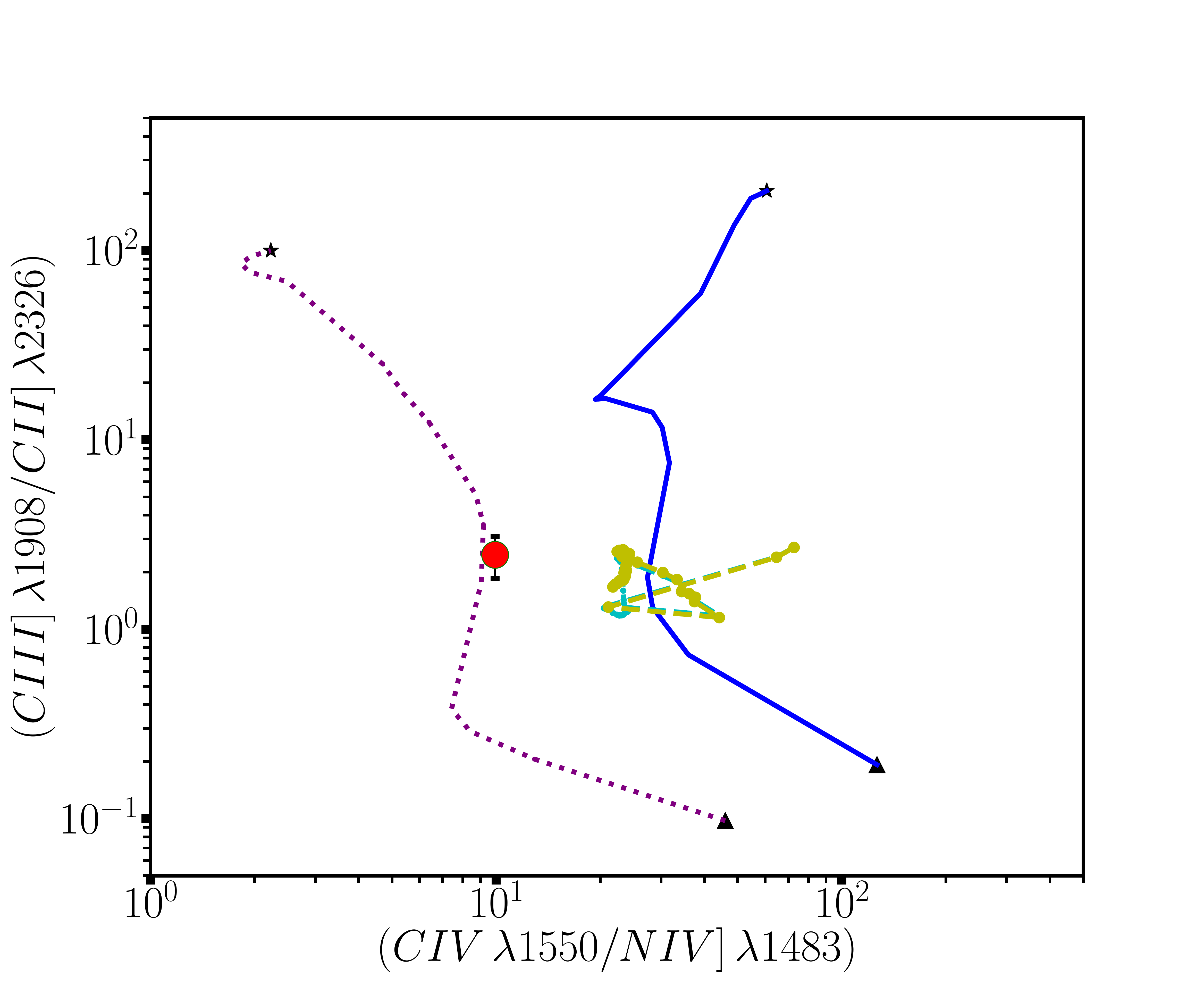}}
	\quad
	\subfloat[$\alpha$ = --1.5]{
		\includegraphics[width=\columnwidth,height=6.8cm,keepaspectratio]{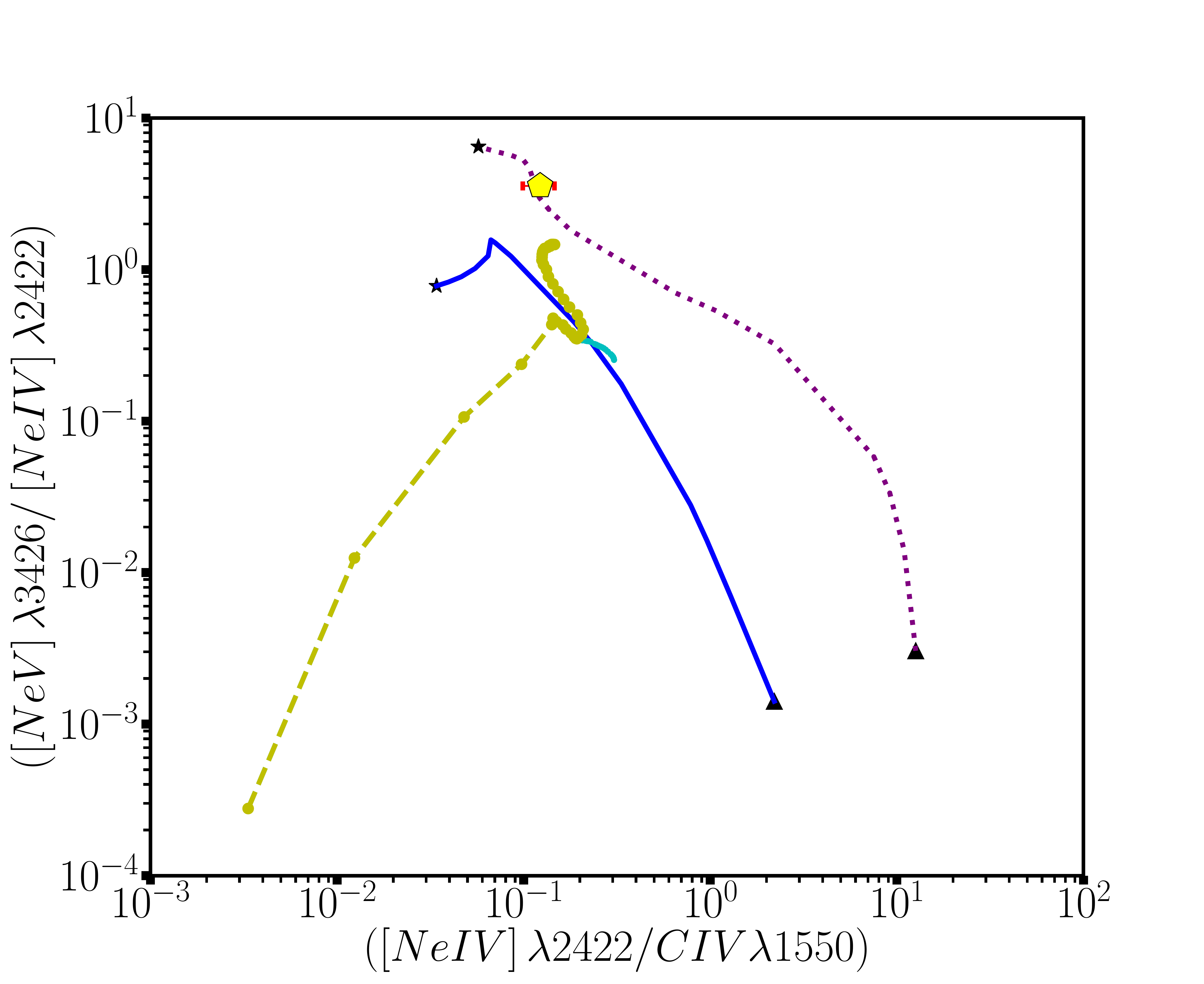}}
	\subfloat[$\alpha$ = --1.0]{
		\includegraphics[width=\columnwidth,height=6.8cm,keepaspectratio]{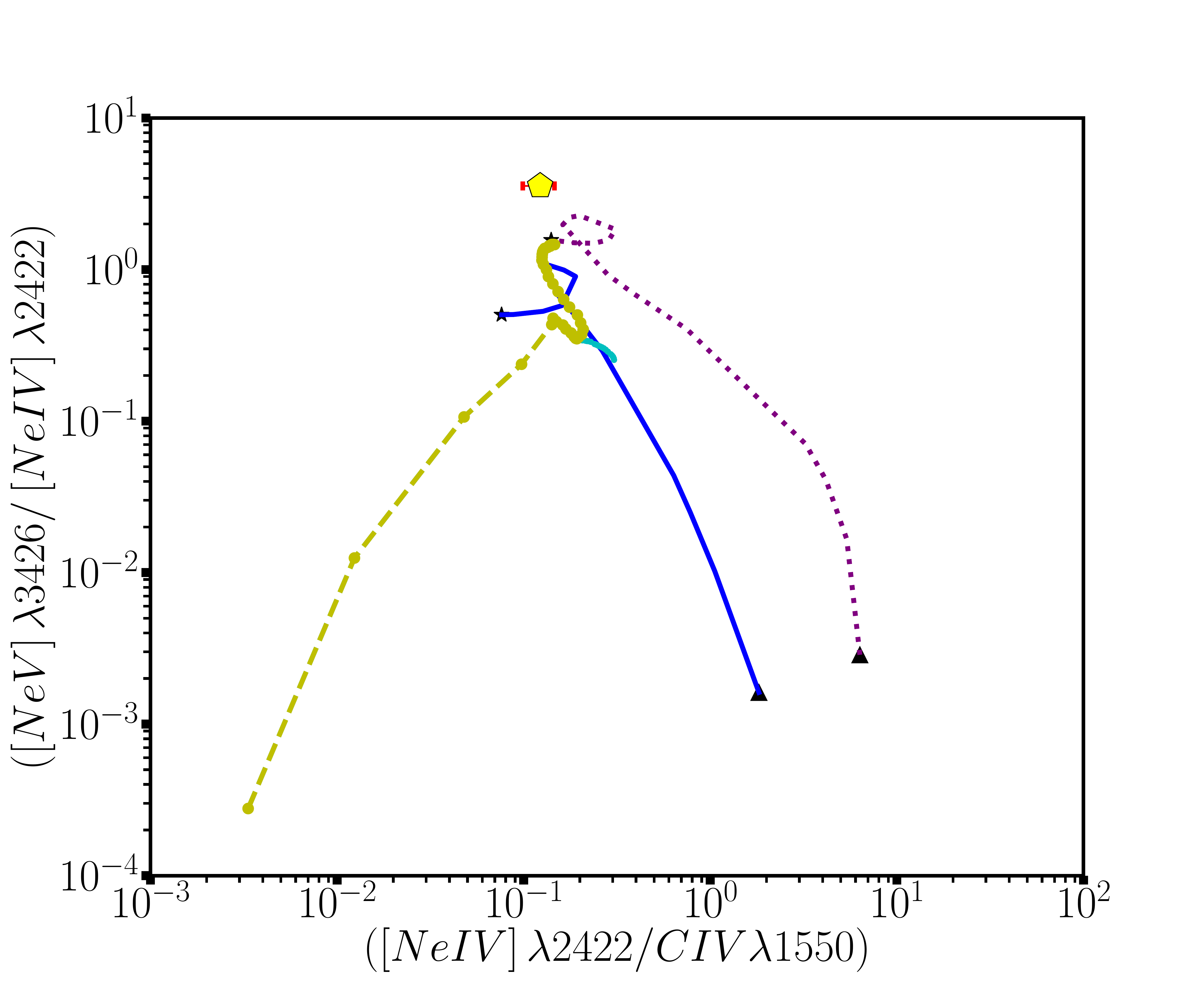}}
	
	\caption{Comparison of the observed emission line ratios using integrated spectra from the MUSE IFU (red clircle), X-SHOOTER long-slit (yellow pentagon) and KECK II LRIS spectra (green triangle) with photoionization, shocks and the composite shock + precursor models shown. The green triangle corresponds to the emission line ratios from the KECK II data from \citet{Ve2001}. See Fig. \ref{models05} for more details.}
	\label{models03}
\end{figure*}

\begin{figure*}
	\subfloat[$\alpha$ = --1.5]{
		\includegraphics[width=\columnwidth,height=6.8cm,keepaspectratio]{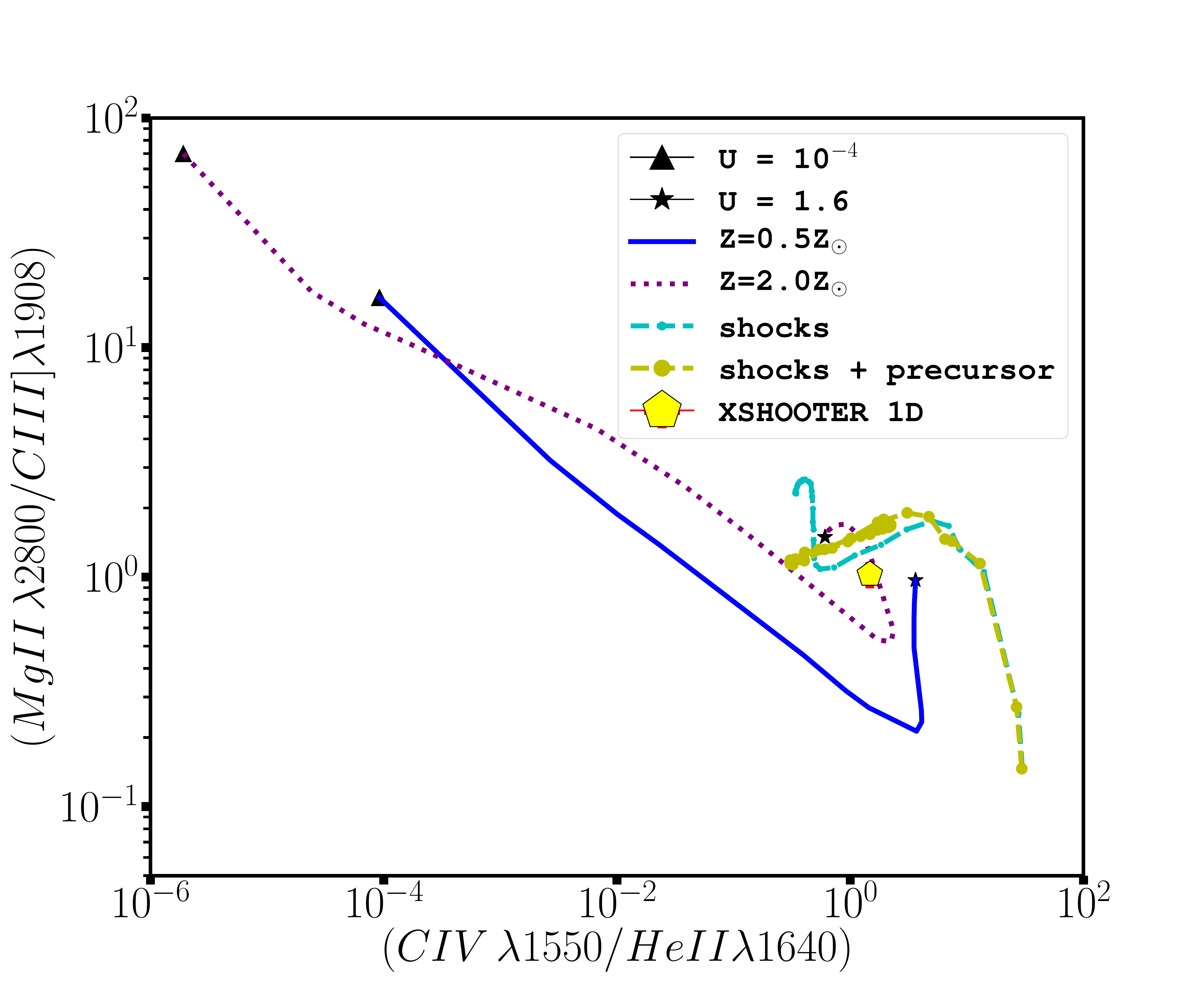}}
	\subfloat[$\alpha$ = --1.0]{
		\includegraphics[width=\columnwidth,height=6.8cm,keepaspectratio]{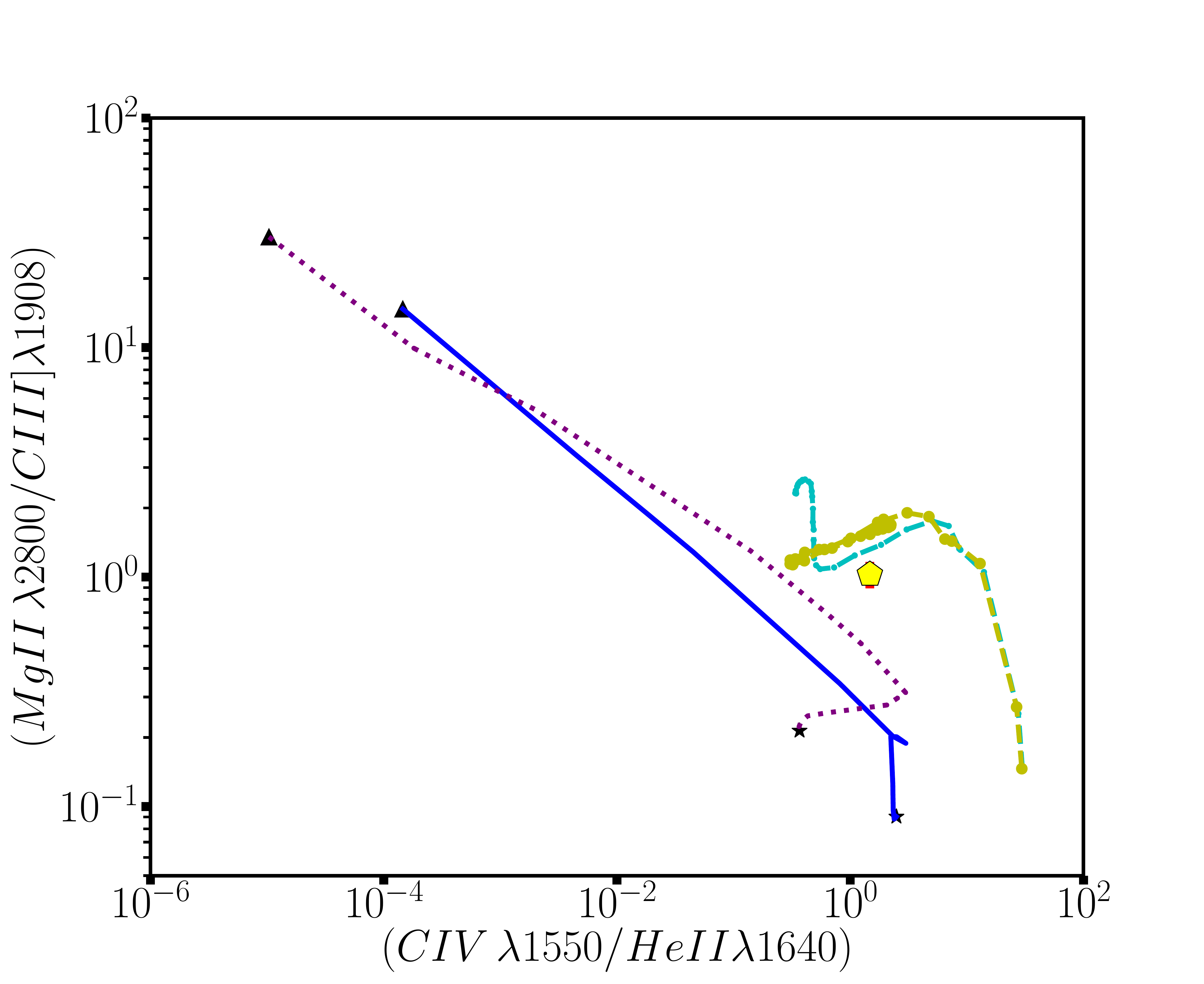}}
	\quad
	\subfloat[$\alpha$ = --1.5]{
		\includegraphics[width=\columnwidth,height=6.8cm,keepaspectratio]{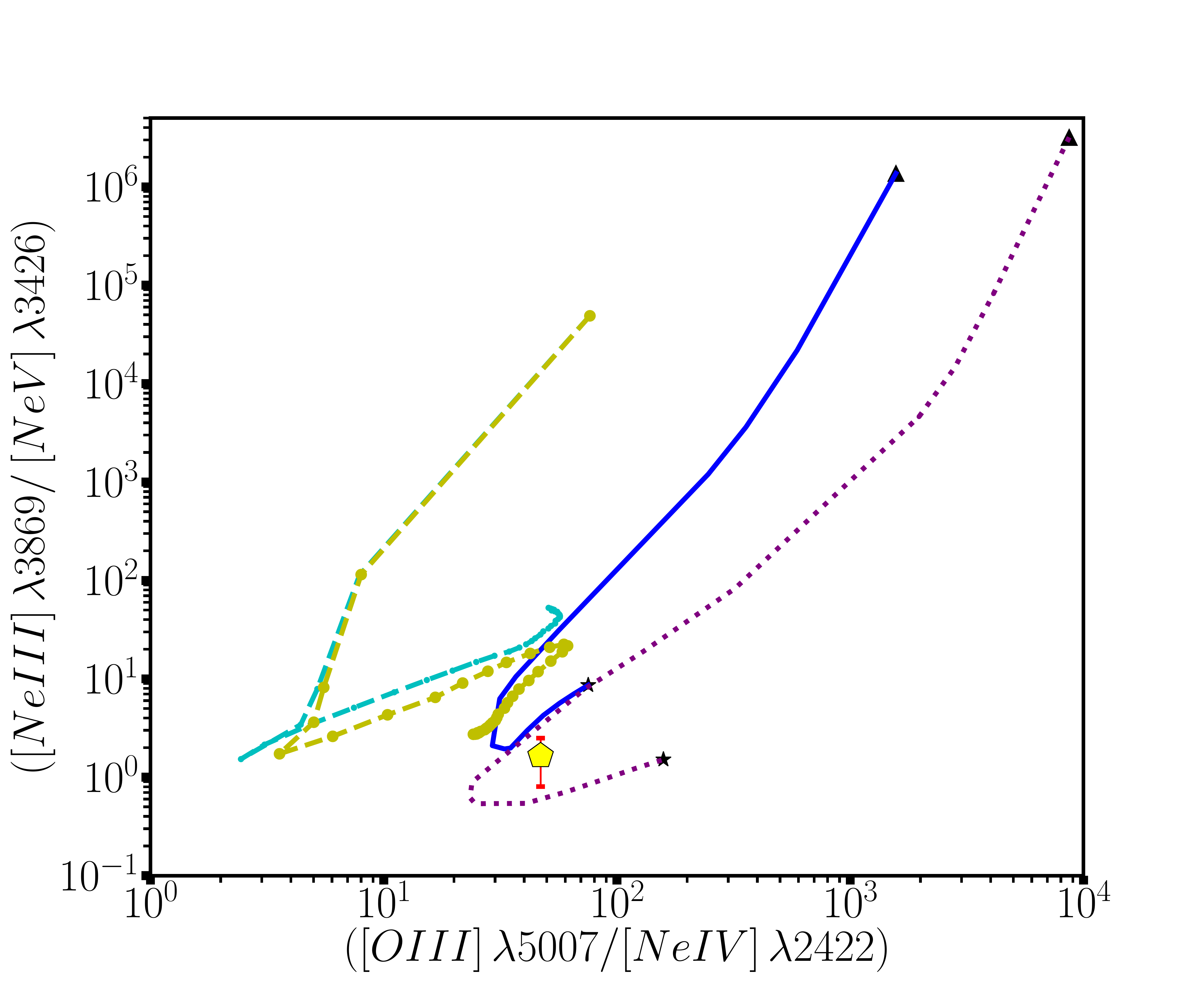}}
	\subfloat[$\alpha$ = --1.0]{
		\includegraphics[width=\columnwidth,height=6.8cm,keepaspectratio]{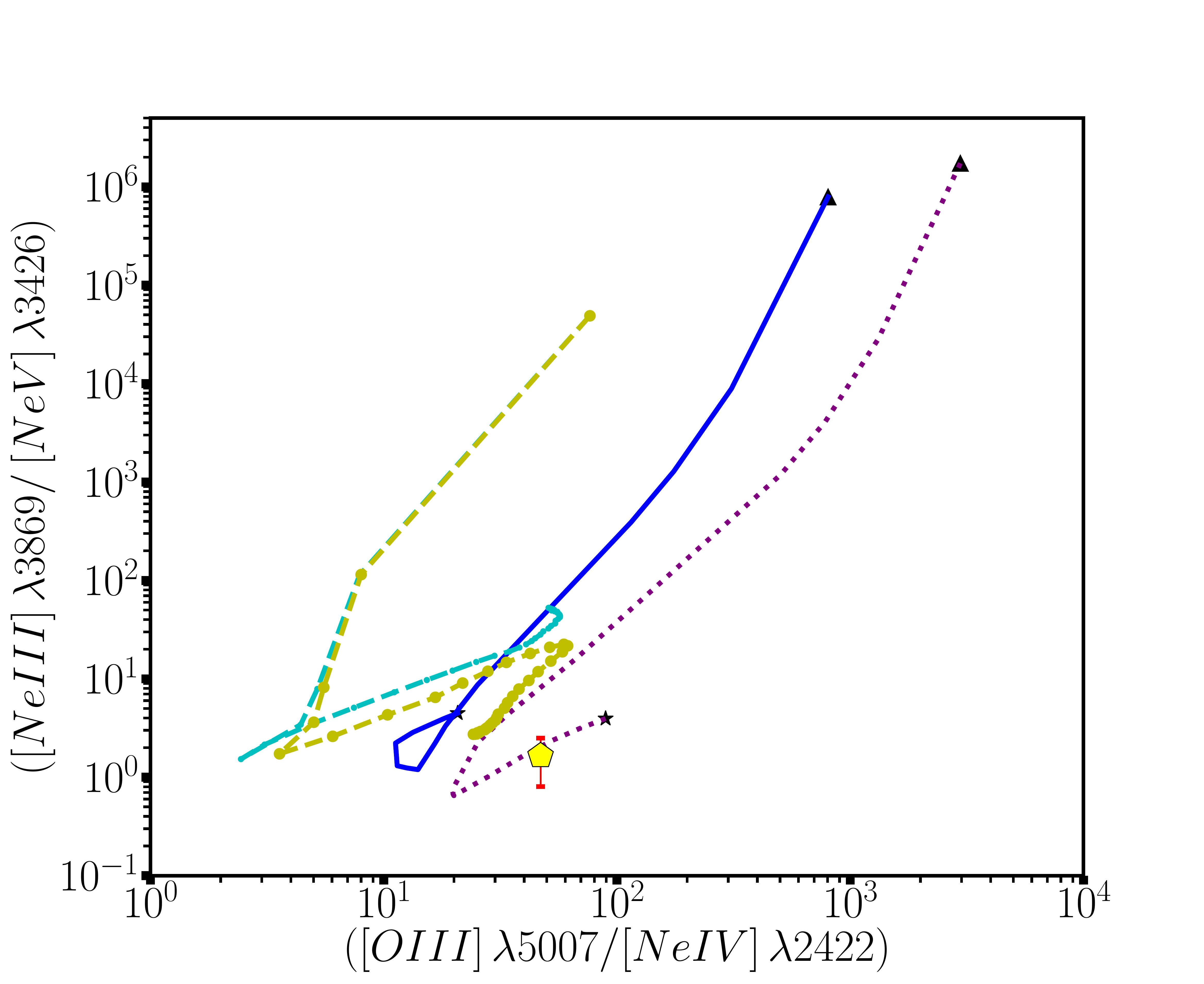}}
	\quad
	\subfloat[$\alpha$ = --1.5]{
		\includegraphics[width=\columnwidth,height=6.8cm,keepaspectratio]{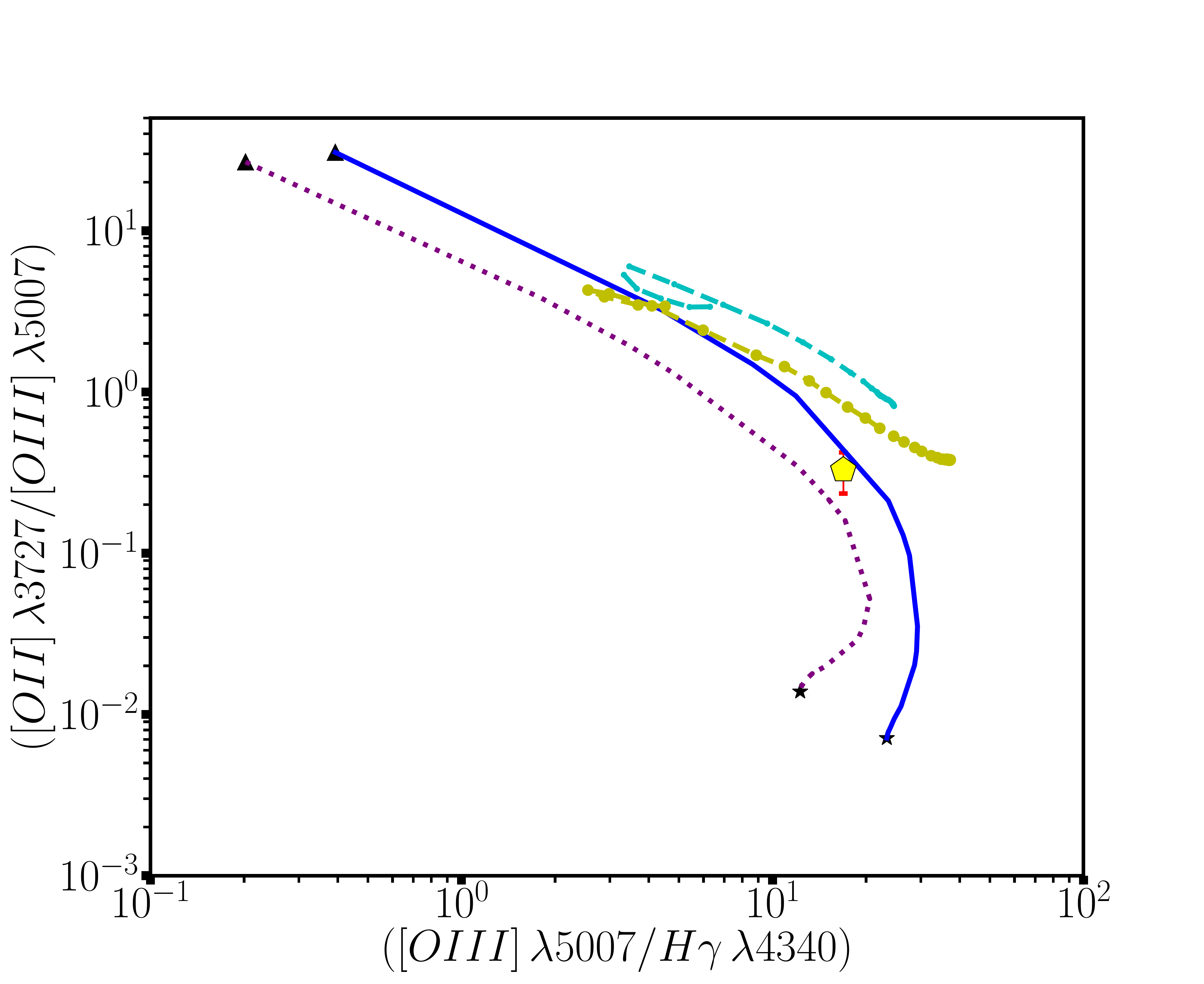}}
	\subfloat[$\alpha$ = --1.0]{
		\includegraphics[width=\columnwidth,height=6.8cm,keepaspectratio]{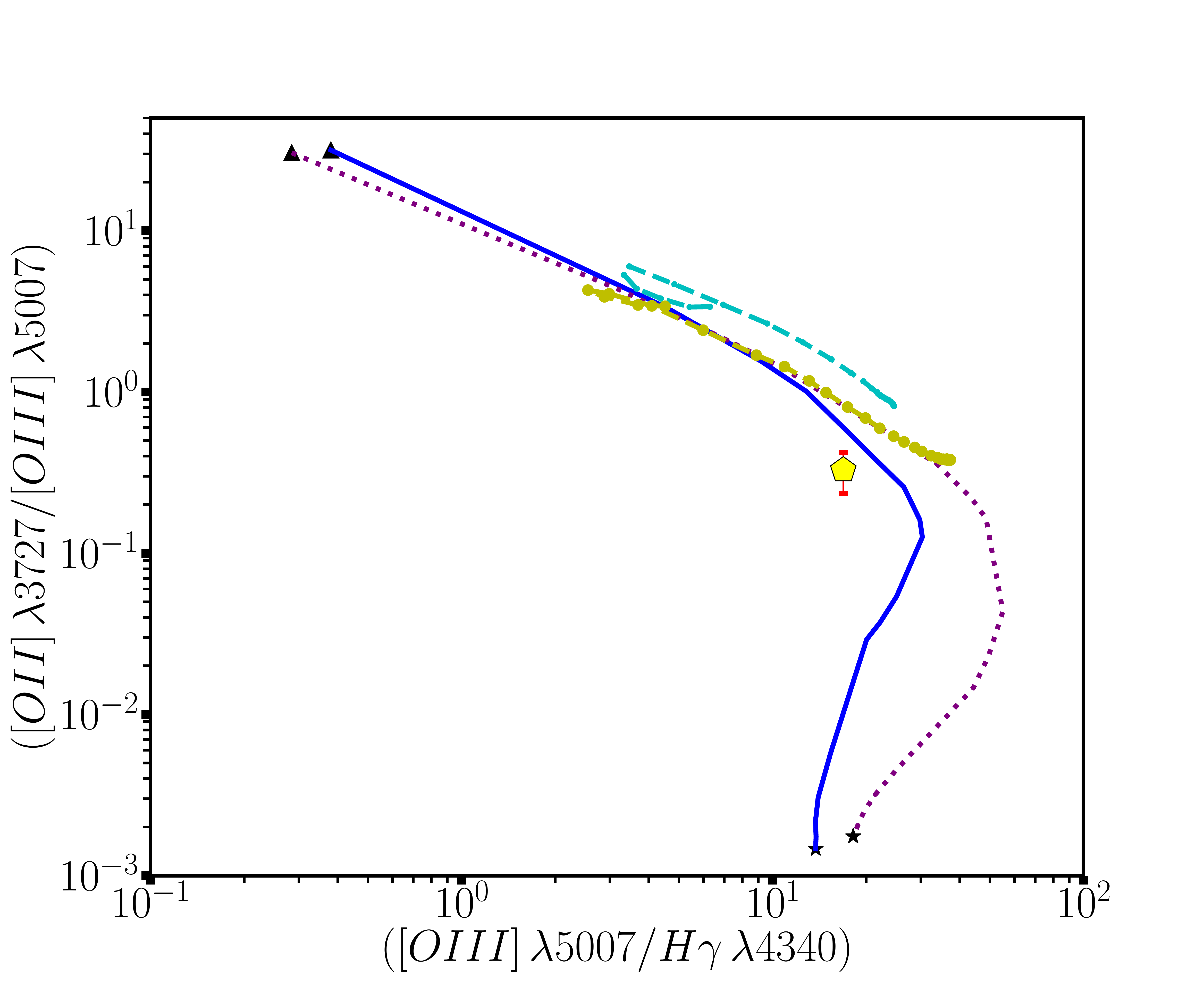}}
	
	\caption{Comparison of the observed emission line ratios using integrated spectra from the X-SHOOTER long-slit (yellow pentagon) with photoionization, shocks and the composite shock + precursor models. See Fig. \ref{models05} for more details.}
	\label{models04}
\end{figure*}

\begin{figure*}
	\subfloat[$\alpha$ = --1.5]{
	\includegraphics[width=\columnwidth,height=6.8cm,keepaspectratio]{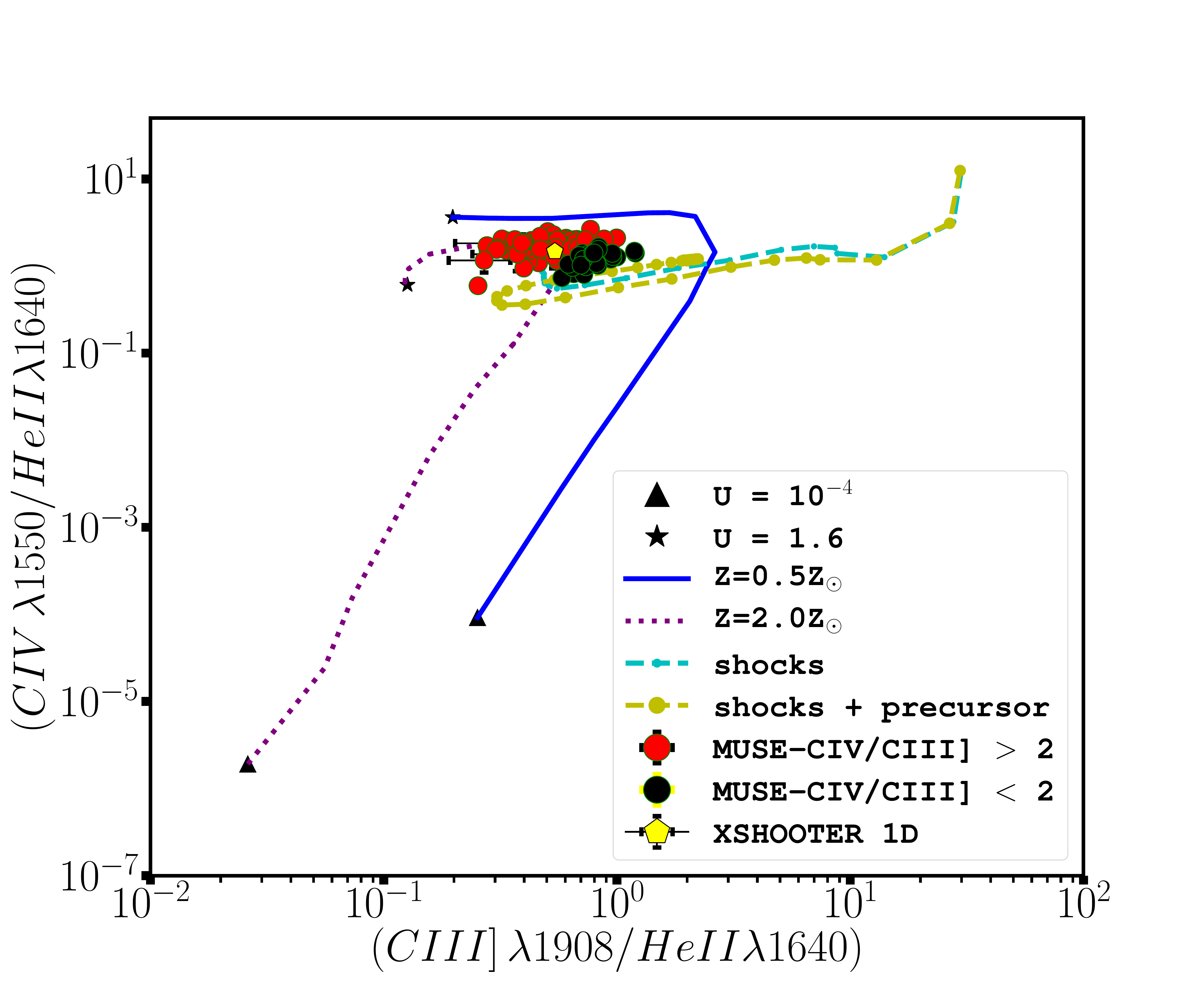}}
	\subfloat[$\alpha$ = --1.0]{
	\includegraphics[width=\columnwidth,height=6.8cm,keepaspectratio]{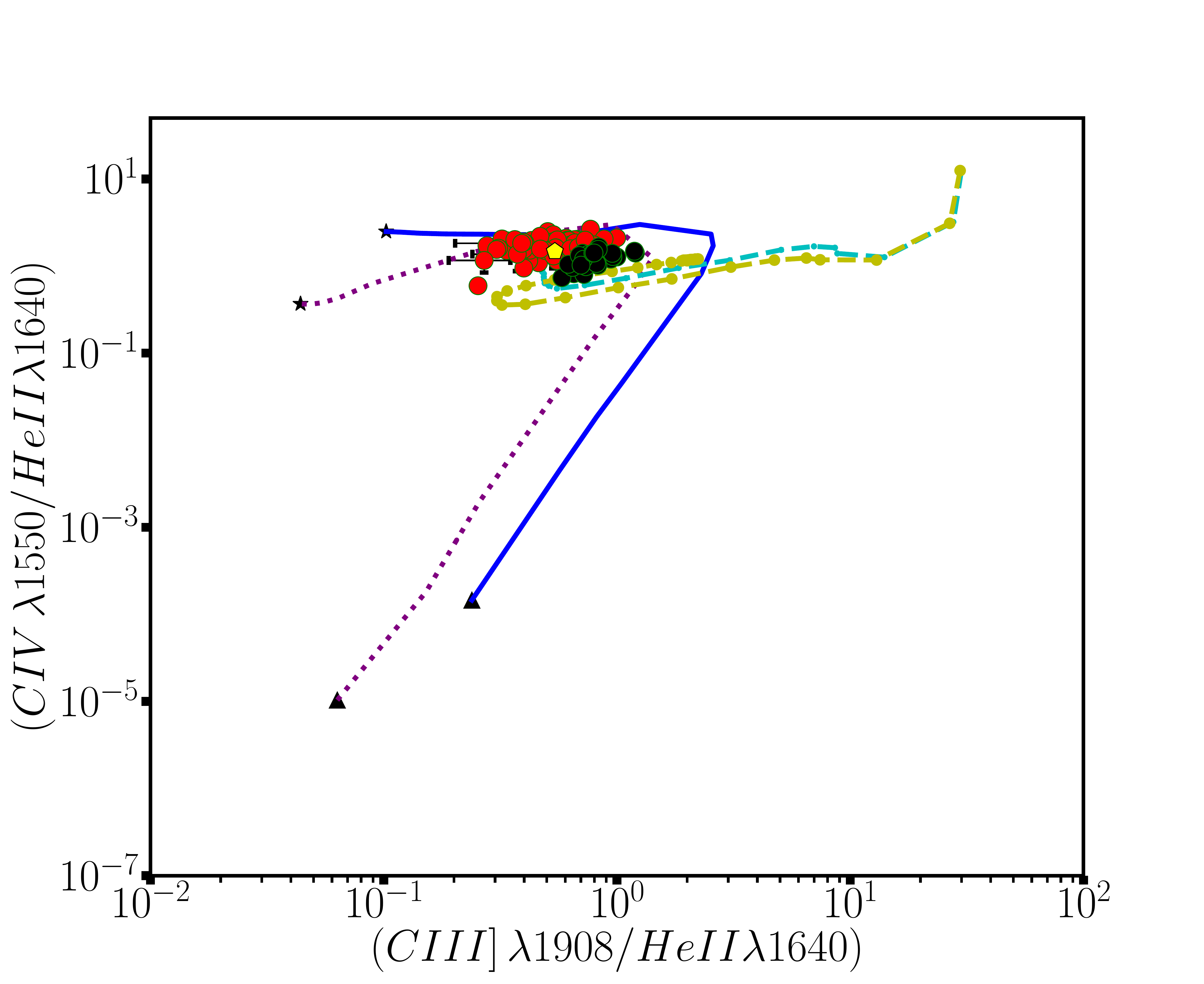}}
	\quad
	\subfloat[$\alpha$ = --1.5]{
	\includegraphics[width=\columnwidth,height=6.8cm,keepaspectratio]{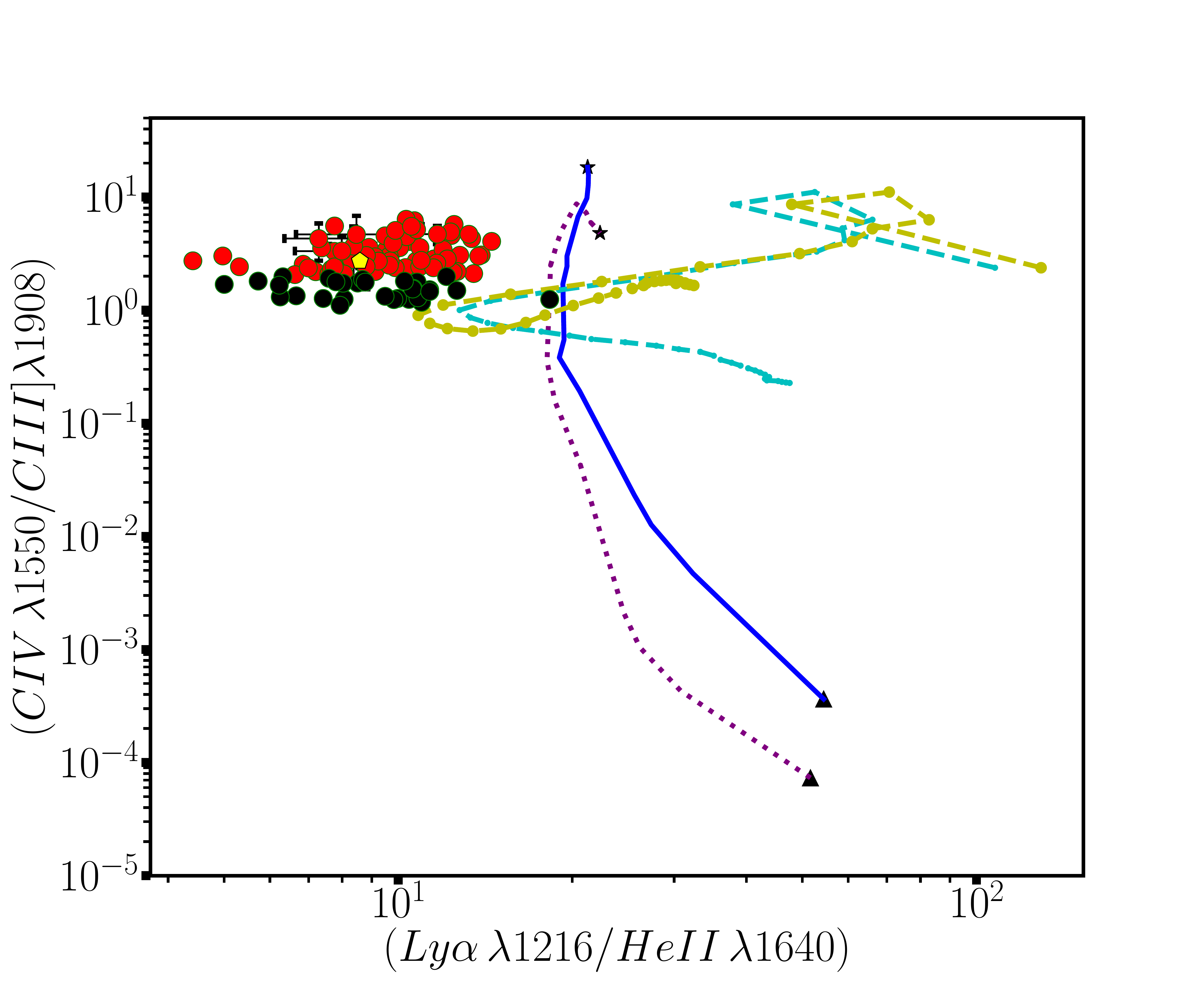}}
	\subfloat[$\alpha$ = --1.0]{
	\includegraphics[width=\columnwidth,height=6.8cm,keepaspectratio]{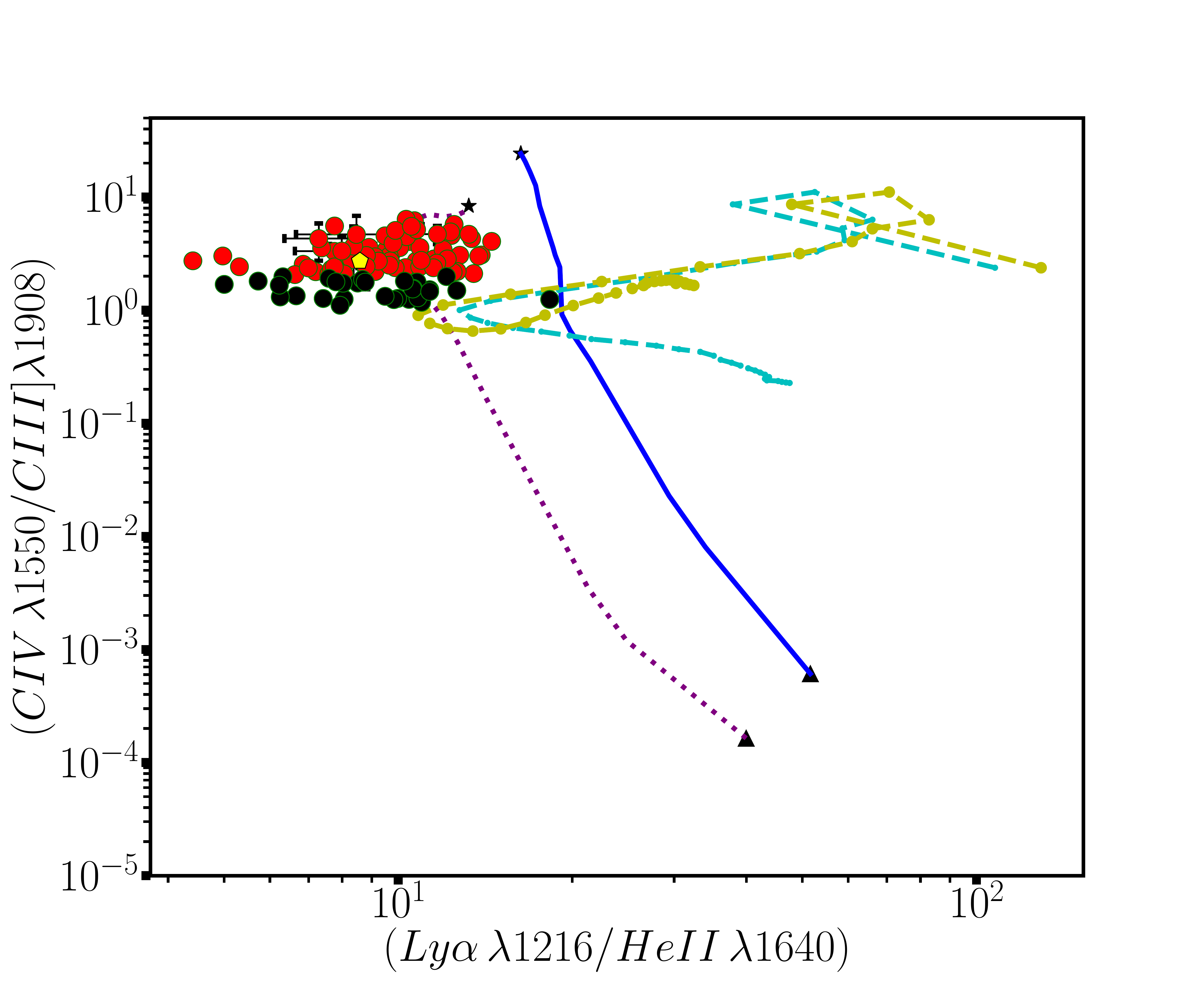}}
	
	\caption{Comparison of the observed spatial variation of the MUSE UV emission line ratios with photoionization models ($0.5Z_{\odot}$ sequence is represented by the solid blue line and the $2.0Z_{\odot}$ sequence by the purple dotted line), pure shock models (blue solid circles connected by a dashed line) and the composite shock + precursor models (large yellow solid circles connected by a dashed line). In the case of the photoionization models, we use ionizing continuum power law index $\alpha$=--1.5 (left side) or $\alpha$=--1.0 (right side). The red circles represent the regions in which \ion{C}{IV}/\ion{C}{III]} $>$ 2, while the black circles represent that regions with \ion{C}{IV}/\ion{C}{III]} $<$ 2 which lie close to the positions of the radio hotspots. The yellow pentagon represent the integrated spectrum extracted from the X-SHOOTER data. At the end of each sequence, a solid black triangle corresponds to the initial value of the ionization parameter ($U$ = 10$^{-4}$) and a solid black star that corresponds to the maximum value of the ionization parameter ($U$ = 1.6). The pure shock and the composite shock + precursor models are from \citet{allen2008}. Both shock model sequences are characterized by hydrogen density 100 cm$^{-3}$, magnetic field 100 $\mu$ G and velocity covering the range $v_{s}$ = 100 up to 1000 km s$^{-1}$ .}
	\label{models01}
\end{figure*}

To assist in understanding the physical conditions and nature of the extended line emitting gas of the radio galaxy, a grid of photoionization models (see Figure \ref{models05} to \ref{models01}) were computed using the multipurpose code MAPPINGS Ie \citep{Bi85,Fe97}.

Our model grid contains two possible values for the ionizing spectral index, $\alpha$ = --1.5 with a high-energy cut-off of $5\times 10^{4}$ eV, and $\alpha$ = --1.0 with a high energy cut-off of $1\times 10^{3}$ eV. The gas chemical abundances were varied between 0.5$Z_{\odot}$ and 3.0$Z_{\odot}$ (with $Z_{\odot}$ being the solar metallicity), with all metals being scaled linearly with O/H except for nitrogen. The nitrogen abundance was varied such that N/H $\propto$ O/H at $Z/Z_{\odot}$$<$0.3 and N/O $\propto$ O/H at $Z/Z_{\odot}$$\ge$0.3, to take into account its expected secondary behaviour at moderate to high metallicity \citep[e.g.][]{villar1999,henry2000}. We adopt the Solar chemical abundances of \citet{asplund2006}. The ionization parameter\footnote{The ionization parameter U is defined as the ratio of ionizing photons to hydrogen atoms and its expression is  $Q$/($4\pi r^{2} n_{H} c$), with $Q$ as the ionizing photon luminosity of the source, r is the distance of the cloud from the ionizing source, $n_{H}$ is the hydrogen density and c is the speed of light.} varies from $U$ = $10^{-4}$ to 1.6. For all models, we use hydrogen density of 100 cm$^{-3}$ and adopt a single-slab, ionization-bounded, isochoric geometry. 

Because MRC 0943--242 is a powerful radio galaxy, it is plausible that the radio jets drive ionizing shocks into the interstellar medium. Thus, we also make use of shock and shock plus precursor models from the literature, computed by \citet{allen2008} using MAPPINGS III. The models used consist of one sequence in which the emission comes solely from shock-heated gas, while the other sequence is a combination of shock heated gas and a photoionized precursor. In order to be as consistent as possible with our MAPPINGS 1e photoionization models, we have selected sequences with solar abundance, shock velocities covering the range $v_{s}$ = 100 up to 1000 km s$^{-1}$ in steps of 25 km s$^{-1}$, hydrogen density $n_{H}$ = 100 cm$^{-3}$ and magnetic field $B$ = 100 $\mu$ G.

The large number of emission lines now detected from this radio galaxy makes it challenging to diagnose the physical conditions using only diagnostic diagrams. For this reason, we have written a simple Python code that searches the assembled grid of ionization models to find the model that best reproduces an ensemble of emission line ratios. The routine includes extinction A$_V$ as a free parameter, and we have assumed R$_{V}$ = 3.1 and the dust extinction curve of \citet{fitz1999}. For this we use all possible line ratios formed using lines detected in our X-SHOOTER long-slit spectrum, with the exception of ratios involving Ly$\alpha$, where transfer effects are likely to be strong. All line ratios included in this analysis are given equal weighting, and the goodness of fit is evaluated using reduced chi-square ($\chi^{2} _{\nu}$).

\begin{figure*}
	\includegraphics[width=11cm,height=11cm,keepaspectratio]{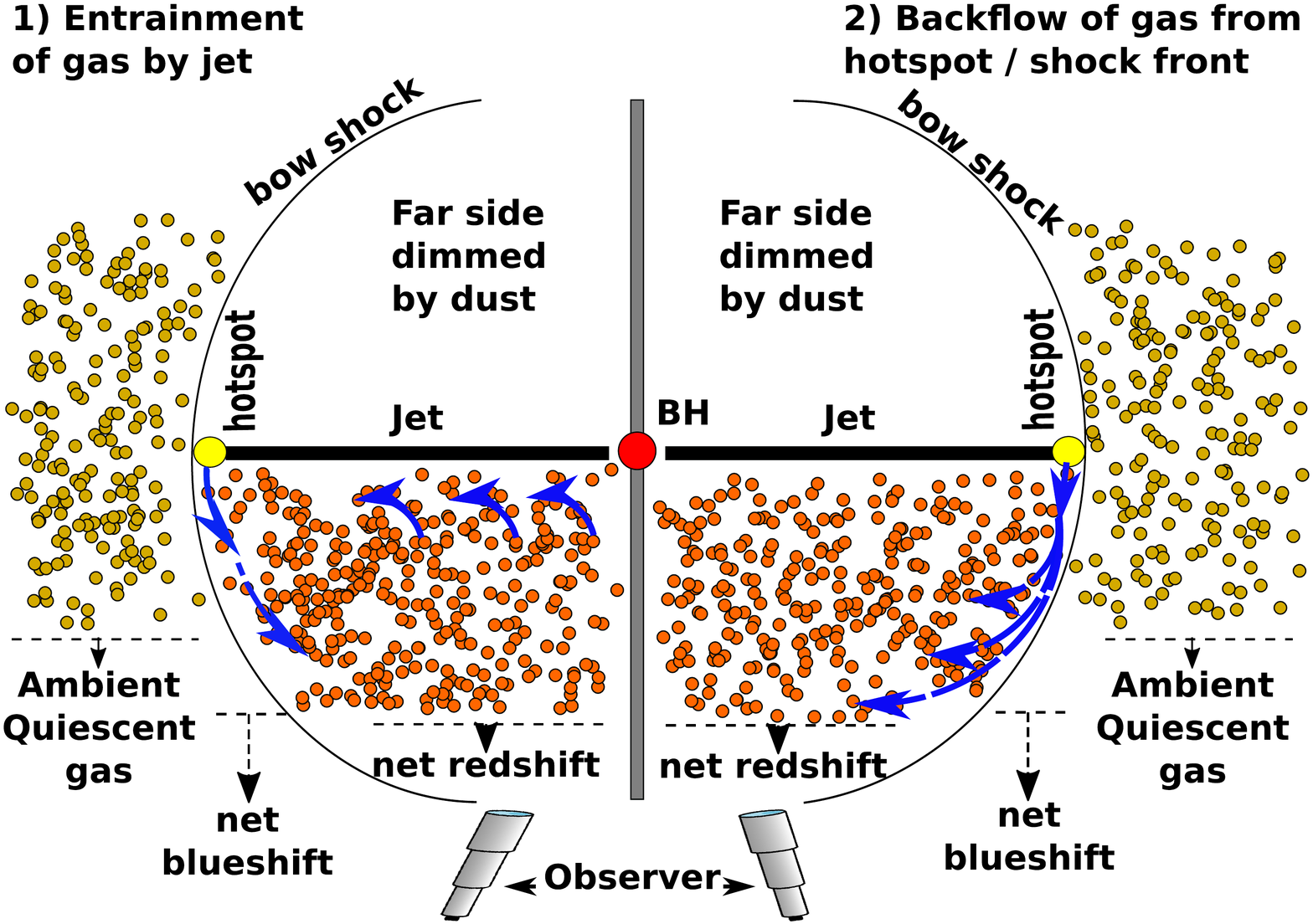}
	\caption{Here we show two simple scenarios to explain the emission line kinematics in MRC 0943--242. (1) The relativistic radio jets are produced in the central active nucleus and propagate outward through the host galaxy, terminating in a hotspot that represents the working surface of the jet against the ambient ISM. The radio plasma cools and diffuses/flows laterally away from the hotspot, carrying with it condensations of warm ionized gas which are then seen as localised blueshifted line emission with relatively large FWHM, closely associated with the hotspots. Beyond the radio cocoon, the ambient ISM remains untouched by the jets and thus shows relatively narrow emission lines. The jets entrain gas from the ISM of the galaxy, dragging gas in towards the jet as well as along the velocity vector of the jet. The inward motion of the entrained gas results in a net redshift when the radio jet axis is viewed side-on. (2) As above, but with a backflow of material away from the head of the radio jet, instead of entrainment, producing the observed redshifting of the kinematically perturbed gas. In both cases, the ISM would need to contain a significant quantity of dust, to dim the emission from gas on the far side of the galaxy.}
	\label{scenario1}
\end{figure*}

\begin{figure*}
	\subfloat[$\alpha$ = --1.5]{
		\includegraphics[width=\columnwidth,height=6.8cm,keepaspectratio]{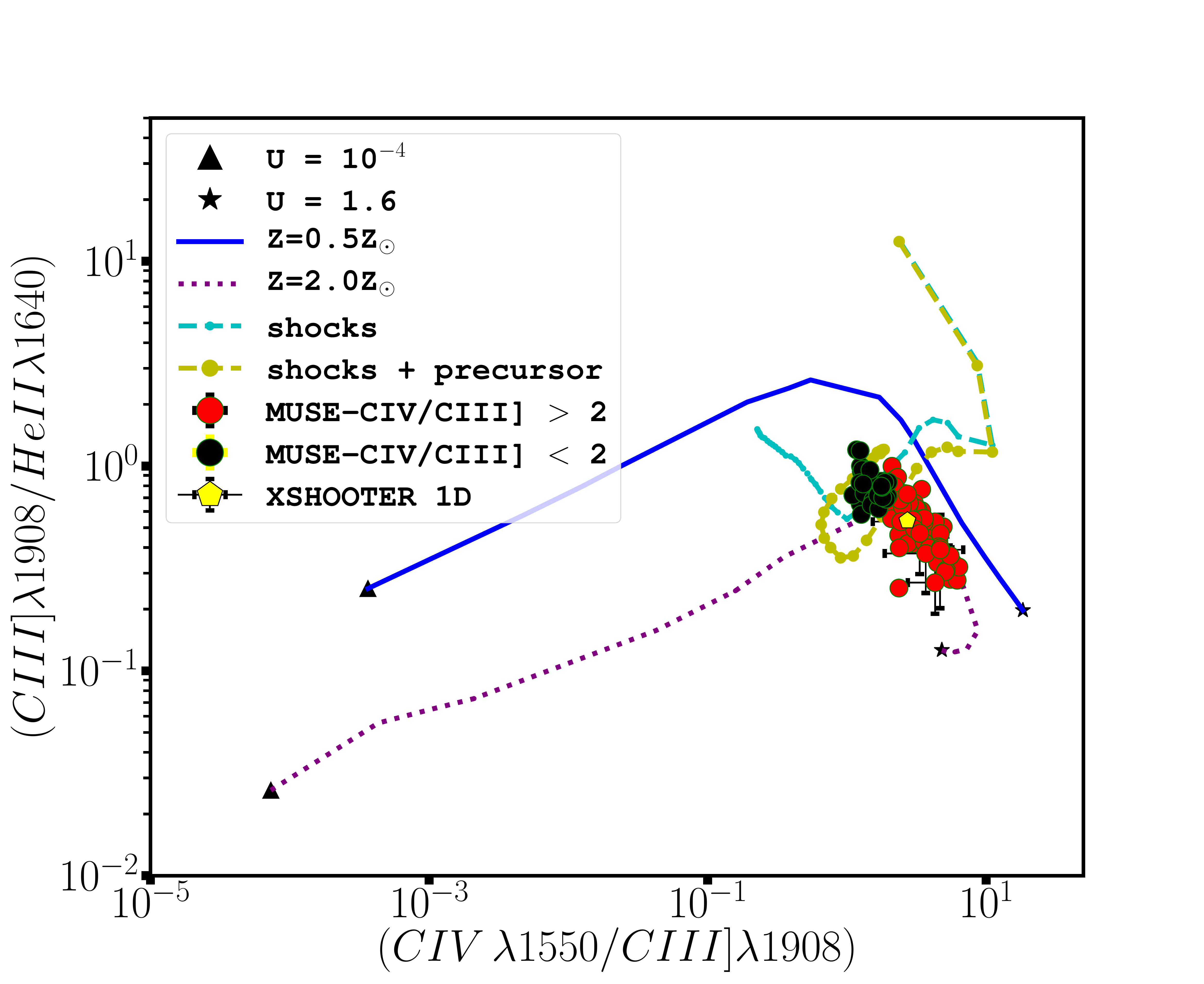}}
	\subfloat[$\alpha$ = --1.0]{
		\includegraphics[width=\columnwidth,height=6.8cm,keepaspectratio]{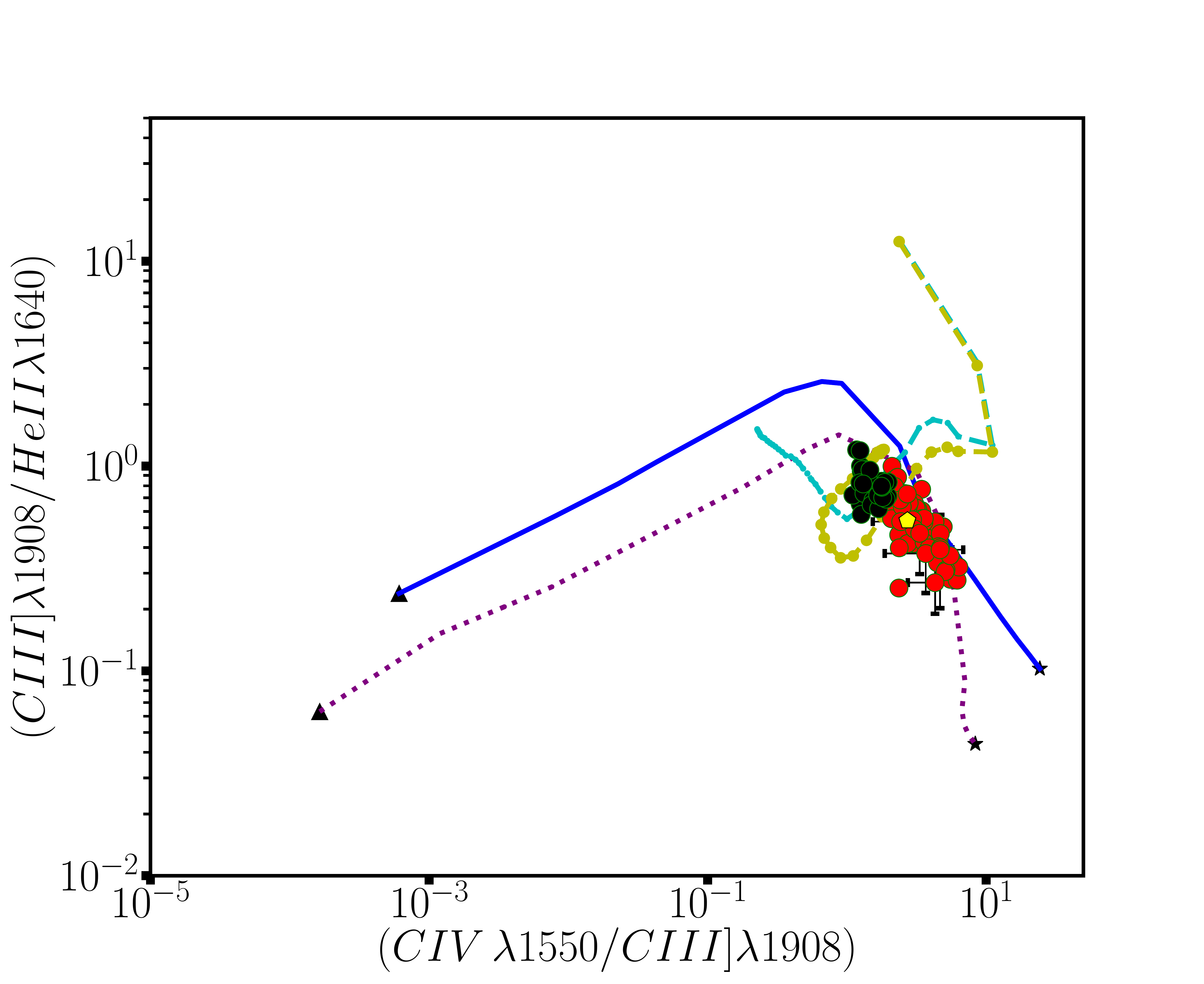}}
	\quad
	\subfloat[$\alpha$ = --1.5]{
		\includegraphics[width=\columnwidth,height=6.8cm,keepaspectratio]{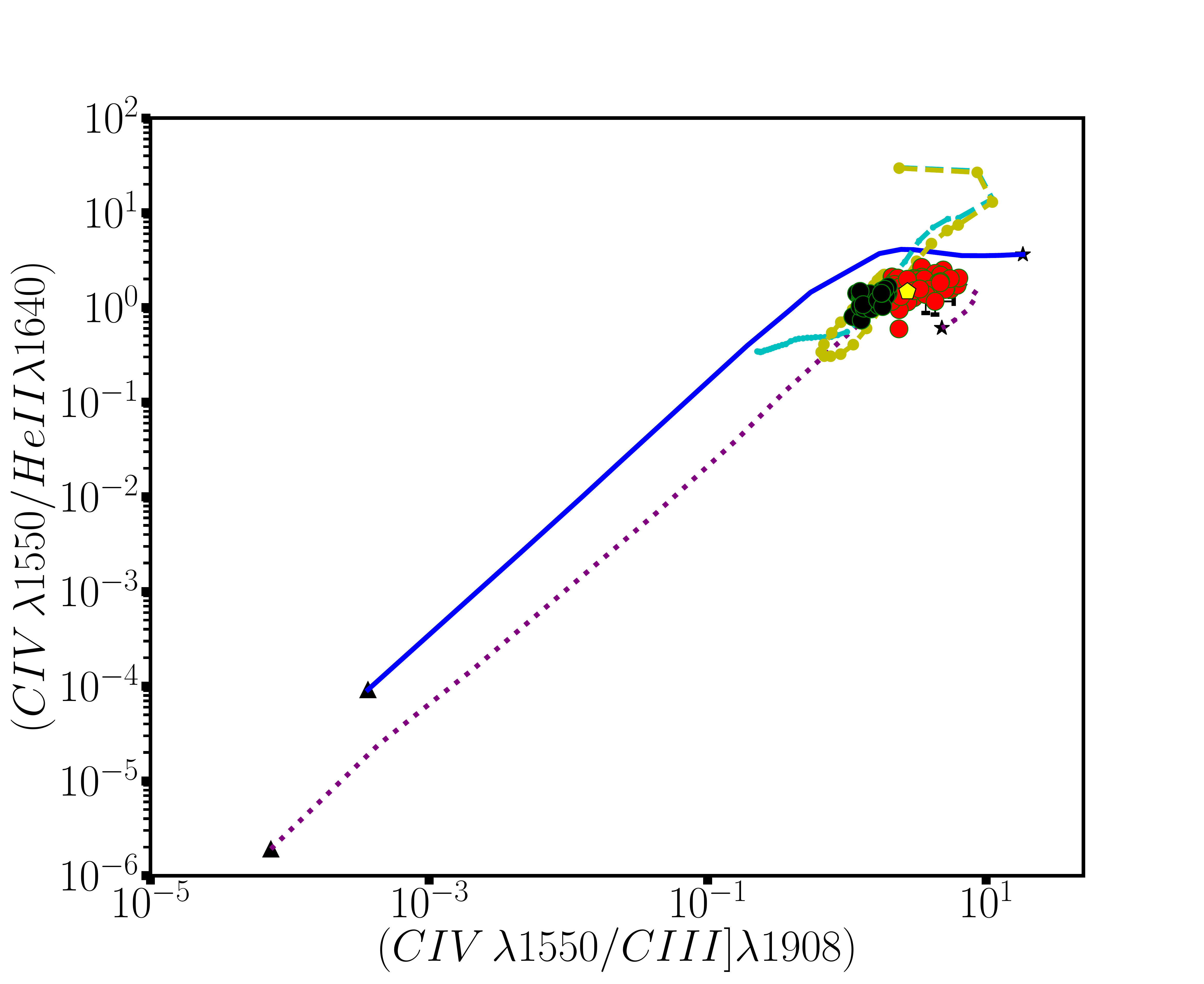}}
	\subfloat[$\alpha$ = --1.0]{
		\includegraphics[width=\columnwidth,height=6.8cm,keepaspectratio]{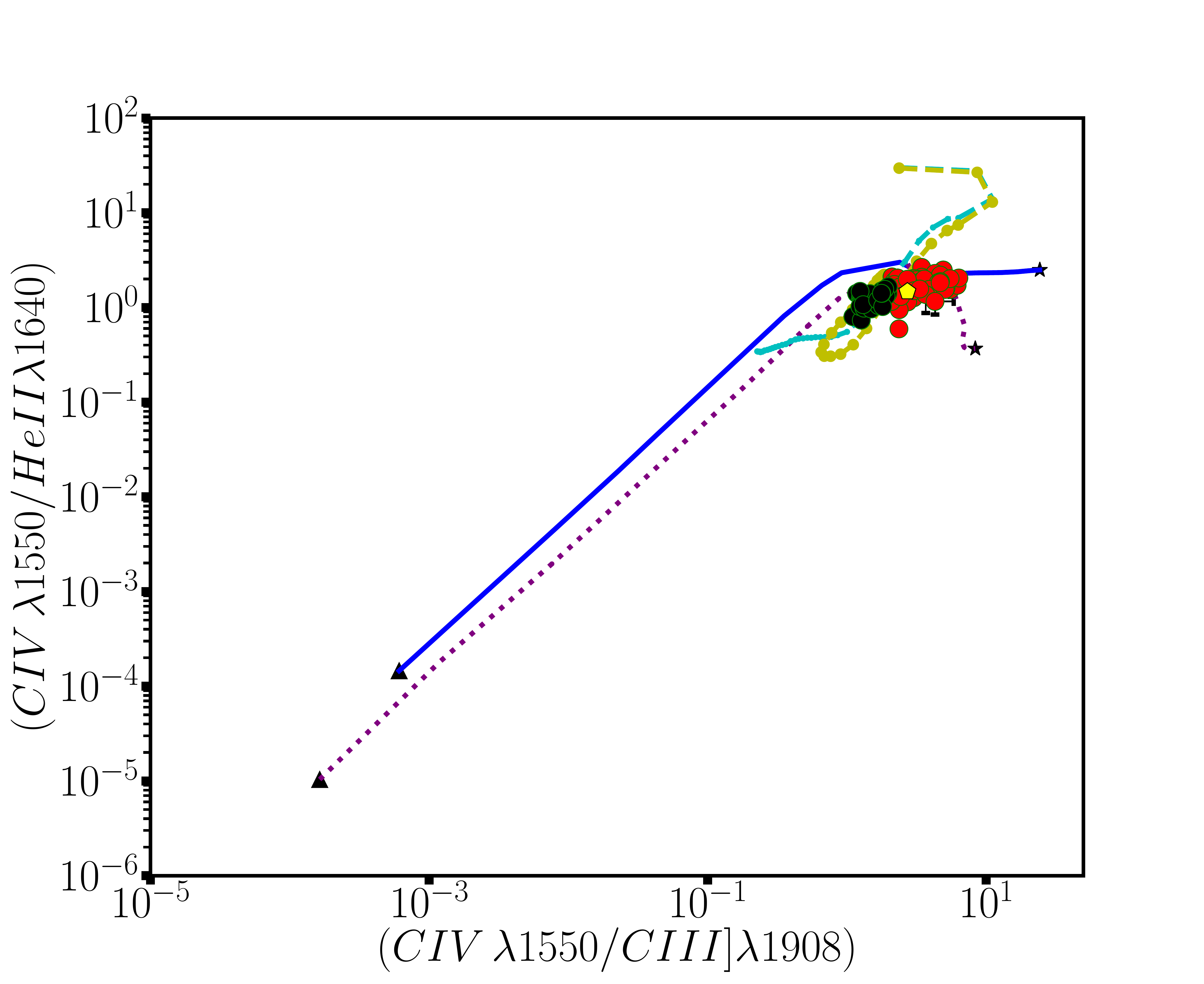}}
	
	\caption{Comparison of the observed spatial variation of the MUSE UV emission line ratios with photoionization models ($0.5Z_{\odot}$ sequence is represented by the solid blue line and the $2.0Z_{\odot}$ sequence by the purple dotted line), pure shock models (blue solid circles connected by a dashed line) and the composite shock + precursor models (large yellow solid circles connected by a dashed line). In the case of the photoionization models, we use ionizing continuum power law index $\alpha$=--1.5 (left side) or $\alpha$=--1.0 (right side). See Fig. ~\ref{models01} for more details.}
	\label{models02}
\end{figure*}

\section{Results and Discussion}
\label{discussion}

 \subsection{Radio mode feedback}
 \label{feedback}

In Fig. \ref{kin}, we show maps of the flux and kinematic properties of the lines Ly$\alpha$, \ion{C}{IV}, \ion{He}{II} and \ion{C}{III]} from our analysis of the MUSE datacube, with the position of the nucleus and the radio hotspots marked. To supplement the MUSE results, we also show kinematic results from our single-slit observations from X-SHOOTER and Keck II in Fig. ~\ref{1dim_kinXSH} and ~\ref{1dim_kinKECK}, respectively. Taking into account differences in spatial resolution, spectral resolution and slit position between the different observations, we find good consistency between the kinematic results from the MUSE, X-SHOOTER and Keck II datasets. 

The pseudo narrow band images show an extended emission line halo with a major axis that is well aligned with the PA of the radio source \citep[see also][]{Gu}, similar to the close alignment found by \citet{Pe1999,Pe2001} using broad-band HST images of this object. 

In agreement with previous kinematic studies of this galaxy \citep{VM1,Hu2}, we find that the extended emission line halo shows a central, extended region of kinematically turbulent gas with relatively high FWHM ($\ga$ 800 km s$^{-1}$) which almost reaches the radio hotspots, with kinematically more quiescent gas ($\la$ 800 km s$^{-1}$) located at larger radii. In addition, we find the most extreme gas kinematics within $\sim$1\arcsec of the position of the West radio hotspot, with all of these lines showing their highest FWHM and greatest relative blueshift there. Along the Keck II PA, the FHWMs of both \ion{C}{IV} and \ion{He}{II} show a modest increase from the nucleus towards the radio hotpots, but peaking $\sim$0.6\arcsec short and sharply falling to smaller FWHM values. 

Our velocity maps (Fig. ~\ref{kin}) and velocity curves (see Fig. ~\ref{1dim_kinXSH} and Fig. ~\ref{1dim_kinKECK}) show that the central, kinematically perturbed gas shows the most redshifted velocities, with a dramatic transition to blueshifted velocities taking place near the positions of the radio hotspots. At the position of the West radio hotspot, spatially coincident with the largest measured FWHM, we find the highest relative blueshift ($\ga$ 300 km s$^{-1}$) in Ly$\alpha$, \ion{C}{IV}, \ion{He}{II} and \ion{C}{III]}.

Thus, there is clear evidence for jet gas interactions strongly affecting the kinematic properties of the extended emission line gas (see Fig. \ref{hst0943}). However, the precise nature of this interaction is not immediately clear. Whereas our data only give us the line of sight velocity, the radio jet motion is likely to be close to the plane of the sky. 

Naively, one might have expected the perturbed gas at radii smaller than the hotspots to be in outflow from the galaxy, and thus blueshifted relative to the ambient quiescent gas. However, with the notable exception of a localised region of gas associated with the West hotspot, the vast majority of the perturbed gas shows a relative {\it redshift} with respect to the quiescent gas. In addition, we find that the quiescent gas that lies beyond the highly perturbed, inner regions shows a net blueshift, a feature that is seen on the east and west sides of the object, in both \ion{C}{IV} and \ion{He}{II}. 

In Fig. ~\ref{scenario1} we illustrate two scenarios that may explain the remarkable kinematic properties observed in MRC 0943--242. In both cases, the relativistic jets of radio plasma are produced in the central active nucleus and propagate outward through the host galaxy, terminating in a hotspot that represents the working surface of the jet against the ambient ISM. Upon reaching the hotspot, the radio plasma cools and diffuses/flows laterally away from the hotspot, carrying with it condensations of warm ionized gas that are seen as localised blueshifted line emission with relatively large FWHM, closely associated with the hotspots. Beyond the radio cocoon, the ambient ISM remains untouched by the radio jets and thus shows relatively narrow emission lines. Within this general framework, the fact that much of the kinematically perturbed gas is {\it redshifted} suggests that we are witnessing the inflow of gas driven by radio-mode feedback, due to (1) gas being pulled in towards the radio jets as part of the entrainment process; or (2) a backflow of material from the radio hotspots which cycles material back into the host galaxy. An interesting implication of the second scenario is that, judging by the fact that this perturbed and blueshifted component dominates the overall line emission from within the radius of the radio source, a significant quantity of gas is funnelled back towards the host galaxy after being entrained/accelerated by the radio jets. 

Although the long-term effect of the radio-loud activity may well be to quench star-formation (hereafter SF) and starve the AGN of fuel, we speculate that MRC 0943--242 may be in a phase of {\it positive} radio-mode feedback where fuel is cycled back into the central regions of the galaxy to form stars and fuel the AGN. Indeed, a number of studies have found a correlation between small radio-jets and a relatively higher star formation rate \citep[e.g.][]{Hu2,humphrey2011}, and this HzRG appears to have a substantial rate of star-formation in the range 200 -- 1400 M$_{\odot}$ yr$^{-1}$ (\citealt{Gu}; see also \citealt{Hu2}). 

The two scenarios described above are not necessarily mutually exclusive; it seems plausible that both could operate simultaneously. In either case, we point out that the ISM would need to contain dust so as to dim the perturbed gas located on the far side of the galaxy from the observer, suggesting that significant quantity of dust is able to survive the shocks driven into the ISM during the passage/growth of the radio source through the host galaxy, consistent with the detection of significant UV continuum polarization \citep{Ve2001}. 

Maps of several important UV line ratios are shown in Fig. ~\ref{lnr} (see also Fig. ~\ref{lnr_plot}). The large variation of \ion{C}{IV}/\ion{C}{III]} in our map suggests there is a substantial range in ionization level across the object. In addition, we find a spatial correlation between the radio hostpots and several of the emission line ratios, with relatively low values of \ion{C}{IV}/\ion{He}{II} and \ion{C}{IV}/\ion{C}{III]}, and high values of \ion{C}{III]}/\ion{He}{II}, spatially associated with both radio hotspots, and qualitatively consistent with lower $U$ in the vicinity of the hostpots. 

The Ly$\alpha$/\ion{He}{II} ratio shows no direct spatial correlation with the positions of the radio hotspots. However, we do find relatively low values for this ratio ($\la$ 6) to the immediate East of the Eastern radio hotspot and $\sim$0.9\arcsec$\,$ South of the Western hotspot, in both cases spatially coincident with a region of kinematically quiescent gas as seen in the \ion{He}{II} and \ion{C}{III]} kinematic maps (Fig. ~\ref{kin}), suggesting that the Ly$\alpha$ escape fraction is much lower in regions that are unaffected by radio mode feedback. Scenarios that may explain this result include: (i) reduction of dust in the kinematically perturbed regions due to jet-driven shocks \citep[e.g.][]{villar2001}; (ii) greater velocity overlap between emitting and absorbing gas phases in the kinematically quiescent regions \citep[e.g.][]{TT}; (iii) the presence of an optically thick shell of gas encasing the expanding radio cocoon \citep[e.g.][]{Bi3}. The ionization/excitation and metallicity of the extended gas will be examined in greater detail in $\S$\ref{emission_line}. 

 \subsection{Ionization and metallicity of the extended emission line gas}
 \label{emission_line}

\begin{table*}
	\centering
	\caption{Comparison of model line ratios with observed line ratios. (1) Emission line ratios. (2) Observed X-SHOOTER line fluxes normalised by \ion{He}{II} $\lambda$1640. (3) Parameters and relative line fluxes produced by our best-fitting MAPPINGS model. (4) Our best-fitting model using $\alpha\,=$ --1.5 instead of $\alpha\,=$ --1.0. Parameters and relative line fluxes produced by our best-fitting shock models (5) and shock + precursor models (6) extracted from \citet{allen2008}.}
	\label{lnr-chi}
	\begin{tabular}{lccccr}
		\hline
		Line ratios & Obs. flux & Model 01 & Model 02 &  Model 03 & Model 04 \\ 
		(1)		&		(2)		&		(3)	&		(4)	 &     (5)   &   (6)	\\
		\hline
		~       &     ~      & $U$ = 0.018  & $U$ = 0.032  &   shocks  & shock + prec. \\ 
		~       &     ~      & $\alpha\,= -1.0$  &  $\alpha\,= -1.5$  &  $v$ = 225 $km/s$ &  $v$ = 725 $km/s$   \\ 
		~       &     ~      & $Z/Z_{\odot}\,= 2.1$  &  $Z/Z_{\odot}\,= 1.2$  &  $Z/Z_{\odot}\,= 1.0$ & $Z/Z_{\odot}\,= 1.0$ \\ 
		~       &     ~      &  $\chi^{2} _{\nu}\,= 2.75$  &   $\chi^{2} _{\nu}\,= 4.66$  & $\chi^{2} _{\nu}\,= 7.15$ & $\chi^{2} _{\nu}\,= 6.53$ \\
		\hline
		Ly$\alpha$ $^{(*)}$/HeII 	& 9.90 $\pm$ 0.43   & 11.30  & 16.56 & 59.15 & 28.50 \\
		(OVI$+$CII)/HeII    		& 0.77 $\pm$ 0.12   & 0.31   & 0.24  & 4.90  & 5.70  \\  
		NV/HeII         			& 0.30 $\pm$ 0.06   & 0.39   & 0.21  & 3.01  & 0.69   \\ 
		CIV/HeII        			& 1.47 $\pm$ 0.11   & 2.14   & 2.05  & 6.99  & 1.48  \\ 
		CIII]/HeII      			& 0.54 $\pm$ 0.04   & 1.24   & 1.01  & 1.68  & 2.21  \\ 
		CII]/HeII       			& 0.27 $\pm$ 0.05   & 0.17   & 0.08  & 1.13  & 2.33   \\ 
		$[$NeIV$]$/HeII     		& 0.18 $\pm$ 0.05   & 0.43   & 0.31  & 1.00  & 1.51  \\ 
		MgII/HeII       			& 0.56 $\pm$ 0.08   & 0.50   & 0.45  & 2.81  & 3.58  \\ 
		$[$NeV$]$/HeII      		& 0.64 $\pm$ 0.06   & 0.58   & 0.43  & 0.48  & 0.49  \\ 
		$[$OII$]$/HeII      		& 2.79 $\pm$ 0.15   & 0.93   & 0.46  & 3.45  & 1.10  \\ 
		$[$NeIII$]$/HeII    		& 1.06 $\pm$ 0.08   & 0.69   & 0.64  & 1.03  & 1.12 \\ 
		H$\gamma$/HeII  			& 0.50 $\pm$ 0.10   & 0.17   & 0.31  & 0.84  & 0.44  \\ 
		$[$OIII$]$4363/HeII 		& 0.19 $\pm$ 0.02   & 0.11   & 0.09  & 0.20  & 0.68  \\ 
		$[$OIII$]$5007/HeII 		& 8.49 $\pm$ 0.33   & 8.79   & 8.45  & 2.90  & 0.95  \\ 
		\hline
		\multicolumn{6}{c}{$^{(*)}$ Ly$\alpha$ was not used in the fitting.} \\
		\hline
	\end{tabular}
\end{table*}

Our X-SHOOTER spectrum of MRC 0943--242 covers a wide range of wavelength and contains numerous emission lines from various species, making it particularly useful for studying the ionization and chemical abundances in the extended ionized gas. From our ionization model grid (see Sect. ~\ref{model_grid}), the single model that provides the best overall fit to the X-SHOOTER emission line spectrum is a photoionization model with $U$ = 0.018, $\alpha =$ --1.0 and gas metallicity $Z/Z_{\odot} =$ 2.1. In addition, our best fitting model has A$_V$ = 0. Table ~\ref{lnr-chi} shows the X-SHOOTER emission lines normalised to the flux of HeII $\lambda$1640. Our best-fitting model parameters are generally consistent with conclusions obtained from smaller subsets of the UV lines by previous authors \citep{Ve2001,Hu4,Gu}.

Although the reduced chi-square is reasonably small ($\chi^2_{\nu}=$2.75) and many line ratios are well reproduced by the model, a number of other line ratios are not well reproduced. In particular, the model produces too low a value of \ion{[O}{III]} $\lambda$4363 / \ion{[O}{III]} $\lambda$5007 (see Figure \ref{models05}), indicating that T$_e$ is too low in the model. This could be due to the presence of some shock-heating in the extended gas in addition to the dominant AGN-photoionized gas \cite[see also][]{tadhunter1989}, as also suggested by \citet{Gu} based on the strength of \ion{C}{II]} relative to \ion{C}{IV}, \ion{C}{III]} and \ion{He}{II} in the integrated MUSE spectrum. 

In addition, we notice that \ion{O}{VI}$+$\ion{C}{II}, \ion{[O}{II]} and \ion{C}{II]} are underpredicted in the model relative to our measurements, suggesting the presence of clouds with a substantial range in U within the extended gas sampled by our X-SHOOTER spectrum \citep[e.g.][]{Hu4}. 

The best-fitting model also produces an \ion{H}{$\gamma$}/\ion{He}{II} $\lambda$1640 ratio that is significantly below the observed value. This suggests that the ionizing SED ($\alpha$=--1.0) may be too hard, or that reddening, for which we have no reliable diagnostic, may be important in this object. 

To supplement our analysis and partially illustrate the above results, we also show selected diagnostic diagrams (Figs. ~\ref{models05}, ~\ref{models03} and ~\ref{models04}). 

In order to study the spatial variation of the UV emission line ratios, we show in Figs. \ref{models01} and \ref{models02} several diagnostics diagrams with UV line ratios measured from the individual spaxels in the MUSE datacube. In these diagrams we use line ratios involving Ly$\alpha$, \ion{C}{IV}, \ion{He}{II} and \ion{C}{III]} only, due to the narrower spectral range of MUSE and the lower chance of detecting the faint lines in the individual spaxels. 

The MUSE spaxels show a substantial dispersion in each of the diagrams (Fig. ~\ref{models01} and Fig. ~\ref{models02}), with a 'centre of gravity' that corresponds approximately with the position of the X-SHOOTER data point (yellow pentagon). 

The distribution of data points in diagrams involving only \ion{C}{IV}, \ion{He}{II} and \ion{C}{III]} is qualitatively consistent with a range in $U$. Of these ratios, \ion{C}{IV}/\ion{C}{III]} is likely to be the most reliable indicator of $U$, and given its large variation ($\sim$0.6 dex) we conclude that there is likely to be a large variation in $U$ (or ionization state) throughout the extended ionized halo, as suggested by our analysis of the 1D X-SHOOTER spectrum above. However, we note that exploring the spatial variation in metallicity and ionization parameter using the X-SHOOTER spectrum with a different aperture (2.1\arcsec) has only a minor effect on these parameters. From our ionization model grid, we find that the single model that provides the best overall fit to the new aperture shows metallicity and ionization parameter affected by changes $<$15\% when compared with that obtained with a smaller aperture (0.8\arcsec)(see Tables \ref{lnr-chi} and \ref{lnr-chiAp}).

We also note the presence of significant scatter perpendicular to the U-sequence loci (see the flux ratio maps (Fig. \ref{lnr}) plotted on the diagrams in Figs. \ref{models01} and \ref{models02}). We suggest this may be due to metallicity inhomogeneities or local differences in the hardness or source of ionization.

The spaxels with relatively low values of \ion{C}{IV}/\ion{C}{III]} (black circles), which are also relatively close to the radio hotspots, clearly have different ionization conditions to the rest of the nebula, but degeneracies between models make it challenging to determine the origin of this difference. Although they are consistent with having among the lowest values of $U$, these spaxels are also close to the shock model loci. Thus, we suggest that the radio source induces a lower ionization parameter where it most strongly interacts with the ISM, perhaps due to compression of gas, or that shock ionization contributes significantly in these specific regions of the Ly$\alpha$ halo. 

The Ly$\alpha$/\ion{He}{II} values show a dispersion within the nebula ($\sim$0.6 in dex), with most points having lower values than the minimum values produced by the plotted photoionization or shock models. This is likely due to a varying impact of transfer effects across the nebula, but it is not clear whether the observed Ly$\alpha$ flux is being suppressed by resonant scattering due to dust or due to redirection of photons into other lines of sight. Interestingly, the locus of photionization models using $\alpha$=--1.0 and $Z/Z_{\odot}$=2.0 passes through the cloud of MUSE points in the \ion{C}{IV}/\ion{C}{III]} vs. Ly$\alpha$/\ion{He}{II} diagram, suggesting that the systematically low Ly$\alpha$/\ion{He}{II} ratios of MRC 0943--242 may be partly due to a relatively hard ionizing SED and relatively high gas metallicity. Unlike the other UV ratios, we find no correlation between Ly$\alpha$/\ion{He}{II} and proximity to the radio hotspots.

 \subsection{Nature of the extended H\textbf{\small I}  Absorber} 
 \label{absorber}
 
 \subsubsection{H column density}
 One of the key findings from our analysis of the Ly$\alpha$ velocity profile of MRC 0943--242 is the strong degeneracy between the column density (N(\ion{H}{I})) and the Doppler width ($b$) of the main absorber, with a broad range in N(\ion{H}{I}) yielding a reasonable fit to the data, and with the best-fitting located near each end of this range: log N(\ion{H}{I}/cm$^{-2}$) = 15.20 and 19.63. The latter value is in agreement with the fits obtained in previous studies of MRC 0943--242 \citep[e.g.][]{Ro95,Bi1,Ja,Gu}. None of the previous studies identified the N-b degeneracy or the second, low column density fit. Clearly, this degeneracy has implications for our understanding of the properties of the absorbing gas where they are derived from N(\ion{H}{I}).
 
 As argued by previous authors, detection of \ion{C}{IV} in the main absorber indicates that the gas is at least partially ionized and contains some metal enriched gas, but the column ratio between \ion{C}{IV} with \ion{H}{I} does not allow a straightforward determination of the ionization structure or metallicity of the gas \citep[e.g.][]{Bi1,Ja}. In addition, we can calculate a lower limit on the total H column density by assuming N$_C$ $\ge$ N$_{\ion{C}{IV}}$ and appropriate limits for the column ratio N$_C$/N$_H$. Assuming N$_C$ $\ge$ N$_{\ion{C}{IV}}$ and C/H $\le$ 3 times Solar, we obtain N$_H$ $\ga$ 5$\times 10^{17}$ cm$^{-2}$. Note that this limit is not dependent on the value of N(\ion{H}{I}). 
 
 \subsubsection{Ionization structure and metallicity}
 
 \begin{table}
 	\caption{Low-ionization absorption lines extracted from the Keck II spectrum at the redshift of the main absorber. (1) Absorption lines; (2) Rest-frame wavelength; (3) Oscillator strength; (4) Rest-frame equivalent width; (5) Column density.}
 	\label{low_ion}
 	\begin{tabular}{lcccc}
 		\hline
 		Line & $\lambda_{rest}$ (\AA) & f & W$_{rest}$ (\AA) & N (cm$^{-2}$) \\
 		(1)   &         (2)   &    (3)     &     (4)     &    (5)     \\
 		\hline
 		\ion{O}{I} & 1302.2  & 0.05  & $<\,$5  &  $<\, 1.5\times10^{15}$ \\
 		\ion{C}{I} & 1277.2  & 0.10  & $<\,$4  &  $<\, 8.0\times10^{14}$ \\
 		\ion{C}{I} & 1328.8  & 0.06  & $<\,$3  &  $<\, 8.2\times10^{14}$ \\
 		\ion{C}{I} & 1560.3  & 0.08  & $<\,$2  &  $<\, 3.0\times10^{14}$ \\
 		\ion{C}{I} & 1656.9  & 0.14  & $<\,$2  &  $<\, 1.5\times10^{14}$ \\
 		\ion{C}{II} & 1334.5  & 0.13  & $<\,$4  &  $<\, 5.0\times10^{14}$ \\
 		\hline
 	\end{tabular}
 \end{table}
 
 To obtain additional constraints on its chemical and ionization properties, we have searched our deep Keck II spectrum for low-ionization absorption lines at the redshift of the main absorber. Our analysis has not revealed any detection of additional absorption lines, and we show the most relevant 3$\sigma$ upper limits in table ~\ref{low_ion}. To convert equivalent width W$_{\lambda}$ to column density N, we use
 
 \begin{equation}
 N = \frac{1.13\times10^{20} W_{\lambda,0}}{f \lambda{_0}{^2}}\, cm^{-2}
 \end{equation}
 
 \noindent where $f$ is the oscillator strength, ${W_{\lambda,0}}$ is the rest-frame equivalent width (in \AA) and ${\lambda_0}$ is the rest-frame wavelength of the line (in \AA) \cite[e.g.][]{Hu1}. 
 
 Interestingly, the fact that N$_{\ion{C}{IV}}$ is at least twice N$_{\ion{C}{I}}$ indicates that the gas is mostly ionized, while N$_{\ion{C}{II}}$ / N$_{\ion{C}{IV}}$ $\la$ 1.3 indicates that the ionized zone itself is highly ionized. By extension, hydrogen is also likely to be mostly ionized, with the total H column density likely to be more than three times that of \ion{H}{I}, independently of which value of N$_{\ion{H}{I}}$ we adopt. 
 
 Provided the \ion{H}{I} absorption is dominated by gas in a neutral zone, it can then be assumed that N$_{\ion{C}{I}}$/N$_{\ion{H}{I}}$ $\sim$ C/H and N$_{\ion{O}{I}}$/N$_{\ion{H}{I}}$ $\sim$ O/H, which would allow constraints to be placed on the gas chemical abundances. In the extreme case where N$_{\ion{H}{I}}$ = 4.3$\times$10$^{19}$ cm$^{-2}$, we would then obtain abundance ratios C/H $\la$ 0.013$Z_{\odot}$ and O/H $\la$ 0.076$Z_{\odot}$, meaning the absorber would be extremely metal poor. Conversely, if the \ion{H}{I} absorption is due to gas in an ionized zone with low, but non-negligible \ion{H}{I} fraction, i.e., if the absorber is 'matter bounded', then the assumption that N$_{\ion{C}{I}}$/N$_{\ion{H}{I}}$ $\sim$ C/H and N$_{\ion{O}{I}}$/N$_{\ion{H}{I}}$ $\sim$ O/H does not necessarily hold.

 \begin{figure*}
 	\subfloat[]{
 		\includegraphics[width=7.0cm,height=4.75cm,keepaspectratio]{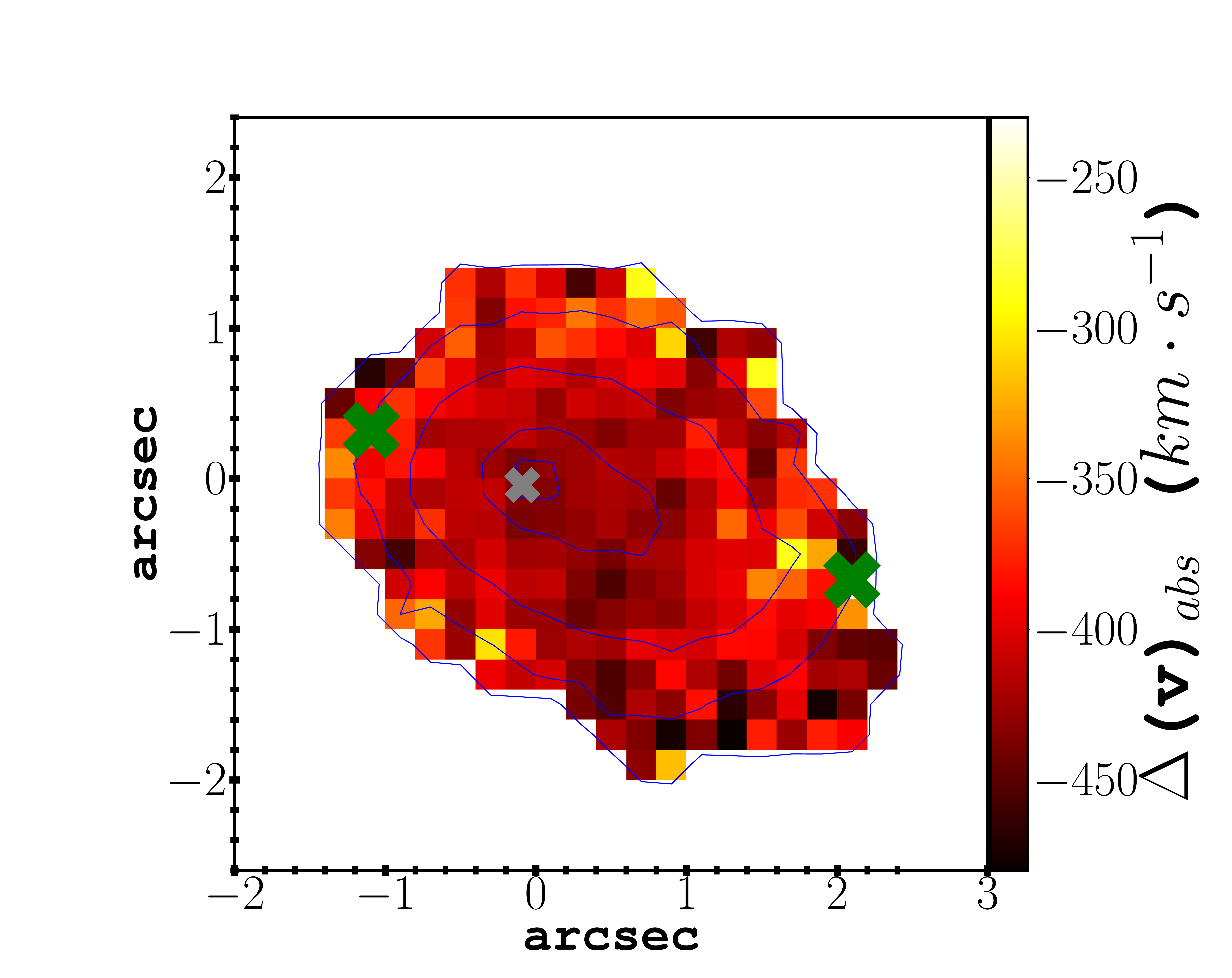}
 		\label{ly-abs}}
 	\subfloat[]{
 		\includegraphics[width=7.0cm,height=4.75cm,keepaspectratio]{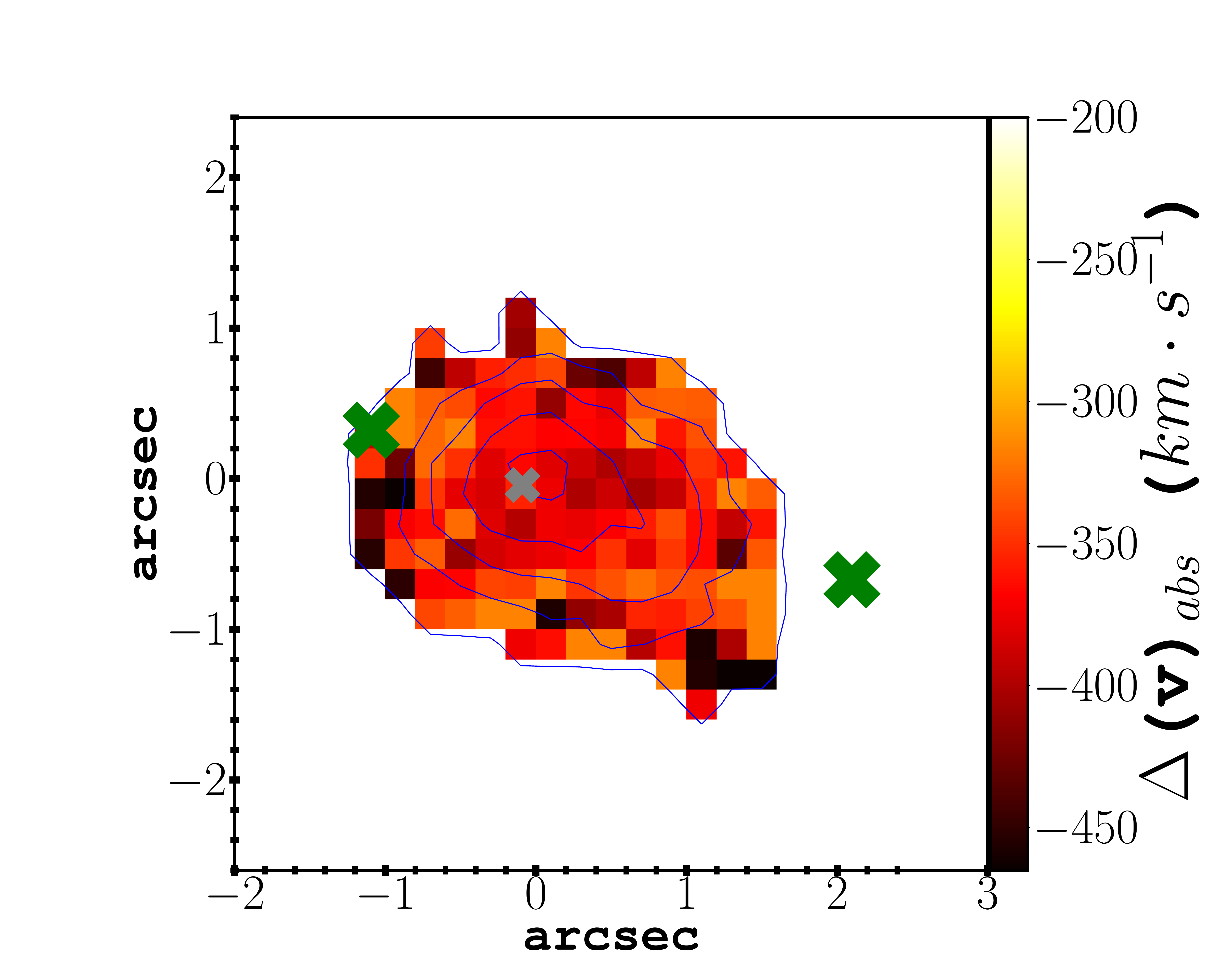}
 		\label{civ-abs}}
 	\subfloat[]{
 		\includegraphics[width=5.9cm,height=4.8cm]{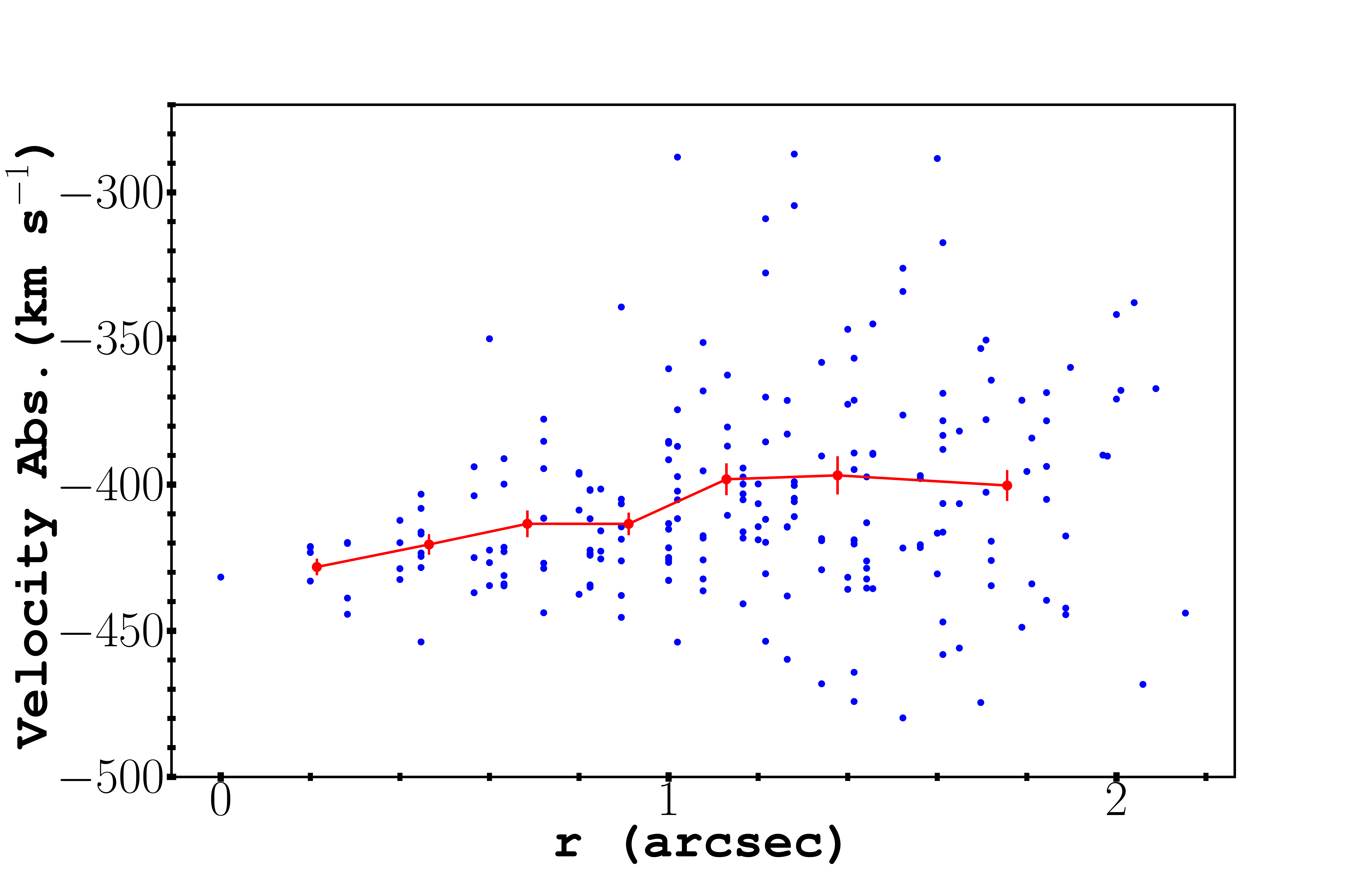}
 		\label{lyplot_abs}}
 	\caption{(a) Velocity offset map of the main absorber in the Ly$\alpha$ profile extracted from the MUSE IFU data. The velocity offset was measured relative to the \ion{He}{II} emission line at the centre of the radio galaxy. The map also shows the contour levels of the emission line intensity recovered by the fitting procedure. Contour levels: (0.3,1.3,3.8,12.5,16.8)$\times 10^{-16}$ erg cm$^{-2}$ s$^{-1}$ arcsec$^{-2}$. (b) Velocity offset map for the absorption trough in \ion{C}{IV} profile extracted from MUSE IFU. The velocity offset was measured relative to the \ion{He}{II} emission line at the centre of the radio galaxy. The map also shows the contour levels of the emission line intensity recovered by the fitting procedure. Contour levels: (0.5,1.0,2.0,3.5,4.3)$\times 10^{-16}$ erg cm$^{-2}$ s$^{-1}$ arcsec$^{-2}$. The green "X" represent the positions of the radio hotspots and the grey "x" the position of the AGN. (c) The diagram shows the velocity offset of the main Ly$\alpha$ absorption as a function of the projected distance from the centre of the galaxy. The black points are the individual pixels, and the red points are bins in distance, with their 1$\sigma$ error bars. Evaluating the strengths of correlation between the velocity offset of the main absorber and the projected distance using Spearman's rho and t-distribution, we find $\rho$ = 0.24 with a p-value = 0.0003 (for unbinned data) and $\rho$ = 0.86 with a p-value = 0.014 (for binned data).}
 \end{figure*}

 \subsubsection{Size, shape and mass}
 The most likely geometry for the main absorber is a shell or bubble of gas surrounding the host galaxy and at least part of its Ly$\alpha$ emitting halo \citep{Bi1}. The simplest way to obtain a lower limit to the radius of the absorbing structure is to determine the maximum observed offset between the projected position of the nucleus, and the most distant pixel or spaxel where the absorber is detected. 
 Measuring from the MUSE data cube, \citet{Gu} obtained a maximum offset from the nucleus of $\gtrsim$ 60 kpc, and we adopt this value as a lower limit to the radius of the shell.
 
 In Fig. \ref{ly-abs} we show a map of the line of sight velocity of the main \ion{H}{I} absorber, measured from the MUSE data. The absorption feature shows a gradient in velocity, with a significant decrease in blueshift outward from the (projected) central region of the galaxy (see also Fig. \ref{1dim_kinXSH}). The presence of this trend is independent of whether we adopt the log N(\ion{H}{I}/cm$^{-2}$) = 15.20 or 19.63 fit to the \ion{H}{I} absorber. Similarly, the \ion{C}{IV} absorption doublet also appears to show a radial decrease in its blueshift (Fig. \ref{civ-abs}). This result is consistent with what one would expect if the absorber is an expanding shell centred on the radio galaxy: At larger projected distances from the centre, the bulk velocity vector of the shell material is at a larger inclination to the line of sight, and thus appears to be less blueshifted.  
 As a consistency check on the radius of the shell, we can also estimate its radius using the velocity gradient of the HI absorption feature. Assuming the shell is spherical and using the velocity gradient of 30 km s$^{-1}$ measured between $r$ = 0 to $r$ = 14 kpc (see Fig. \ref{ly-abs} and Fig. \ref{lyplot_abs}) we obtain $R$ $\sim$ 38 kpc, which is much smaller than the value derived by \citet{Gu} using the maximum observed spatial offset of the absorber (60 kpc), perhaps indicating that the shell is not spherically symmetric. 
 
 Assuming the absorber is a spherical shell with uniform column density, and assuming its covering factor is 1 based on the fact that the absorption feature is black at its centre, we calculate the mass of the absorber using the expression 
 \begin{align}
 M_{H} \gtrsim 3.6\times10^{9}(R/60\,kpc)^{2}(N_{H}/10^{19} cm^{-2}) M_{\sun},  
 \end{align} 
 where $R$ is the radius of the absorption system in kpc and N$_{H}$ is the H column density in cm$^{-2}$. Assuming $R$ $\gtrsim$ 60 kpc and N$_H$ $\ga$ 5$\times$10$^{17}$ cm$^{-2}$, we obtain the hard lower limit log (M$_H$/M$_{\sun}$) $\gtrsim$ 8.3. 
 
 Clearly, because this is a lower limit, the mass of the shell could be even more massive than log (M$_H$/M$_{\sun}$) $\gtrsim$ 8.3. Indeed, if we were instead to use the value from our high column density fit (log N(\ion{H}{I}/cm$^{-2}$) = 19.63) and assume that the absorber is entirely neutral, we would then obtain log (M$_H$/M$_{\sun}$) $\gtrsim$ 10.2. If one were to assume the gas is partially ionized would lead to an even higher limit. 
 
\subsection{On the evolutionary status of MRC 0943-242}

With a stellar mass of $\sim$10$^{11.2}$ M$_{\odot}$ \citep{seymour2007}, MRC 0943--242 is remarkably massive for its redshift of 2.92. Significant star formation activity is present in the host galaxy ($\sim$200 M$_{\odot}$ yr$^{-1}$) and in companion galaxies ($\sim$1400 M$_{\odot}$ yr$^{-1}$; \citealt{Gu}), and the powerful active nucleus appears to be subjecting its associated extended gas reservoir to substantial radio- and quasar-mode feedback. In the long-run, this feedback activity may quench the modest star formation activity found in the host galaxy, but our kinematic results suggest that processes related to the radio-mode feedback are drawing gas deeper into the potential well of the galaxy, and we speculate this could give rise to a short-lived enhancement in star formation and the fueling of the AGN. Indeed, this would be consistent with the reported anti-correlation between the luminosity of young stellar populations in HzRGs and the age of their radio sources \citep{Hu2}. 

The very high metallicity of the ionized gas within the radio galaxy demonstrates the precence of an already highly enriched gas, compatible with the scenario that this system could evolve into a passive spheroidal system characterised by supersolar metallicity (e.g. \citealt{lonoce2015} and references therein; see also \citealt{Ve2001}). 

Interestingly, the high gas metallicity appears inconsistent with the scenario of smooth accretion of pristine gas, because in that scenario the gas metallicity should be low \citep[e.g.][]{dekel2009}. Although there is evidence for filamentary gas accretion into MRC 0943--242, the filamentary material appears to be dusty and metal-enriched \citep{Gu} and thus is likely to have been stripped from companion galaxies, rather than being low metallicity gas from the cosmic web. In fact, there appears to be no convincing evidence for smooth accretion of pristine gas in MRC 0943--242, although we cannot rule out the presence of faint, filamentary accretion structures of the kind detected near the $z = 3.1$ radio galaxy MRC 0316--257 \citep{vernet2017}.

In summary, we suggest that we are witnessing MRC 0943--242 after the bulk of gas accretion has taken place, and the galaxy is now moving towards transformation into a spheroidal galaxy where the AGN feedback will probably, ultimately quench the star-formation. 
 
 \section{Conclusions}
 \label{conclusions}
 
 Making use of observations from MUSE, X-SHOOTER and other instruments we have studied the kinematic, chemical and excitation properties of the giant Ly$\alpha$ emitting halo and the giant \ion{H}{I} absorber associated with the $z = 2.92$ radio galaxy MRC 0943--242. The main conclusions of this study are summarized as follows:
 
 \begin{itemize}
 	\item We find clear evidence for jet gas interactions affecting the kinematic properties of the Ly$\alpha$ nebula of MRC 0943--242. The MUSE datacube reveals a region of kinematically turbulent gas with relatively high FWHM extending from the nucleus out to the positions of the radio hotspots, where we see the most extreme kinematic properties, beyond which the gas has rather more quiescent kinematics in agreement with the long-slit study of \citet{VM1}. The gas most closely associated with the radio hotspots shows a blueshift relative to other regions, consistent with jet-induced outflows. However, at smaller radii the emission line gas shows a relative redshift, even compared to the kinematically quiscent gas, which we suggest might signal the presence of a feedback-driven {\it inflow} due to gas being pulled inwards as part of the entrainment process, and/or a backflow of material from the radio hotspots. 
 	\item We have computed a grid of photoionization models, to which we have added shock ionization models from the literature \citep{allen2008}, and have searched for the model that best reproduces the complete ensemble of emission line relative fluxes measured from our X-SHOOTER spectrum. We find that a photoionization model with moderate ionization parameter ($U$ = 0.018), a relatively hard ionizing SED ($\alpha$ = --1.0) and high gas metallicity ($Z/Z_{\odot}$ = 2.1) provide the best overall fit. However, we note that the \ion{[O}{III]} $\lambda$4363 / \ion{[O}{III]} $\lambda$5007 flux ratio is not well reproduced by this model, suggesting the presence of some shock-heating in the extended gas. In addition, the apparent inability to simultaneously reproduce lines of all ionization states suggests the presence of clouds with a substantial range in $U$, a result that is validated by our analysis of the 2D ionization properties. 
 	\item We find a substantial range in ionization level across the object, with a clear spatial correlation between the radio hotspots and UV emission line ratios indicative of relatively low ionization which we suggest may be due to shock heating and/or the compression of gas by the radio source. We also find evidence for a lower Ly$\alpha$ escape fraction in regions unaffected by radio mode feedback, and speculate scenarios that may explain this result: (i) reduction of dust in the kinematically perturbed regions due to jet-driven shocks \citep[e.g.][]{villar2001}; (ii) greater velocity overlap between emitting and absorbing gas phases in the kinematically quiescent regions \citep[e.g.][]{TT}; (iii) the presence of an optically thick shell of gas encasing the expanding radio cocoon \citep[e.g.][]{Bi3}.
 	\item Our Ly$\alpha$/\ion{He}{II} flux ratio map reveals a large range of values across the nebula, with most spaxels having values lower than the ionization in our grid. We attribute this to an inhomogeneity in the impact of transfer effects such as quenching by dust and/or the scattering of Ly$\alpha$ photons into different sight-lines.  
 	\item We identify and explore a strong degeneracy between column density and Doppler width of the strong, blueshifted HI Ly$\alpha$ absorber. We have been able to obtain a reasonable fit to the absorption feature across the range log N(\ion{H}{I}/cm$^{-2}$) = 15.20 and 19.63, with the best-fitting occurring near the extreme ends of this range. Independently of our fits to Ly$\alpha$, we use N$_{\ion{C}{IV}}$ to obtain a lower limit to the total H column density of N$_H$ $\ga$ 5$\times 10^{17}$ cm$^{-2}$. Given the lower limit of the spatial extent of the \ion{H}{I} obtained with the absorber velocity map, we assume a spherical shell of gas with radius $\gtrsim$60 kpc obtained by \citet{Gu}. Assuming N$_H$ $\ga$ 5$\times10^{17}$ cm$^{-2}$, the shell would have log (M$_H$/M$_{\sun}$) $\gtrsim$ 8.3.
 	
 \end{itemize}

\section*{Acknowledgements}

Marckelson Silva acknowledges support from the National Council of Research and Development (CNPq) under the process of number 248617/2013-3. AH acknowledges FCT support through fellowship SFRH/BPD/107919/2015. P.L. acknowledges support by the FCT through the grant SFRH/BPD/72308/2010. MVM  acknowledges support from the Spanish Ministerio de Econom\'\i a y Competitividad through the grant AYA2015-64346-C2-2-P. MS, AH, PL and SM acknowledge project FCOMP-01-0124-FEDER-029170 (Reference FCT PTDC/FIS-AST/3214/2012), funded by the FEDER programme. We also thank Bitten Gullberg and Carlos De Breuck for making available the reduced MUSE datacube of MRC 0943--242.




\bibliographystyle{mnras}
\bibliography{example} 



\appendix

\section{Additional material}

\begin{figure*}
	\textsc{High N(\ion{H}{I}) solution} \hspace{2cm} \textsc{Low N(\ion{H}{I}) solution}
		\subfloat[X-SHOOTER IFU]{
			\includegraphics[width=\columnwidth,height=4.5cm,keepaspectratio]{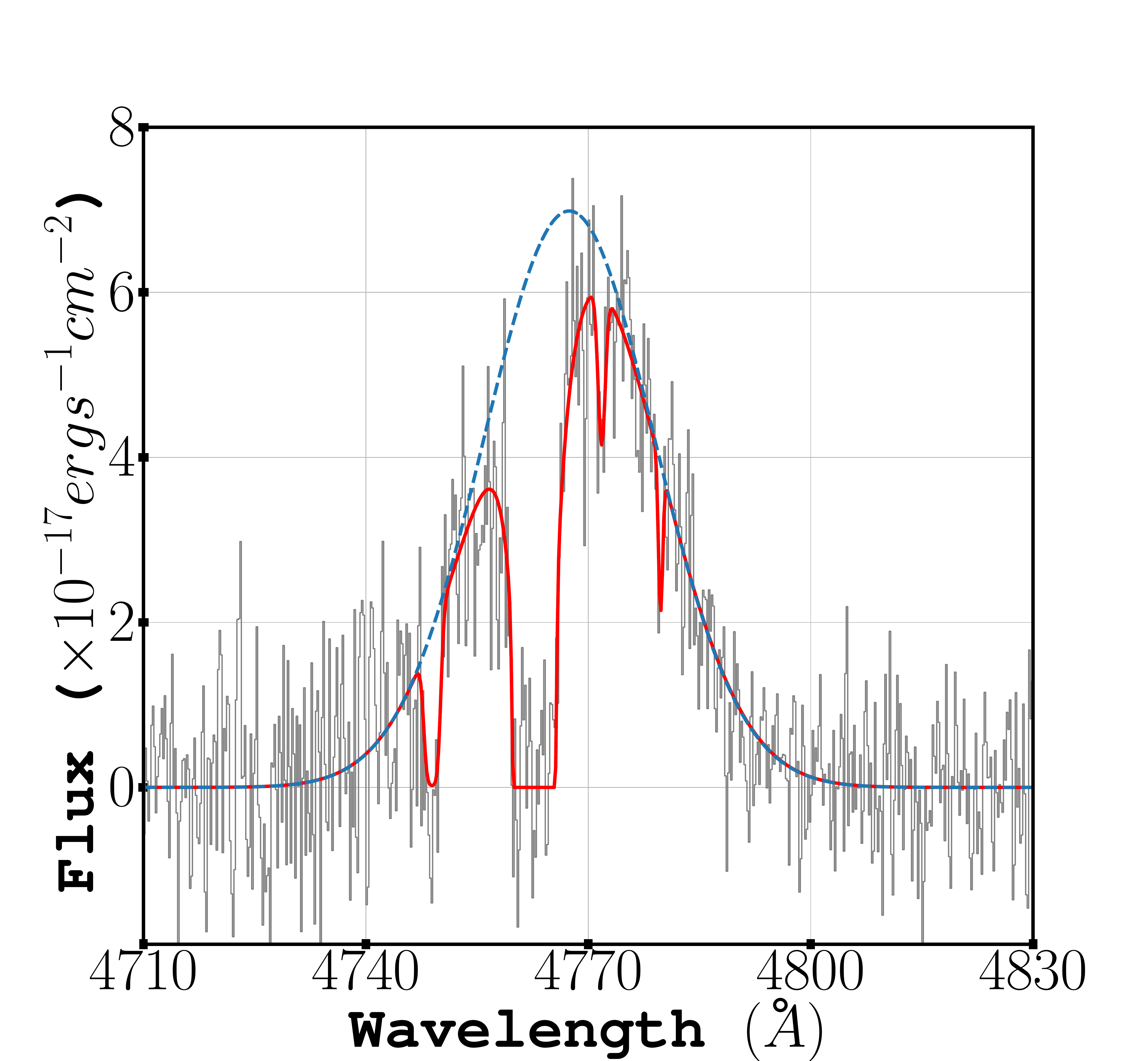}
			\includegraphics[width=\columnwidth,height=4.5cm,keepaspectratio]{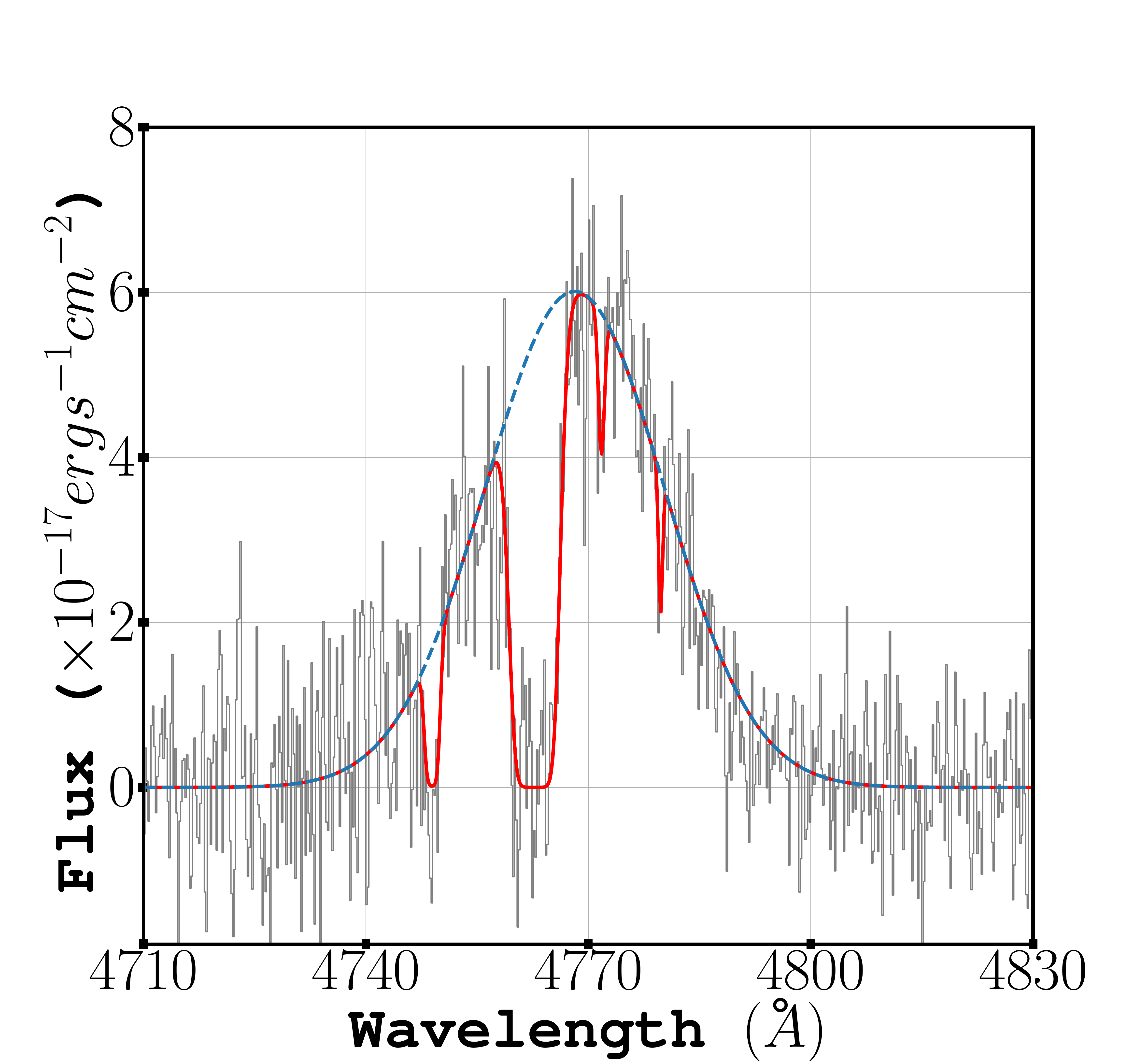}
			\label{ly-ifu}}
		\quad
		\subfloat[VLT MUSE]{
			\includegraphics[width=\columnwidth,height=4.5cm,keepaspectratio]{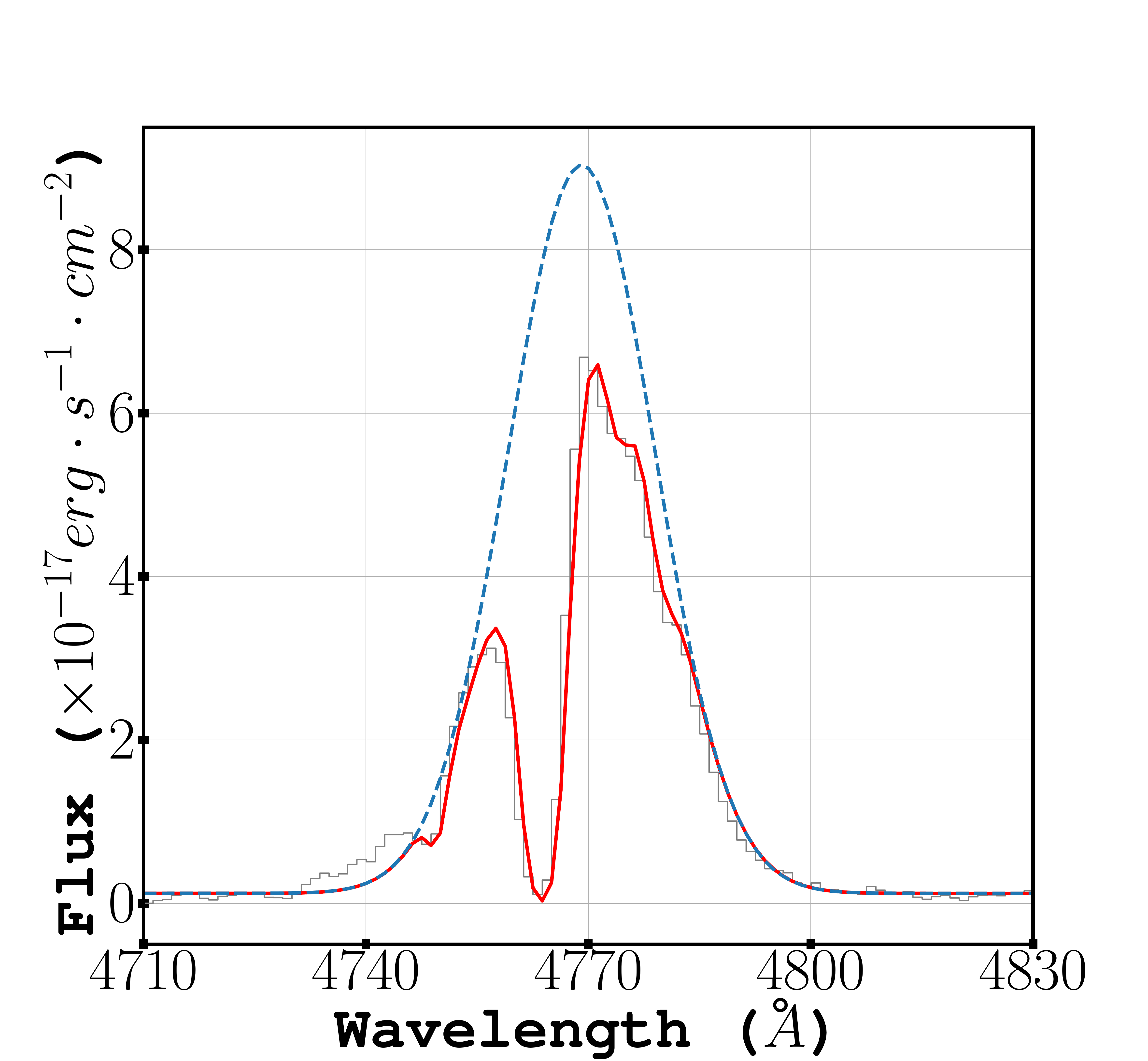}
			\includegraphics[width=\columnwidth,height=4.5cm,keepaspectratio]{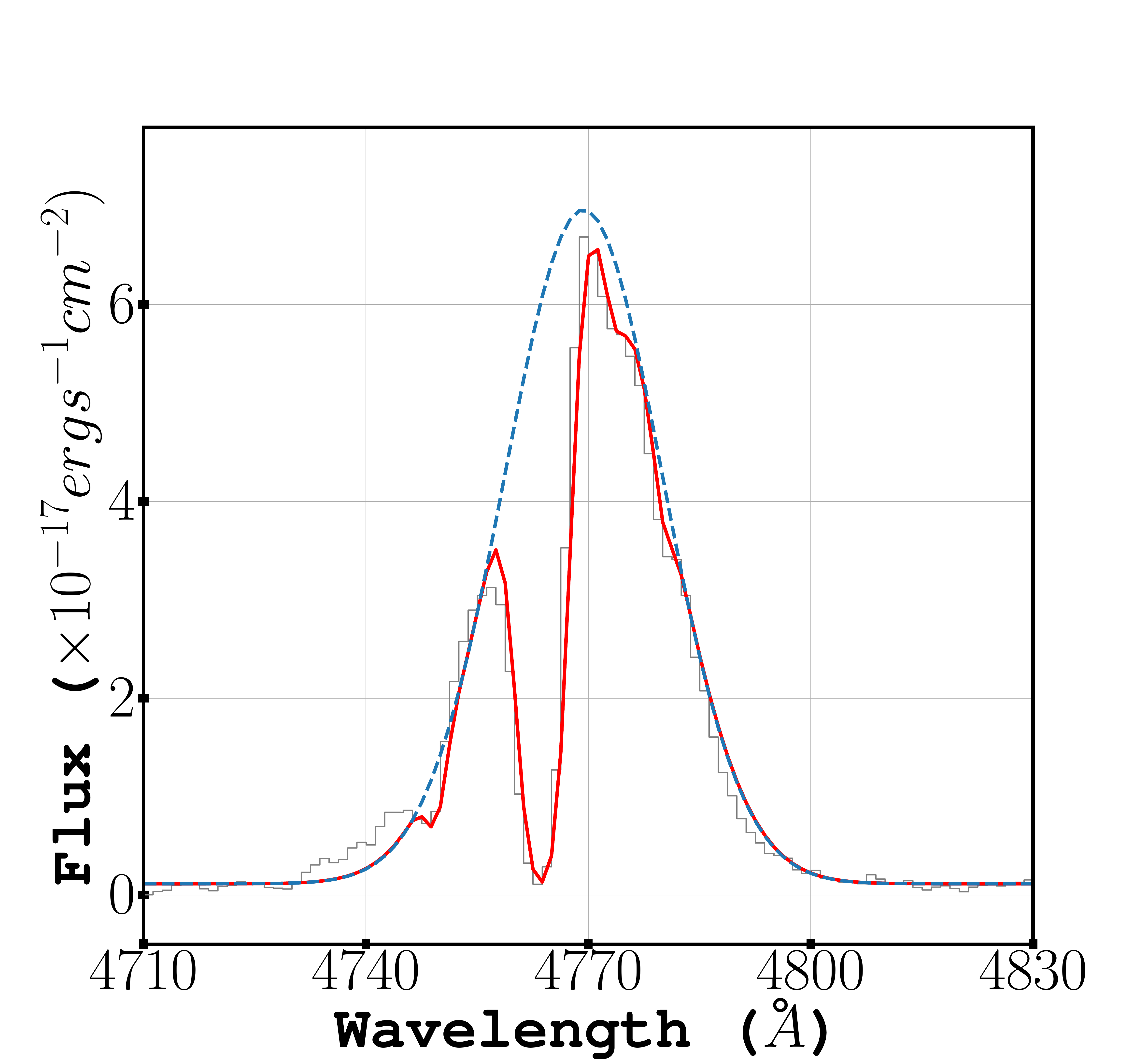}
			\label{ly-muse}}
		\quad
		\subfloat[VLT UVES]{
			\includegraphics[width=\columnwidth,height=4.5cm,keepaspectratio]{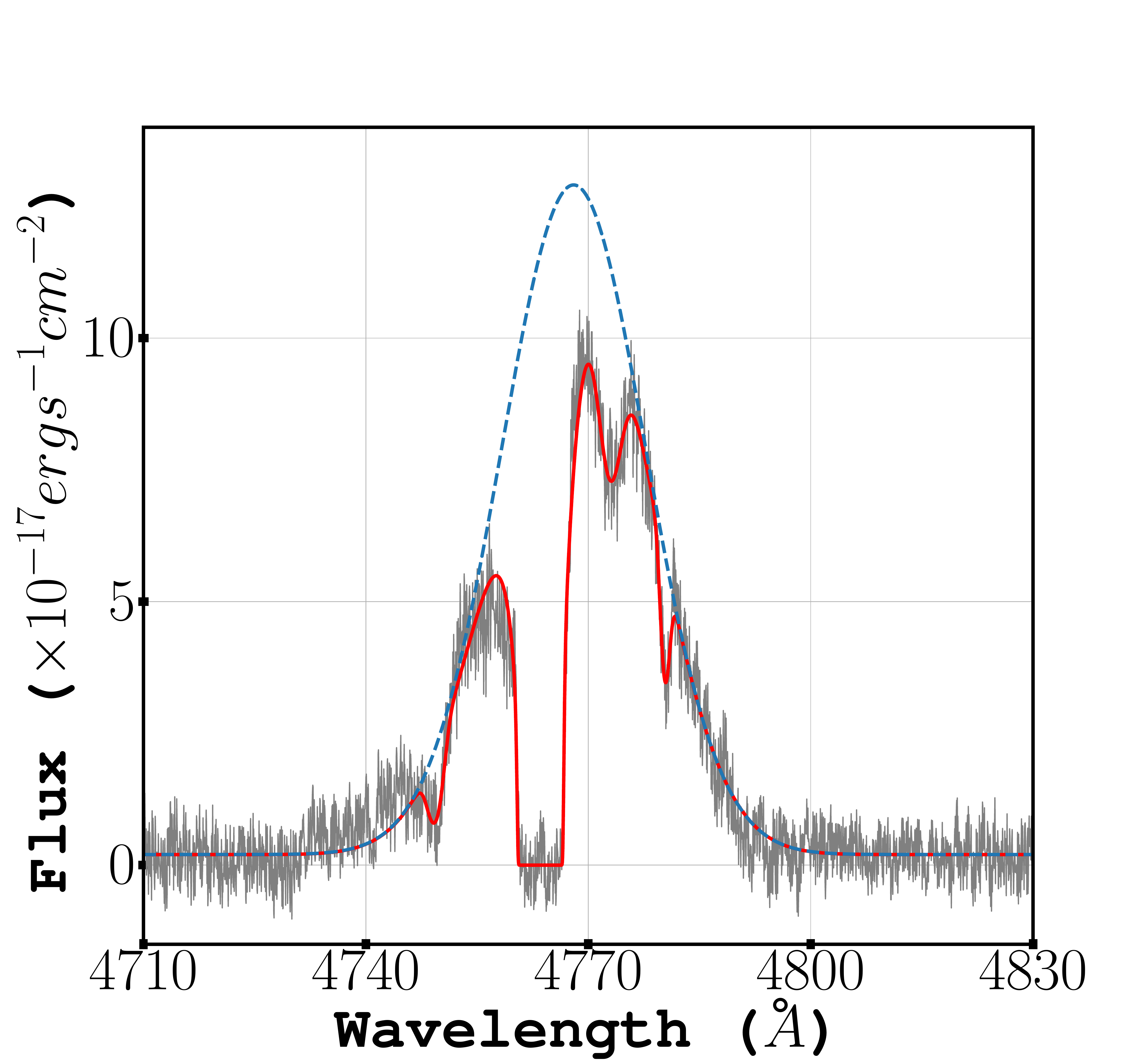}
			\includegraphics[width=\columnwidth,height=4.5cm,keepaspectratio]{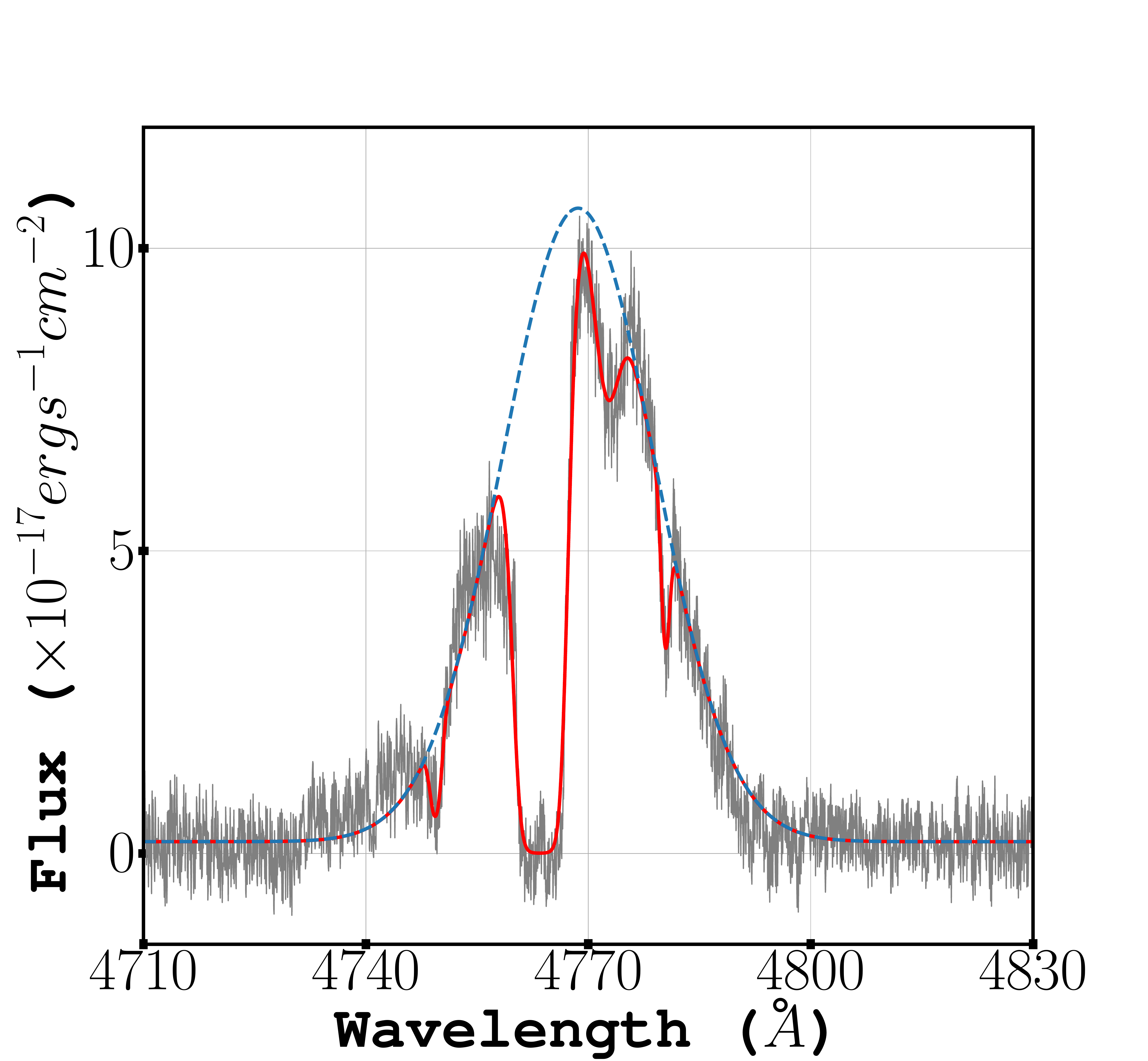}
			\label{ly-uves}}
		\quad
		\subfloat[AAT]{	
			\includegraphics[width=\columnwidth,height=4.5cm,keepaspectratio]{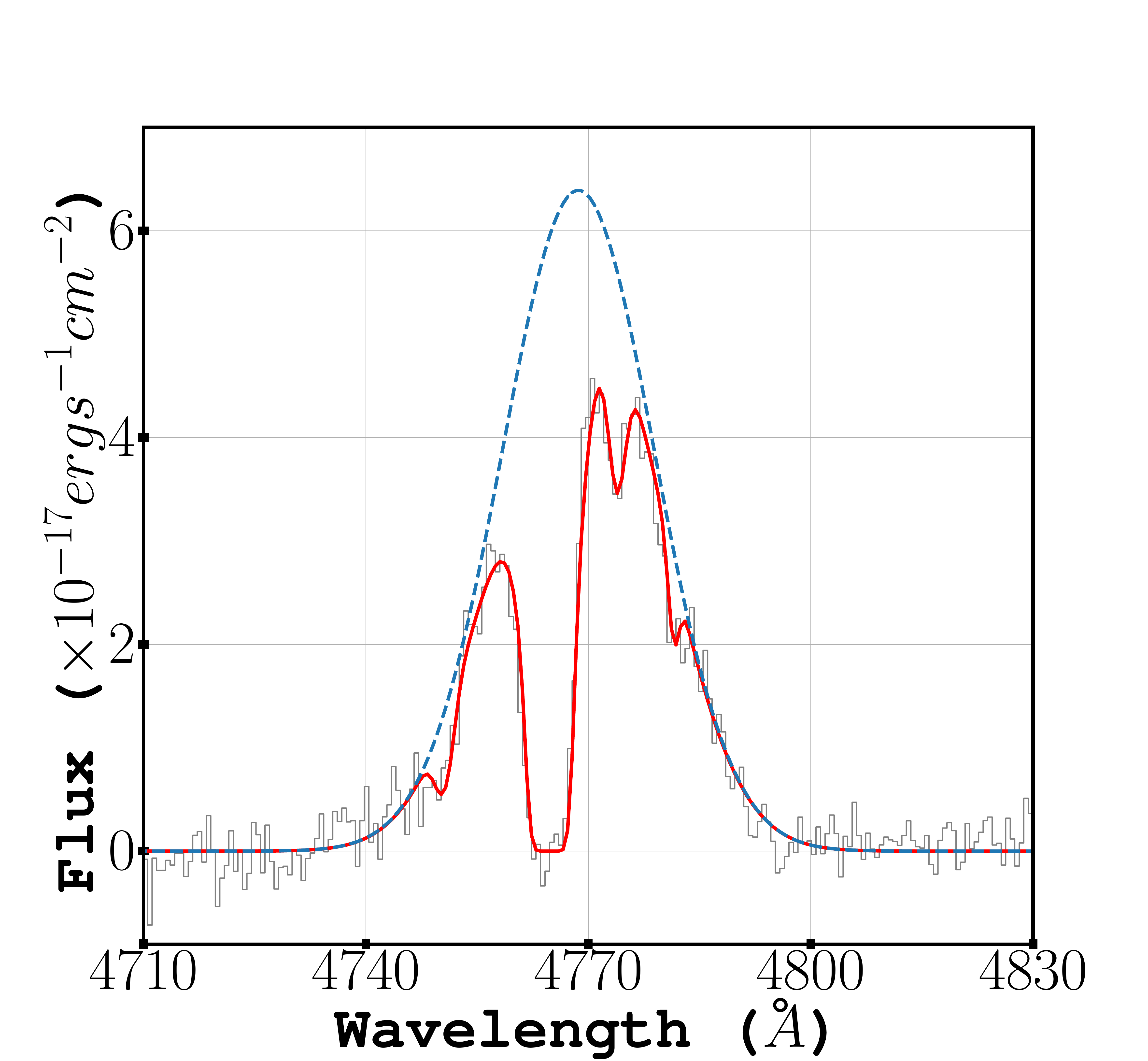}
			\includegraphics[width=\columnwidth,height=4.5cm,keepaspectratio]{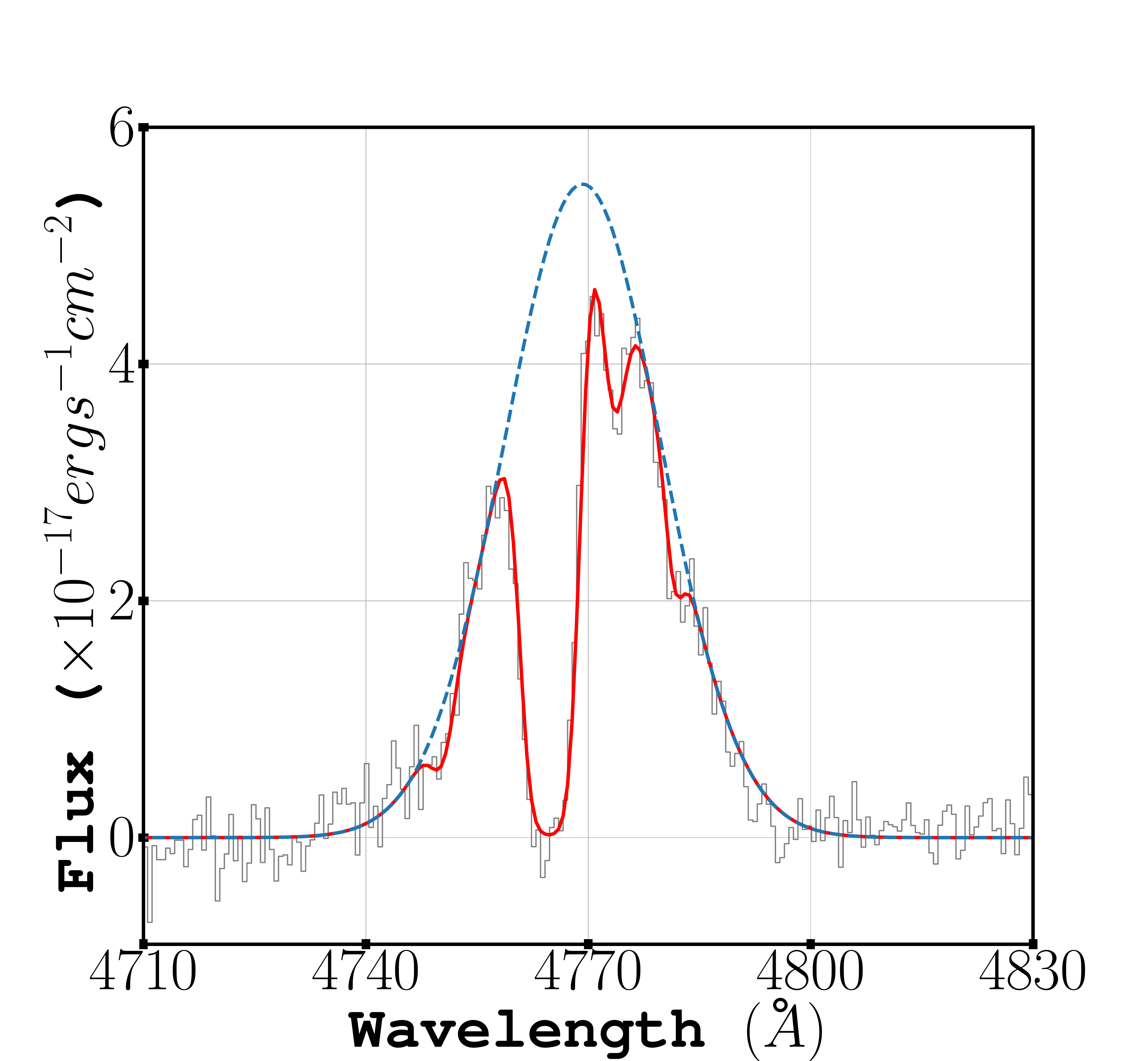}
			\label{ly-aat}}
	\caption{Ly$\alpha$ profile of MRC 0943--242 extracted from different instruments, with the Gaussian emission component (dashed blue line) plus absorption model overlaid (red line). The left and right columns show the high and low column density best-fitting, respectively. The Ly$\alpha$ profile from the MUSE IFU was extracted using a circular aperture of 0.8\arcsec.}
	\label{ly_ap}
\end{figure*}

\begin{figure*}
	\subfloat[VLT MUSE]{
		\includegraphics[width=5.2cm,height=5.2cm]{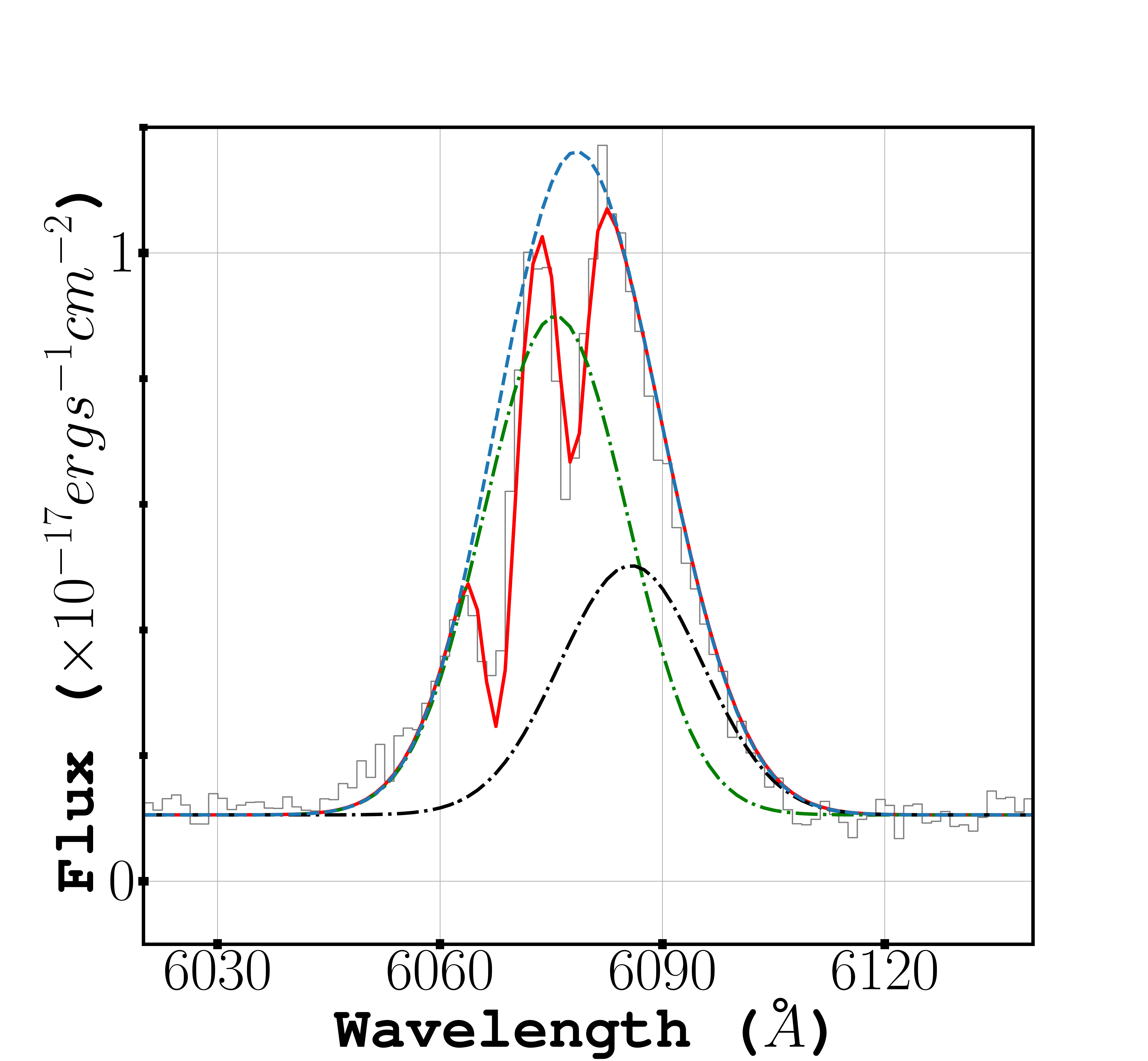}
		\label{muse}}
	\subfloat[VLT UVES]{
		\includegraphics[width=5.2cm,height=5.2cm]{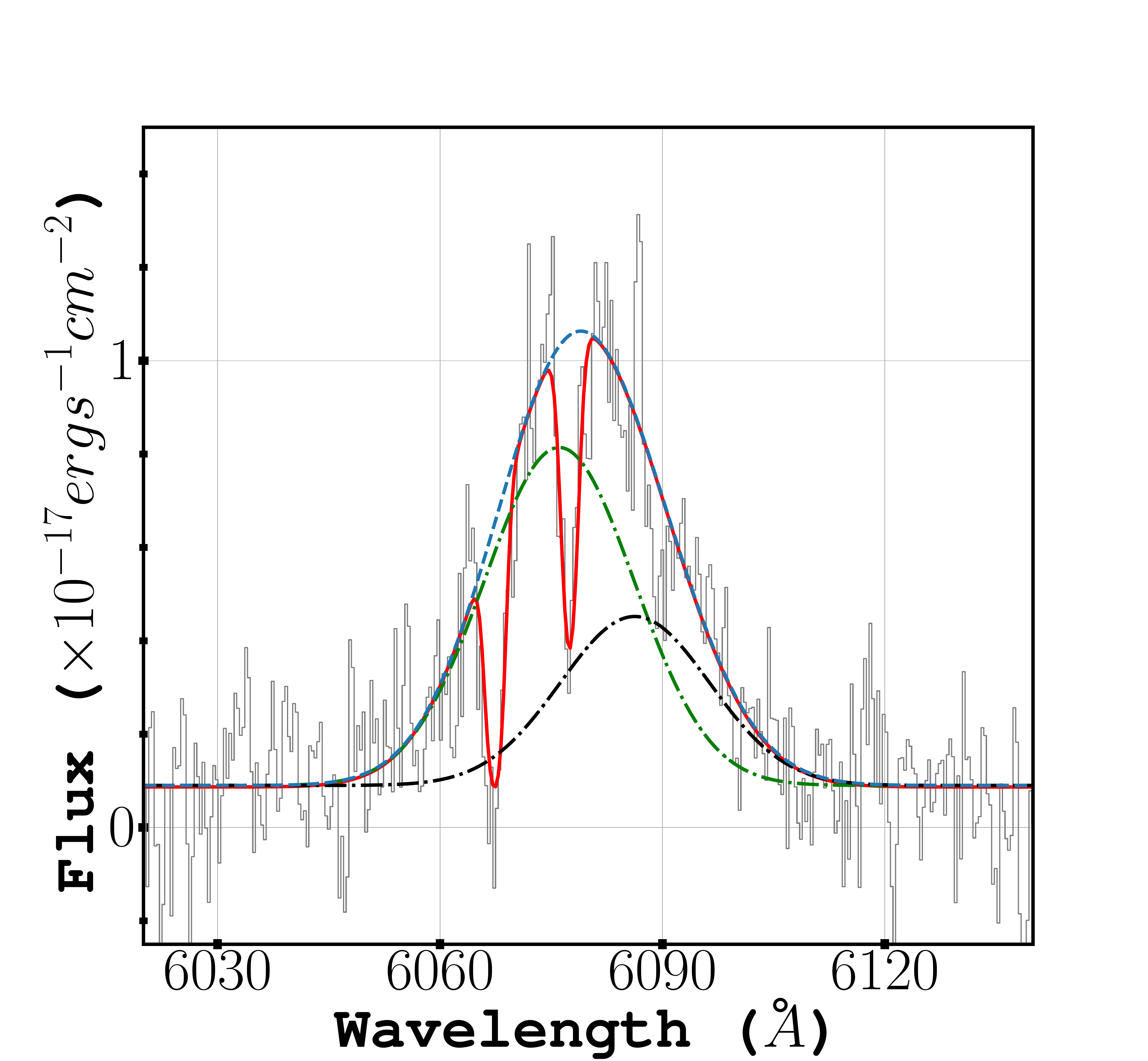}
		\label{uves}}
	\quad
	\subfloat[AAT]{
		\includegraphics[width=5.2cm,height=5.2cm]{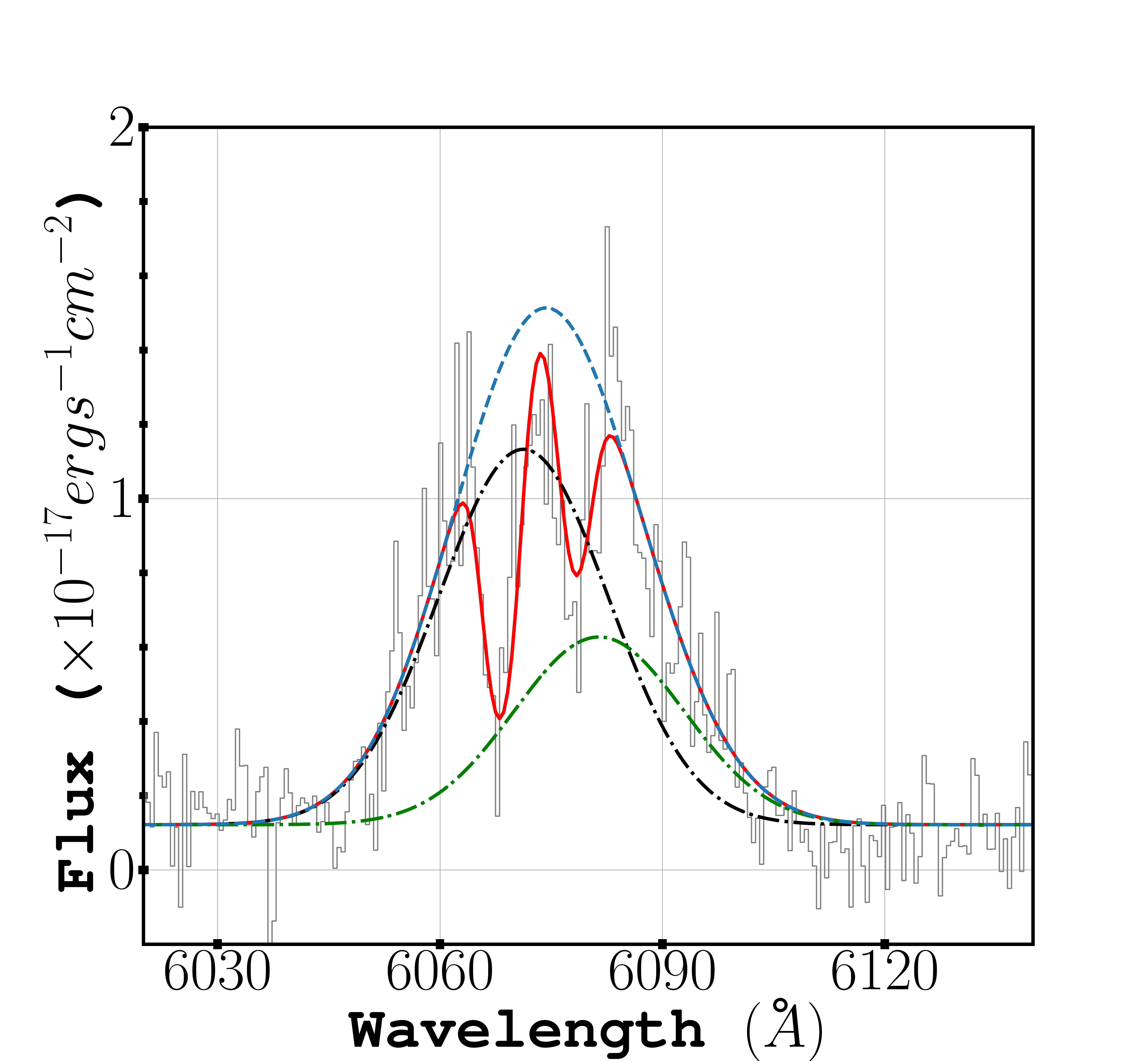}
		\label{aat}}
	\caption{The \ion{C}{IV} profile of MRC 0943--242, with the Gaussian emission component (dashed blue line) and emission plus absorption model overlaid (red line). The two individual doublet components are also shown.}
	\label{civ-ap}
\end{figure*}

\begin{table*}
	\centering
	\caption{Measurements of the rest-frame UV and optical emission lines obtained with the fitting routine. Ly$\alpha _{15}$ and Ly$\alpha _{19}$ correspond to the lower (log N(\ion{H}{I}/cm$^{-2}$) $\sim$ 15.20) and higher (log N(\ion{H}{I}/cm$^{-2}$) $\sim$ 19.63) column density results, respectively.}
	\label{instru_ap}
	\begin{tabular}{lccccr} 
		\hline
		Line & $\lambda _{rest}$ & $\lambda _{obs}$ & Line Flux & FWHM  &  $\Delta$v \\
		& \AA		      &		\AA		     & ($\times10^{-16}$ erg cm$^{-2}$ s$^{-1}$) & (km s$^{-1}$)  &  (km s$^{-1}$) \\
		\hline
		\multicolumn{5}{c}{VLT UVES} \\
		\hline
		Ly$\alpha _{15}$ & 1215.7 & 4768.6 $\pm$ 0.1 & 26.95 $\pm$ 0.33  &  1516 $\pm$ 9  & -161 $\pm$ 3 \\
		Ly$\alpha _{19}$ & 1215.7 & 4768.0 $\pm$ 0.1 & 31.03 $\pm$ 0.43  &  1439 $\pm$ 8  & -199 $\pm$ 3 \\
		\ion{C}{IV} & 1548.2,1550.8 & 6076.1 $\pm$ 0.2, 6086.3 $\pm$ 0.1 & 2.76 $\pm$ 0.08 & 1173 $\pm$ 29  &  -7 $\pm$ 11 \\
		\hline
		\multicolumn{5}{c}{AAT} \\
		\hline
		Ly$\alpha _{15}$ & 1215.7 & 4769.2 $\pm$ 0.2 & 14.36 $\pm$ 0.65  &  1529 $\pm$ 35  & -11 $\pm$ 10 \\
		Ly$\alpha _{19}$ & 1215.7 & 4768.7 $\pm$ 0.2 & 16.46 $\pm$ 0.80  &  1486 $\pm$ 31  & -45 $\pm$ 10 \\
		\ion{C}{IV} & 1548.2,1550.8 & 6071.3 $\pm$ 0.4, 6081.5 $\pm$ 0.1 & 4.36 $\pm$ 0.34 & 1331 $\pm$ 66  &  -136 $\pm$ 18 \\
		\ion{He}{II} & 1640.4 & 6434.2 $\pm$ 0.6 & 1.41 $\pm$ 0.14 & 885 $\pm$ 69 & 0 $\pm$ 27 \\
		\hline
		\multicolumn{5}{c}{KECK II} \\
		\hline
		Ly$\alpha$ & 1215.7 & 4773.1 $\pm$ 0.2 & 19.76 $\pm$ 0.47  &  1695 $\pm$ 19  & 156 $\pm$ 9 \\
		\ion{N}{V}	  & 1238.8, 1242.8	& 4864.1 $\pm$ 0.5, 4879.7 $\pm$ 0.5 &  0.43 $\pm$ 0.05 & 1318 $\pm$ 125 & 162 $\pm$ 29 \\
		\ion{N}{IV]}  & 1483.3, 1486.5	& 5820.1 $\pm$ 1.8, 5832.6 $\pm$ 1.8 & 0.11 $\pm$ 0.04 & 757 $\pm$ 193 & -44 $\pm$ 93 \\
		\ion{C}{IV} & 1548.2,1550.8 & 6076.6 $\pm$ 0.2, 6086.8 $\pm$ 0.2 & 2.75 $\pm$ 0.07 & 1183 $\pm$ 18  &  51 $\pm$ 12 \\
		\ion{He}{II} & 1640.4 & 6435.8 $\pm$ 0.2 & 1.94 $\pm$ 0.05 & 1098 $\pm$ 20 & 0 $\pm$ 8 \\
		\ion{O}{III]}  & 1660.8, 1666.1	& 6512.8 $\pm$ 1.3, 6533.7 $\pm$ 1.3 & 0.30 $\pm$ 0.05 & 1320 $\pm$ 203 & -215 $\pm$ 62 \\
		\ion{C}{III]}  & 1906.7, 1908.7	& 7480.9 $\pm$ 0.3, 7488.9 $\pm$ 0.3 & 1.16 $\pm$ 0.04 & 1008 $\pm$ 32 & -58 $\pm$ 11 \\
		\hline		
	\end{tabular}
\end{table*}

\begin{table*}
	\centering
	\caption{Best fit parameters for the Ly$\alpha$ absorption features, for different instruments. Column (1) gives the redshift for the Ly$\alpha$ emission Gaussian. Column (2) gives the redshift for each Ly$\alpha$ absorption. Column (3) gives the column density (N({\ion{H}{I}})). Column (4) gives the Doppler width b. Column (5) gives the velocity shift of the main absorber with respect to \ion{He}{II} emission in the same spectrum. Note: The \ion{He}{II} emission line was outside the spectral range covered by the red arm of VLT UVES and thus we do not give the velocity shift for this instrument.}
	\label{instru02_ap}
	\begin{tabular}{lcccr} 
		\hline
		Ly$\alpha$ emission redshift & Absorption redshift & Column Density & Doppler b Parameter &  $\Delta$v \\
		($z_{em}$)  		    &      ($z_{abs}$)	     &	  (cm$^{-2}$)      & (km s$^{-1}$) & (km s$^{-1})$  \\
		\hline	
		\multicolumn{5}{c}{VLT UVES} \\
		\hline
		& 2.90674 $\pm$ 0.00006 & (1.02 $\pm$ 0.09)$\times10^{14}$ & 81 $\pm$ 6  &     \\
		2.92213 $\pm$ 0.00004 & 2.91845 $\pm$ 0.00001 & (4.29 $\pm$ 0.11)$\times10^{19}$ & 54 $\pm$ 1  &     \\
		& 2.92618 $\pm$ 0.00004 & (5.80 $\pm$ 0.32)$\times10^{13}$ & 128 $\pm$ 5  &     \\
		& 2.93229 $\pm$ 0.00003 & (3.33 $\pm$ 0.20)$\times10^{13}$ & 50 $\pm$ 3  &     \\
		\hline	
		& 2.90682 $\pm$ 0.00005 & (8.65 $\pm$ 0.89)$\times10^{13}$ & 56 $\pm$ 5  &     \\
		2.92263 $\pm$ 0.00004 & 2.91840 $\pm$ 0.00001 & (1.58 $\pm$ 0.07)$\times10^{15}$ & 158 $\pm$ 2  &     \\
		& 2.92581 $\pm$ 0.00005 & (4.99 $\pm$ 0.38)$\times10^{13}$ & 138 $\pm$ 8  &     \\
		& 2.93232 $\pm$ 0.00003 & (2.94 $\pm$ 0.20)$\times10^{13}$ & 45 $\pm$ 3  &     \\
		\hline
		\multicolumn{5}{c}{AAT} \\
		\hline
		& 2.90776 $\pm$ 0.00683 & (9.43 $\pm$ 2.83)$\times10^{13}$ & 82 $\pm$ 30  &     \\
		2.92268 $\pm$ 0.00014 & 2.91947 $\pm$ 0.00003 & (5.11 $\pm$ 0.47)$\times10^{19}$ & 52 $\pm$ 2  &  -290 $\pm$ 3   \\
		& 2.92696 $\pm$ 0.00013 & (4.79 $\pm$ 0.80)$\times10^{13}$ & 100 $\pm$ 39  &     \\
		& 2.93321 $\pm$ 0.00018 & (2.85 $\pm$ 0.81)$\times10^{13}$ & 50 $\pm$ 26  &     \\
		\hline	
		& 2.90770 $\pm$ 0.00029 & (6.72 $\pm$ 2.66)$\times10^{13}$ & 65 $\pm$ 33  &     \\
		2.92313 $\pm$ 0.00014 & 2.91944 $\pm$ 0.00005 & (1.36 $\pm$ 0.15)$\times10^{15}$ & 185 $\pm$ 8  &  -293 $\pm$ 4   \\
		& 2.92660 $\pm$ 0.00014 & (6.16 $\pm$ 1.30)$\times10^{13}$ & 128 $\pm$ 22  &     \\
		& 2.93328 $\pm$ 0.00020 & (2.81 $\pm$ 0.88)$\times10^{13}$ & 56 $\pm$ 3  &     \\
		\hline
	\end{tabular}
\end{table*}

\begin{table*}
	\centering
	\caption{Best fit parameters for the \ion{C}{IV} absorption features, for different instruments. Column (1) gives the redshift for the \ion{C}{IV} emission. Column (2) gives the redshift for each \ion{C}{IV} absorption. Column (3) gives the column density (N(\ion{C}{IV})). Column (4) gives the Doppler width b. Column (5) gives the velocity shift of the main absorber with respect to \ion{He}{II} emission in the same spectrum. Note: The \ion{He}{II} emission line was outside the spectral range covered by the red arm of VLT UVES and thus we do not give the velocity shift for this instrument.}
	\label{instru03_ap}
	\begin{tabular}{lcccr} 
		\hline
		CIV emission redshift & Absorption redshift & Column Density & Doppler Parameter &  $\Delta$v \\
		($z_{em}$)  		    &      ($z_{abs}$)	     &	  (cm$^{-2}$)      & (km s$^{-1}$) & (km s$^{-1}$)  \\
		\hline
		\multicolumn{5}{c}{VLT UVES} \\
		\hline
		2.92459 $\pm$ 0.00015 & 2.91899 $\pm$ 0.00001 & (3.04 $\pm$ 0.24)$\times10^{14}$ & 67 $\pm$ 5  &  \\ 
		\hline
		\multicolumn{5}{c}{AAT} \\
		\hline
		2.92149 $\pm$ 0.00008 & 2.91950 $\pm$ 0.00003 & (3.89 $\pm$ 0.54)$\times10^{14}$ & 140 $\pm$ 15  &  -288 $\pm$ 8   \\
		\hline
	\end{tabular}
\end{table*}

\begin{figure*}
	\includegraphics[width=8.0cm,height=10.0cm,keepaspectratio]{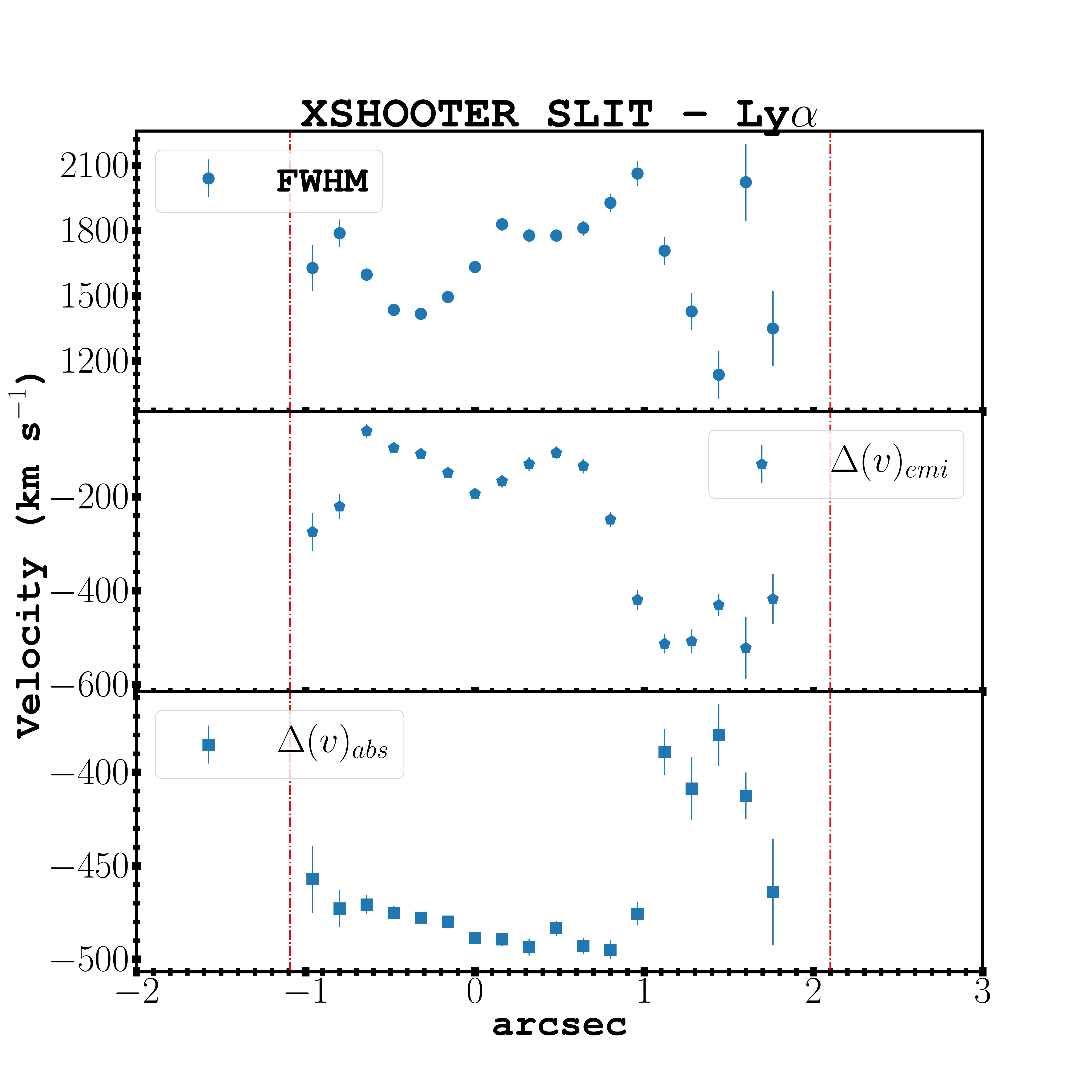}
	\includegraphics[width=8.0cm,height=10.0cm,keepaspectratio]{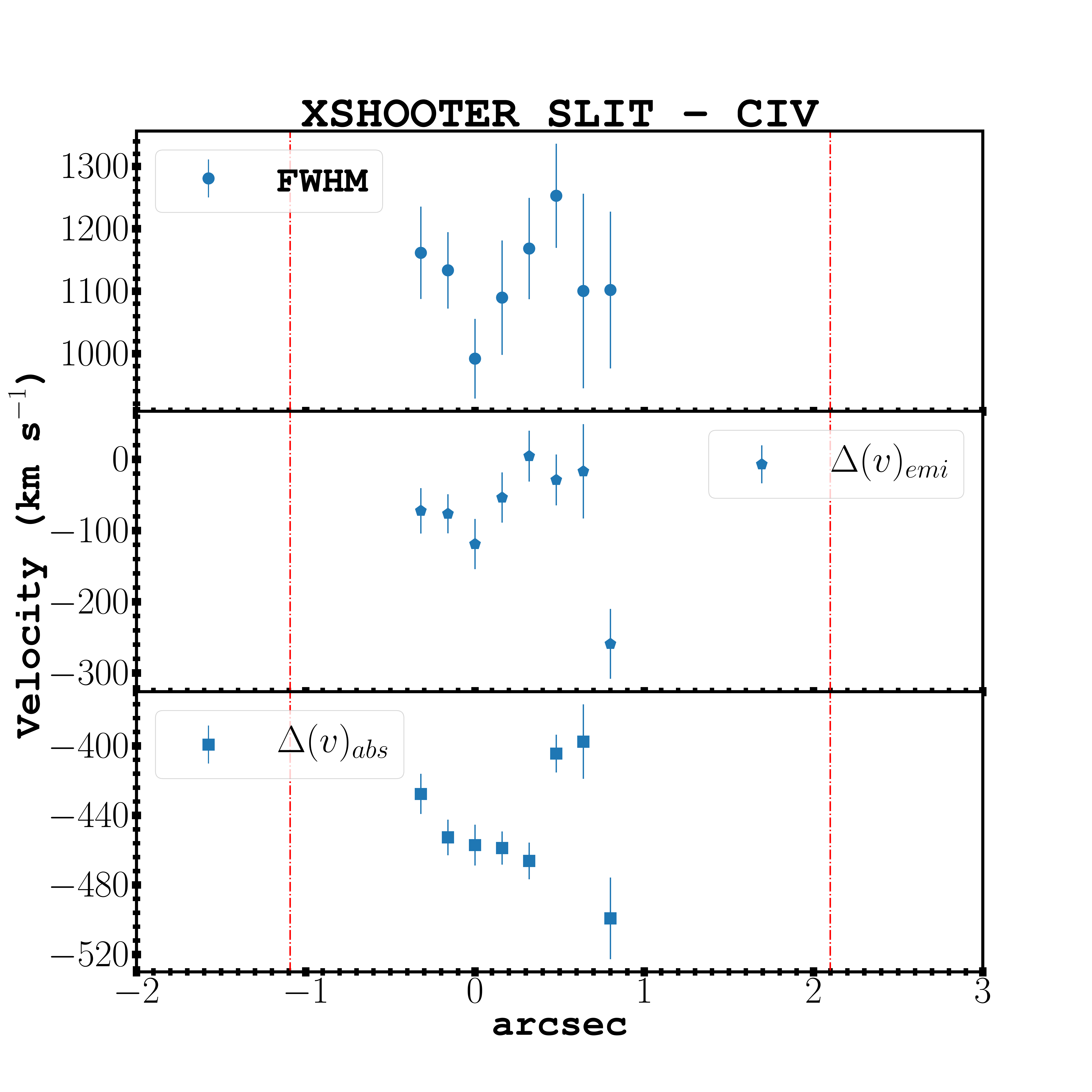}
	\caption{Kinematic properties of the MRC 0943--242 radio galaxy extracted from the XSHOOTER SLIT data. We present the velocity dispersion for the emission-lines and velocity offset for the emission and absorption lines as a function of the position along the slit in arc seconds. The red dashed-dot lines represent the positions of the radio hotspots.}
	\label{1dim_kinXSH}
\end{figure*}

\begin{figure*}
	\includegraphics[width=\columnwidth,height=6.0cm]{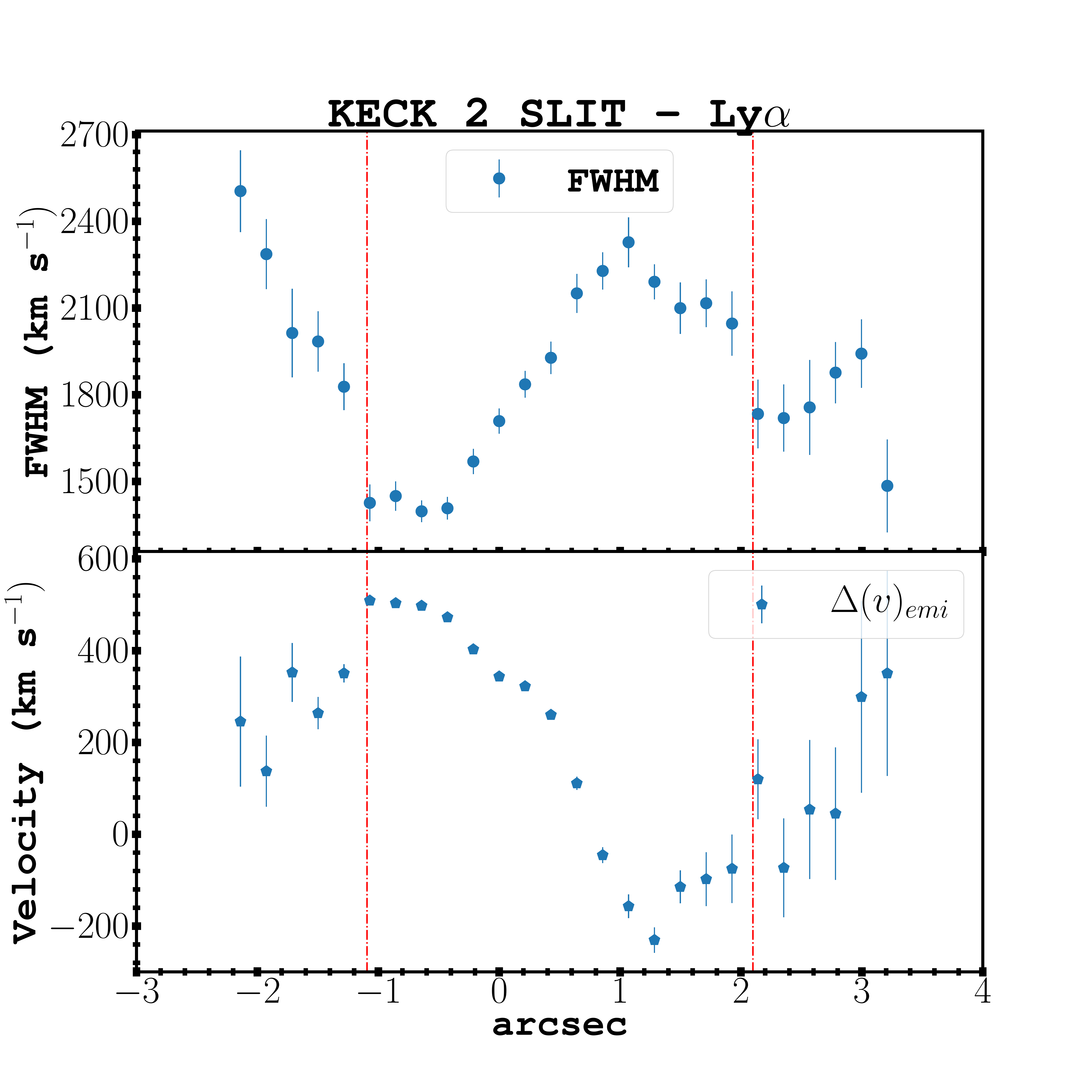}
	\includegraphics[width=\columnwidth,height=6.0cm]{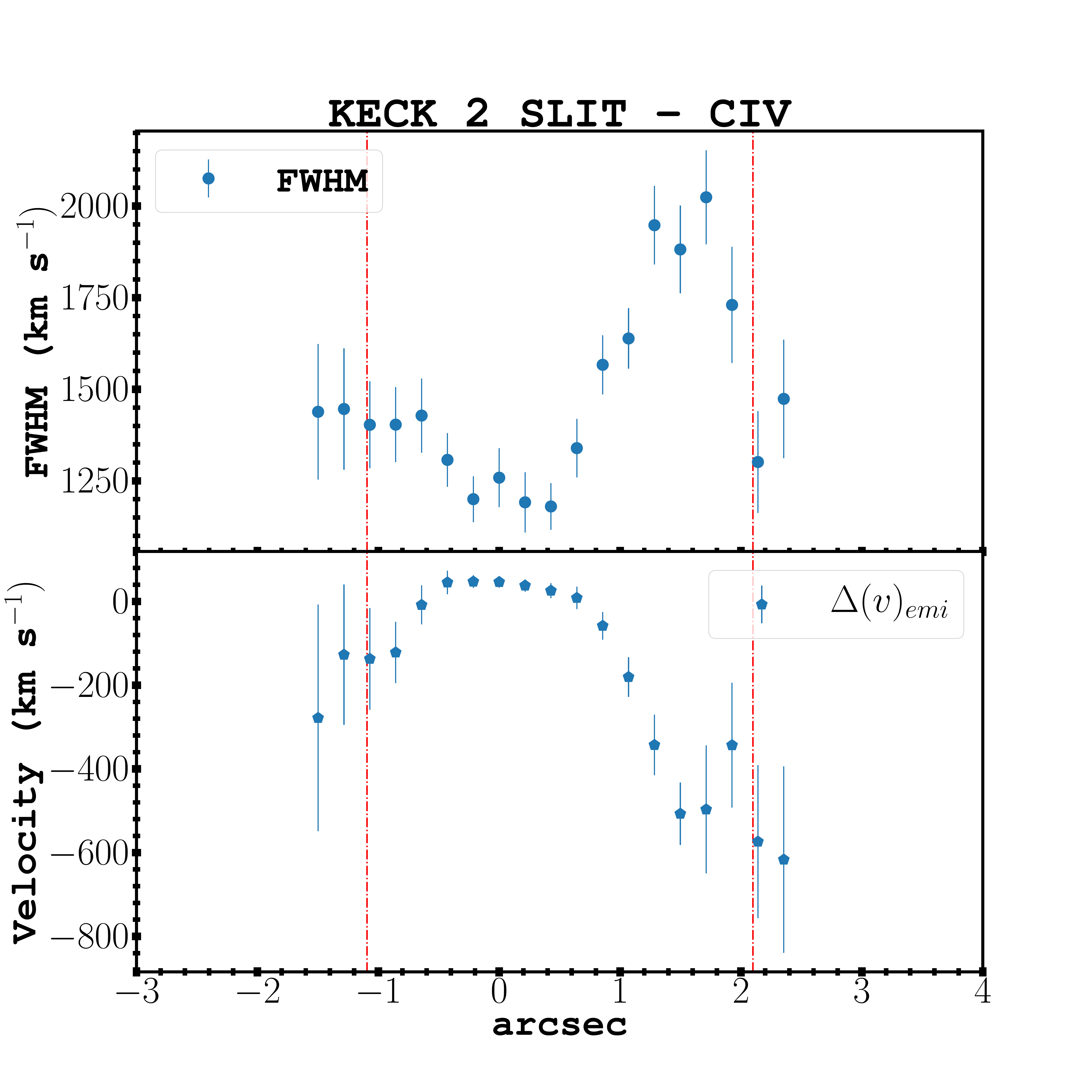}
	\includegraphics[width=\columnwidth,height=6.0cm]{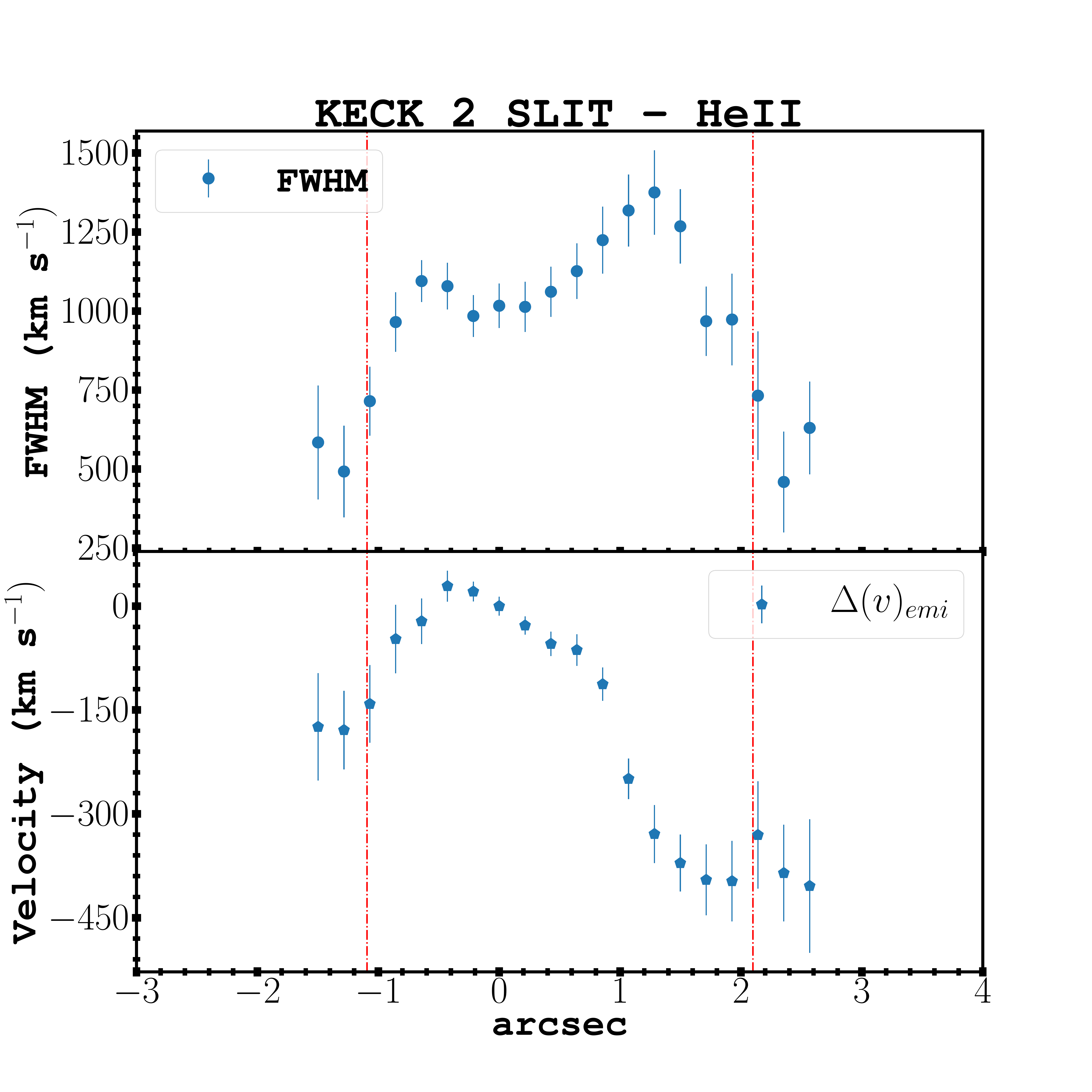}
	\caption{Kinematic properties of the MRC 0943--242 radio galaxy extracted from KECK II long-slit. The red dashed-dot lines represent the positions of the radio hotspots.}
	\label{1dim_kinKECK}
\end{figure*}

\begin{figure*}
	\includegraphics[width=\columnwidth,keepaspectratio]{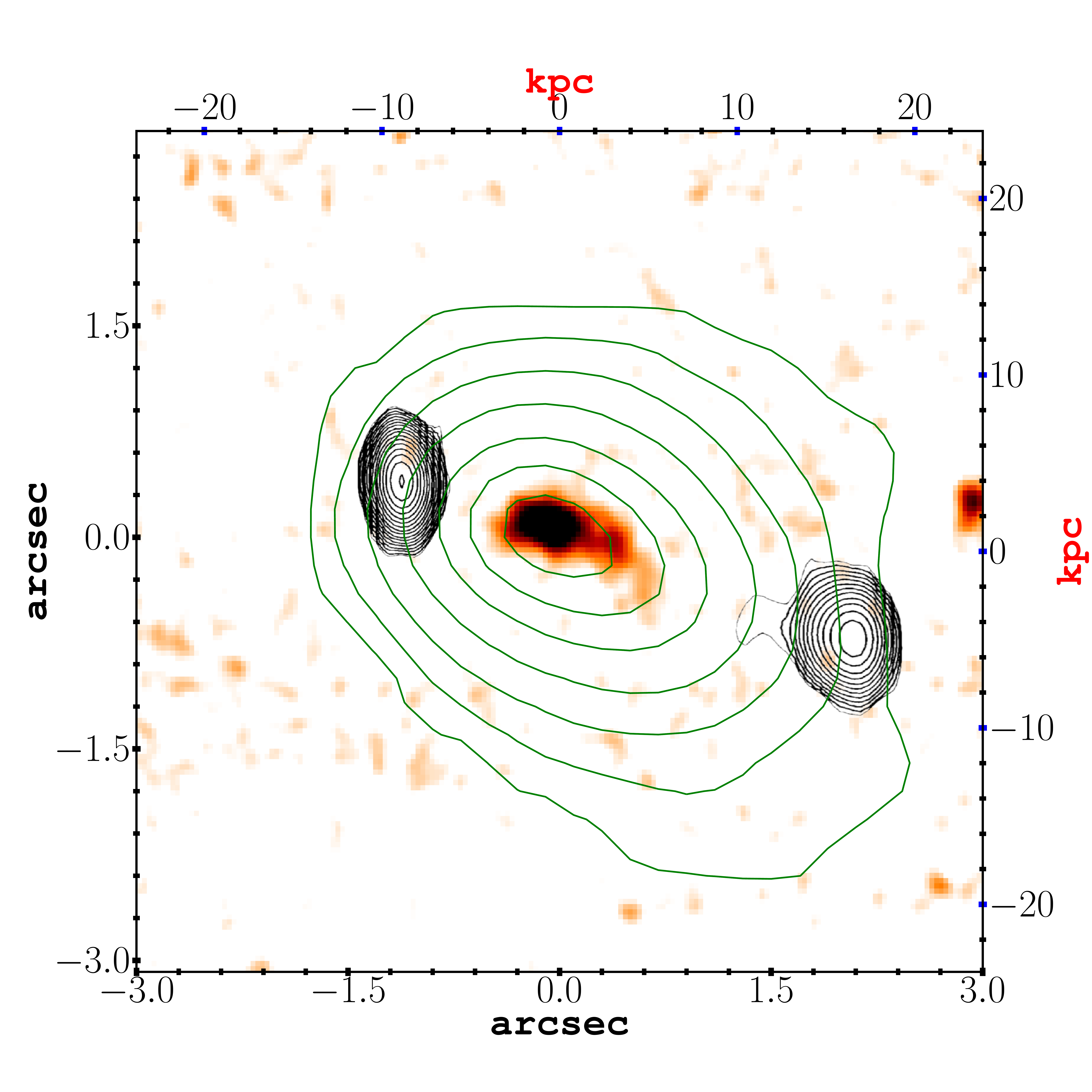}
	\caption{A HST NICMOS continuum image of MRC 0943--242 at $z$ = 2.92 with VLA radio contours (black) and Ly$\alpha$ contours (green; see Fig. \ref{kin-ly}) superimposed.}
	\label{hst0943}
\end{figure*}

\begin{figure*}
	\includegraphics[width=\columnwidth,keepaspectratio]{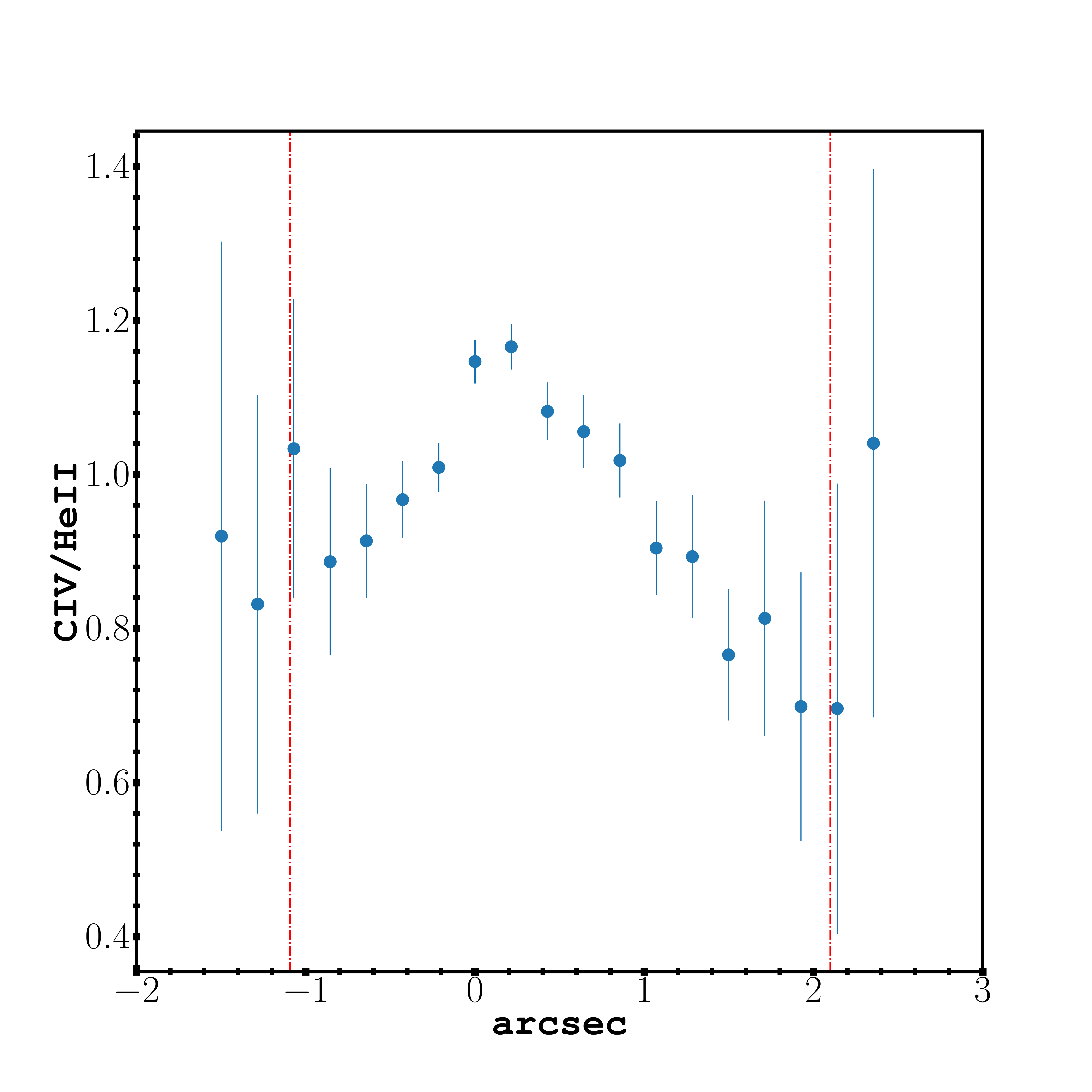}
	\caption{Flux ratio of the MRC 0943--242 radio galaxy extracted from the KECK II slit data. The red dashed-dot lines represent the positions of the radio hotspots.}
	\label{lnr_plot}
\end{figure*}

\begin{table*}
	\centering
	\caption{Comparison of model line ratios with observed line ratios. (1) Emission line ratios. (2) Observed X-SHOOTER line fluxes normalised by \ion{He}{II} $\lambda$1640 using a larger (2.1\arcsec) aperture. (3) Parameters and relative line fluxes produced by our best-fitting MAPPINGS model. (4) Our best-fitting model using $\alpha\,=$ --1.5 instead of $\alpha\,=$ --1.0. Parameters and relative line fluxes produced by our best-fitting shock models (5) and shock + precursor models (6) extracted from \citet{allen2008}.}
	\label{lnr-chiAp}
	\begin{tabular}{lccccr}
		\hline
		Line ratios & Obs. flux & Model 01 & Model 02 &  Model 03 & Model 04 \\ 
		(1)		&		(2)		&		(3)	&		(4)	 &     (5)   &   (6)	\\
		\hline
		~       &     ~      & $U$ = 0.019  & $U$ = 0.035  &   shocks  & shock + prec. \\ 
		~       &     ~      & $\alpha\,= -1.0$  &  $\alpha\,= -1.5$  &  $v$ = 200 km s$^{-1}$  &  $v$ = 750 km s$^{-1}$    \\ 
		~       &     ~      & $Z/Z_{\odot}\,= 1.8$  &  $Z/Z_{\odot}\,= 1.1$  &  $Z/Z_{\odot}\,= 1.0$ & $Z/Z_{\odot}\,= 1.0$ \\ 
		~       &     ~      &  $\chi^{2} _{\nu}\,= 3.11$  &   $\chi^{2} _{\nu}\,= 6.44$  & $\chi^{2} _{\nu}\,= 7.41$ & $\chi^{2} _{\nu}\,= 8.24$ \\
		\hline
		Ly$\alpha$ $^{(*)}$/HeII 	& 7.88 $\pm$ 0.26   & 10.53  & 16.47 & 58.65 & 29.06 \\
		(OVI$+$CII)/HeII        	& 0.87 $\pm$ 0.10   & 0.48  & 0.31 & 3.70  & 4.51  \\  
		NV/HeII         			& 0.70 $\pm$ 0.03   & 0.47  & 0.23 & 3.78  & 0.33   \\ 
		CIV/HeII        			& 1.21 $\pm$ 0.11   & 2.47  & 2.25 & 8.60  & 2.22  \\ 
		CIII]/HeII      			& 0.55 $\pm$ 0.02   & 1.34  & 1.10 & 1.62  & 1.21  \\ 
		CII]/HeII       			& 0.24 $\pm$ 0.01   & 0.16  & 0.08 & 1.14  & 0.64   \\ 
		$[$NeIV$]$/HeII     		& 0.21 $\pm$ 0.02   & 0.45  & 0.31 & 1.22  & 0.28  \\ 
		MgII/HeII       			& 0.42 $\pm$ 0.03   & 0.48  & 0.45 & 2.16 & 2.01  \\ 
		$[$NeV$]$/HeII      		& 0.66 $\pm$ 0.04   & 0.63  & 0.45 & 0.53 & 0.34  \\ 
		$[$OII$]$/HeII      		& 2.84 $\pm$ 0.08   & 0.84  & 0.44 & 3.25 & 3.02  \\ 
		$[$NeIII$]$/HeII    		& 0.99 $\pm$ 0.04   & 0.71  & 0.85 & 0.81 & 1.20 \\ 
		H$\gamma$/HeII  			& 0.38 $\pm$ 0.04   & 0.17  & 0.31 & 0.83 & 0.22  \\ 
		$[$OIII$]$4363/HeII 		& 0.27 $\pm$ 0.02   & 0.13  & 0.10 & 0.23 & 0.13  \\ 
		$[$OIII$]$5007/HeII 		& 7.97 $\pm$ 0.21  	& 8.94  & 8.65 & 2.75 & 8.02  \\ 
		\hline
		\multicolumn{6}{c}{$^{(*)}$ Ly$\alpha$ was not used in the fitting.} \\
		\hline
	\end{tabular}
\end{table*}


\bsp	
\label{lastpage}
\end{document}